\RequirePackage{ifpdf}
\ifpdf 
\documentclass[pdftex]{sigma}
\else
\documentclass{sigma}
\fi

\newcommand{\La}[1]{L^{({#1})}}

\newcommand{\B}[2]{B^{({#1})}_{#2}}

\newcommand{\T}[3]{T^{({#1})}_{{#2}}\big({#3}\big)}
\newcommand{\fun}[3]{{#1}^{({#2})}_{{#3}}}
\newcommand{\tu}[1]{\tau^{(#1)}}
\def\p{\partial}
\def\pq{\p_q^{-1}}
\def\t{\theta}

\numberwithin{equation}{section}

\begin{document}
\allowdisplaybreaks

\renewcommand{\PaperNumber}{060}

\FirstPageHeading

\ShortArticleName{$q$-Deformed KP Hierarchy and $q$-Deformed
Constrained KP Hierarchy}

\ArticleName{$\boldsymbol{q}$-Deformed KP Hierarchy\\ and
$\boldsymbol{q}$-Deformed Constrained KP Hierarchy}

\Author{Jingsong HE~$^{\dag}$$^{\ddag}$, Yinghua LI~$^{\dag}$ and
Yi CHENG~$^{\dag}$}

\AuthorNameForHeading{J.S.~He, Y.H.~Li and  Y.~Cheng}

\Address{$^\dag$~Department of  Mathematics, University of Science
and Technology of China, Hefei,\\
$\phantom{^\dag}$~230026 Anhui, P.R. China}
\EmailD{\href{mailto:jshe@ustc.edu.cn}{jshe@ustc.edu.cn},
\href{mailto:chengy@ustc.edu.cn}{chengy@ustc.edu.cn}}

\Address{$^\ddag$~Centre for Scientific Computing, University of
Warwick, Coventry CV4 7AL, United Kingdom}

\ArticleDates{Received January 27, 2006, in f\/inal form April 28,
2006; Published online June 13, 2006}

\Abstract{Using the determinant representation of gauge transformation
 operator, we have shown that the general form of  
 $\tau$ function of the $q$-KP hierarchy is a $q$-deformed generalized Wronskian,
 which includes  the $q$-deformed Wronskian
  as a special case. On the basis of these, we  study  the $q$-deformed
 constrained KP ($q$-cKP) hierarchy,
 i.e.\ $l$-constraints of $q$-KP hierarchy.
 Similar to the ordinary constrained KP (cKP) hierarchy, a large class of
 solutions of $q$-cKP hierarchy can be represented by $q$-deformed Wronskian
 determinant  of functions satisfying a set of linear $q$-partial dif\/ferential equations with constant
 coef\/f\/icients. We obtained additional conditions for these functions imposed by the
 constraints. In particular, the ef\/fects of
 $q$-deformation ($q$-ef\/fects) in single $q$-soliton from the simplest $\tau$
 function of the $q$-KP hierarchy and in  multi-$q$-soliton   from one-component $q$-cKP
 hierarchy, and their dependence of $x$ and $q$, were also presented.
 Finally, we observe
 that $q$-soliton   tends to the usual
 soliton of the KP equation when $x\rightarrow 0$ and $q\rightarrow  1$, simultaneously.}


\Classification{37K10; 35Q51; 35Q53; 35Q55}

\section{Introduction}

Study of the quantum calculus (or $q$-calculus) \cite{ks,kc} has a
long history, which may go  back to the beginning of the twentieth
century.  F.H.~Jackson was the f\/irst mathematician who  studied
the $q$-integral and $q$-derivative in a systematic way starting
about 1910 \cite{exton,andrews}\footnote{ Detailed notes on the
initial research of $q$-integral, $q$-derivative of Jackson and
wide applications of $q$-series are easily available in the
text.}. Since 1980's, the quantum calculus was re-discovered in
the research of quantum group inspired by the studies on quantum
integrable model that used the quantum inverse scattering method
\cite{jimbo} and  on noncommutative geometry \cite{connes}. In
particular,  S.~Majid derived  the $q$-derivative from the braided
dif\/ferential calculus \cite{majid1,majid2}.

The $q$-deformed integrable system (also called $q$-analogue or
$q$-deformation  of classical integrable system) is def\/ined by
means of $q$-derivative $\partial_q $ instead of usual derivative
$\partial $ with respect to~$x$  in a classical system. It reduces
to a classical integrable system as $q \rightarrow1$. Recently,
the $q$-deformation of the following three stereotypes for
integrable systems attracted more attention. The f\/irst type is
$q$-deformed $N$-th KdV ($q$-NKdV or $q$-Gelfand--Dickey
hierarchy) \cite{zhang,tl}, which is reduced to the $N$-th KdV
(NKdV or Gelfand--Dickey) hierarchy when $q\rightarrow 1$. The
$N$-th $q$-KdV hierarchy becomes $q$-KdV hierarchy for $N=2$. The
$q$-NKdV hierarchy  inherited several integrable structures from
classical $N$-th KdV hierarchy, such as inf\/inite conservation
laws \cite{wzz}, bi-Hamiltonian structure \cite{fr,frenkel},
$\tau$ function \cite{hl,ahv}, B\"acklund
transformation~\cite{tsl}. The second type is the $q$-KP hierarchy
\cite{klr,tu}. Its $\tau$ function, bi-Hamiltonian structure and
additional symmetries have already been reported in
\cite{ilievb,ilievc,ms,tu}. The third type is the $q$-AKNS-D
hierarchy, and its bilinear identity and $\tau$ function were
obtained in~\cite{wwwy}.

In order to get the Darboux--B\"acklund transformations, the two
elementary types of  gauge transformation operators,
dif\/ferential-type denoted by $T$ (or $T_D$) and integral-type
denoted by~$S$ (or $T_I$), for $q$-deformed $N$-th KdV hierarchy
were introduced in~\cite{tsl}. Tu et al.\  obtained not only the
$q$-deformed Wronskian-type but also binary-type representations
of $\tau$ function of $q$-KdV hierarchy. On the basis of their
results, He et al.~\cite{hlc2}  obtained the determinant
representation of gauge transformation operators $T_{n+k}$ $(n
\geq k)$ for $q$-Gelfand--Dickey hierarchy, which is a~mixed
iteration of $n$-steps of $T_D$ and then k-steps of $T_I$. Then,
they obtained a~more general form of $\tau$~function for $q$-KdV
hierarchy, i.e., generalized $q$-deformed Wronskian
($q$-Wronskian) $IW^q_{n+k}$~\cite{hlc2}. It is important to note
that for $k=0$ $IW^q_{n+k}$ reduces to $q$-deformed Wronskian and
for $k=n$ to binary-type determinant \cite{tsl}. On the other
hand, Tu introduced the $q$-deformed constrained KP ($q$-cKP)
hierarchy~\cite{tu} by means of symmetry constraint of $q$-KP
hierarchy, which is a $q$-analogue of constrained KP (cKP)
hierarchy \cite{kss, aratyn2}.

The purpose of this paper is to construct the  $\tau$ function of
$q$-KP and $q$-cKP hierarchy, and then explore the $q$-ef\/fect in
$q$-solitons. The main tool is the determinant representation of
gauge transformation operators \cite{csy,og,hlc,hlc1}. The paper
is organized as follows: In Section~2 we introduce some basic
facts on the  $q$-KP hierarchy, such as Lax operator, Z-S
equations, the existence of $\tau$ function. On the basis of the
\cite{tsl},  two kinds of elementary gauge transformation
operators\ for $q$-KP hierarchy and changing rule of $q$-KP
hierarchy  under it are presented in Section~3. In Section~4, we
establish the determinant representation of gauge transformation
operator $T_{n+k}$ for the $q$-KP hierarchy and then obtain the
general form of $\tau$ function $\tau^{(n+k)}_q=IW^q_{n+k}$. In
particular, by taking $n=1$, $k=0$ we will show $q$-ef\/fect of
single $q$-soliton solution of $q$-KP hierarchy. A brief
description of the sub-hierarchy of $q$-cKP hierarchy is presented
in Section~5, from the viewpoint of the symmetry constraint. In
Section~6, we show that the $q$-Wronskian  is one kind of forms of
$\tau$ function of $q$-cKP if the functions in the $q$-Wronskian
satisfy some restrictions. In Section~7 we consider an example
which illustrates the procedure reducing $q$-KP to $q$-cKP
hierarchy. $q$-ef\/fects of the $q$-deformed multi-soliton are
also discussed. The conclusions and discussions are given in
Section~8. Our presentation is similar to the relevant papers of
classical KP and cKP hierarchy \cite{csy,hlc,dkjm,dl1,os}.

At the end of this section, we shall collect some useful formulae
for reader's convenience.

The $q$-derivative $\partial_q$ is def\/ined by
\begin{gather*}
   \partial_q(f(x))=\frac{f(qx)-f(x)}{x(q-1)} 
   \end{gather*}
and the $q$-shift operator is given by
\begin{gather*}
  \theta(f(x))=f(qx).
  \end{gather*}
Let $\partial_q^{-1}$ denote the formal inverse of $\partial_q$.
We should note that $\theta $ does not commute with $\partial_q$,
\begin{gather*}
(\partial_q \theta^k(f))=q^k\theta^k(\partial_q f), \qquad k\in
\mathbb{Z}.
\end{gather*}
In general, the following $q$-deformed Leibnitz rule holds:
\begin{gather}\label{qleibniz}
     \partial_q^n \circ f=\sum_{k\ge0}\binom{n}{k}_q\theta^{n-k}(\partial_q^kf)\partial_q^{n-k},\qquad n\in
     \mathbb{Z},
     \end{gather}
where the $q$-number and the $q$-binomial are def\/ined by
\begin{gather*}
  (n)_q=\frac{q^n-1}{q-1},
\qquad
   \binom{n}{k}_q=\frac{(n)_q(n-1)_q\cdots(n-k+1)_q}{(1)_q(2)_q\cdots(k)_q},\qquad \binom{n}{0}_q=1,
\end{gather*}
and  ``$\circ $'' means composition of operators, def\/ined by
$\partial_q \circ f=(\partial_q \cdot f)+ \theta(f)\partial_q $.
In the remainder of the paper for any function $f$ ``$\cdot$'' is
def\/ined by $\partial_q \cdot f=\partial_q(f)\triangleq
(\partial_q f)$. For a $q$-pseudo-dif\/ferential operator
($q$-PDO) of the form
$P=\sum\limits_{i=-\infty}^np_i\partial_q^i$, we decompose $P$
into the dif\/ferential part $P_+=\sum\limits
_{i\ge0}p_i\partial_q^i$ and the integral
part$P_-=\sum\limits_{i\leq-1}p_i\partial_q^i$. The conjugate
operation ``$*$'' for $P$ is def\/ined by
$P^*=\sum\limits_i(\partial_q^*)^ip_i$ with
$\partial_q^*=-\partial_q\theta^{-1}=-\frac{1}{q}\partial_{\frac{1}{q}}$,
$(\partial_q^{-1})^*=(\partial_q^*)^{-1}=-\theta
\partial_q^{-1}$. We can write out  several explicit forms of (\ref{qleibniz}) for  $q$-derivative
$\partial_q$, as
\begin{gather}
\partial_q \circ f=(\partial_q f)+\theta(f)\partial_q,\label{qleibnizdifferentialfirst}\\
\partial_q^2 \circ f=(\partial_q^2 f)+ (q+1)\theta(\partial_q
f)\partial_q +\theta^2(f)\partial_q^2, \label{qleibnizdifferentialtwo}\\
 \partial_q^3\circ f=(\partial_q^3 f)+ (q^2 +q
+1)\theta(\partial_q^2f)\partial_q + (q^2 +q
+1)\theta^2(\partial_q f)\partial^2_q +
\theta^3(f)\partial_q^3,\label{qleibnizdifferentialthree}
\intertext{and  $\partial_q^{-1}$}
 \partial_q^{-1}\circ  f=
\theta^{-1}(f)\partial^{-1}_q-q^{-1} \theta^{-2}(\partial_q
f)\partial^{-2}_q+q^{-3} \theta^{-3}(\partial_q^2
f)\partial^{-3}_q-q^{-6} \theta^{-4}(\partial_q^3
f)\partial^{-4}_q \notag \\
\phantom{\partial_q^{-1}\circ  f=}{}
+\dfrac{1}{q^{10}}\theta^{-5}(\p_q^4 f)\p_q^{-5} +\cdots + (-1)^k
q^{-(1+2+3+\cdots+k)}
\theta^{-k-1}(\partial_q^k f)\partial_q^{-k-1}+\cdots,  \label{qleibnizminusonedifferentiala}\\
\partial_q^{-2} \circ f=
\theta^{-2}(f)\partial_q^{-2}-\dfrac{1}{q^2}(2)_q\theta^{-3}(\partial_q
f
)\partial_q^{-3}+\dfrac{1}{q^{(2+3)}}(3)_q\theta^{-4}(\partial_q^2
f )\partial_q^{-4} \notag \\
\phantom{\partial_q^{-2} \circ
f=}{}-\dfrac{1}{q^{(2+3+4)}}(4)_q\theta^{-5}(\partial_q^3 f
)\partial_q^{-5}+\cdots\nonumber\\
\phantom{\partial_q^{-2} \circ f=}{}
+\dfrac{(-1)^k}{q^{(2+3+\cdots+k+1)}}(k+1)_q\theta^{-2-k}(\partial_q^k
f )\partial_q^{-2-k}+\cdots.\label{qleibnizminustwodifferential}
\end{gather}
More explicit expressions of $\p_q^n\circ f$ are given in
Appendix~A. In particular, $\partial_q^{-1}\circ f$ has
dif\/ferent forms,
\begin{gather*}
\partial^{-1}_q \circ f= \theta^{-1}(f)\partial^{-1}_q+ \partial_q^{-1}\circ (\partial_q^*
f)\circ \partial^{-1}_q,\\ 
\partial_q^{-1}\circ  f\circ  \partial_q^{-1}=
(\partial_q^{-1}f)\partial^{-1}_q -\partial^{-1}_q \circ
\theta(\partial^{-1}_q f),
\end{gather*}
which will be used in the following sections.  The $q$-exponent
$e_q(x)$ is def\/ined as follows
\[
e_q(x)=\sum_{i=0}^{\infty}\dfrac{x^n}{(n)_q!}, \qquad (n)_q!=(n)_q
(n-1)_q (n-2)_q\cdots (1)_q.
\]
Its equivalent expression is of the form
\begin{gather}\label{q-exponent}
  e_q(x)=\exp\left(\sum_{k=1}^{\infty}\frac{(1-q)^k}{k(1-q^k)}x^k\right).
\end{gather}
The form (\ref{q-exponent}) will play a crucial role in  proving
the existence \cite{ilievb} of $\tau$ function of $q$-KP
hierarchy.

\section[$q$-KP hierarchy]{$\boldsymbol{q}$-KP hierarchy}

 Similarly to the general way of describing
the classical KP hierarchy \cite{dkjm,dl1}, we shall give a~brief
introduction of $q$-KP based on \cite{ilievb}. Let $L$ be one
$q$-PDO given by
\begin{gather}\label{qlaxoperator}
L=\partial_q+ u_0 +
u_{-1}\partial_q^{-1}+u_{-2}\partial_q^{-2}+\cdots,
\end{gather}
which is called Lax operator of $q$-KP hierarchy. There exist
inf\/inite $q$-partial dif\/ferential equations relating to
dynamical variables $\{u_i(x,t_1, t_2, t_3,\ldots),\, i=0,-1,-2,
-3, \ldots \}$, and they can be deduced from generalized Lax
equation,
\begin{gather}\label{qlaxequation}
\dfrac{\partial L}{\partial t_n}=[B_n, L], \qquad n=1, 2, 3,
\ldots,
\end{gather}
which are called $q$-KP hierarchy. Here
$B_n=(L^n)_+=\sum\limits_{i=0}^n b_i\partial_q^i$ means the
positive part of $q$-PDO, and we will use $L^n_-=L^n-L^n_+$ to
denote the negative part. By means of the formulae given in
(\ref{qleibnizdifferentialfirst})--(\ref{qleibnizminustwodifferential})
and in Appendices A and B, the f\/irst few f\/lows in
(\ref{qlaxequation}) for dynamical variables
$\{u_0,u_{-1},u_{-2},u_{-3}\}$ can be written out as follows. The
f\/irst f\/low is
\begin{gather*}
\p_{t_1}u_0=\theta(u_{-1})-u_{-1},\\ 
\p_{t_1}u_{-1}=(\p_q
u_{-1})+\theta(u_{-2})+u_0u_{-1}-u_{-2}-u_{-1}\theta^{-1}(u_0),\\ 
\p_{t_1}u_{-2}=(\p_q
u_{-2})+\theta(u_{-3})+u_0u_{-2}-u_{-3}-u_{-2}\theta^{-2}(u_0)+\dfrac{1}{q}u_{-1}\theta^{-2}(\p_q
u_0),\\ 
\p_{t_1}u_{-3}=(\p_q
u_{-3})+\theta(u_{-4})+u_0u_{-3}-u_{-4}-\dfrac{1}{q^3}u_{-1}\theta^{-3}(\p_q^2
u_0)\notag \\
\phantom{\p_{t_1}u_{-3}=}{}+\dfrac{1}{q^2}(2)_qu_{-2}\theta^{-3}(\p_q
u_0)-u_{-3}\theta^{-3}(u_0).
\end{gather*}
The second f\/low is
\begin{gather*}
\p_{t_2}u_0=\theta(\p_q
u_{-1})+\theta^2(u_{-2})+\theta(u_0)\theta(u_{-1})+u_0\theta(u_{-1}) \notag\\
\phantom{\p_{t_2}u_0=}{}-\big( (\p_q
u_{-1})+u_{-1}u_0+u_{-1}\theta^{-1}(u_0)+u_{-2} \big),\\
\p_{t_2}u_{-1}=q^{-1}u_{-1}\theta^{-2}(\p_q u_0)+u_{-1} (\p_q
u_0)+ (\p_q^2 u_{-1})+ \big(\theta(u_0)+u_0 \big)(\p_q u_{-1})
\notag \\
\phantom{\p_{t_2}u_{-1}=}{}+(q+1)\theta(\p_q
u_{-2})+\theta(u_0)\theta(u_{-2})+u_0\theta(u_{-2})+\theta(u_{-1})u_{-1}+u_0^2u_{-1}\notag\\
\phantom{\p_{t_2}u_{-1}=}{}-
u_{-1}\theta^{-1}(u_0^2)-u_{-1}\theta^{-1}(u_{-1})
-u_{-2}\theta^{-1}(u_0)-u_{-2}\theta^{-2}(u_0)+\theta^3(u_{-3})-u_{-3},\notag\\
\p_{t_2}u_{-2}=(\p_q^2 u_{-2})+(q+1)\theta(\p_q u_{-3})+(\p_q
u_{-2})v_1+\theta^2(u_{-4})+\theta(u_{-3})v_1+ u_{-2}v_0 \notag \\
\phantom{\p_{t_2}u_{-2}=}{}-\big(q^{-3}u_{-1}\theta^{-3}(\p^2_q
v_1)-q^{-1}u_{-1}\theta^{-2}(\p_q v_0)
-q^{-2}(2)_qu_{-2}\theta^{-3}(\p_q v_1)\notag\\
\phantom{\p_{t_2}u_{-2}=}{}+u_{-2}\theta^{-2}(v_0)+u_{-3}\theta^{-3}(v_1)+u_{-4}\big), \\
\p_{t_2}u_{-3}=(\p_q^2 u_{-3})+(q+1)\theta(\p_q u_{-4})+(\p_q
u_{-3})v_1+ \theta^2(u_{-5})+\theta(u_{-4})v_1+u_{-3}v_0 \notag\\
\phantom{\p_{t_2}u_{-3}=}{}-\big( -q^{-6}\theta^{-4}(\p_q^3
v_1)+q^{-3}u_{-1}\theta^{-3}(\p_q^2
v_0)+q^{-5}(3)_qu_{-2}\theta^{-4}(\p_q^2 v_1)\notag \\
\phantom{\p_{t_2}u_{-3}=}{}- q^{-2}(2)_qu_{-2}\theta^{-3}(\p_q
v_0)-q^{-3}(3)_qu_{-3}\theta^{-4}(\p_q
v_1)+u_{-3}\theta^{-3}(v_0)\\
\phantom{\p_{t_2}u_{-3}=}{}+u_{-4}\theta^{-4}(v_1)+u_{-5}
 \big).
\end{gather*}
The third f\/low is
\begin{gather*}
\p_{t_3}u_0=(\p_q^3 u_0)+(3)_q\theta(\p_q^2
u_{-1})+\tilde{s}_2(\p_q^2
u_0)+(3)_q\theta^2(\p_q u_{-2})+(2)_q\theta(\p_q u_{-1})\tilde{s}_2 \notag \\
\phantom{\p_{t_3}u_0=}{}+(\p_q
u_0)\tilde{s}_1+\theta^3(u_{-3})+\theta^2(u_{-2})\tilde{s}_2+\theta(u_{-1})\tilde{s}_1+u_0\tilde{s}_0
\notag \\
\phantom{\p_{t_3}u_0=}{}-\big( -q^{-1}\theta^{-2}(\p_q
\tilde{s}_2)u_{-1}+u_0\tilde{s}_0+u_{-1}\theta^{-1}(\tilde{s}_1)+u_{-2}\theta^{-2}(\tilde{s}_2)+u_{-3}+(\p_q
\tilde{s}_0) \big), \\
\p_{t_3}u_{-1}=(\p_q^3
u_{-1})+(3)_q\theta(\p_q^2u_{-2})+\tilde{s}_2 (\p_q^2
u_{-1})+(3)_q\theta^2(\p_q u_{-3})+(2)_q\tilde{s}_2\theta(\p_q
u_{-2})\notag \\
\phantom{\p_{t_3}u_{-1}=}{}+\tilde{s}_1(\p_q
u_{-1})+\theta^3(u_{-4})+\tilde{s}_2\theta^3(u_{-3})+\tilde{s}_1\theta(u_{-2})+\tilde{s}_0u_{-1}
\notag \\
\phantom{\p_{t_3}u_{-1}=}{}-\big( q^{-3}u_{-1}\theta^{-3}(\p_q^2
\tilde{s}_2)-q^{-1}u_{-1}\theta^{-2}(\p_q
\tilde{s}_1)-q^{-2}(2)_qu_{-2}\theta^{-3}(\p_q \tilde{s}_2)\notag \\
\phantom{\p_{t_3}u_{-1}=}{}+u_{-1}\theta^{-1}(\tilde{s}_0)+u_{-2}\theta^{-2}(\tilde{s}_1)
+u_{-3}\theta^{-3}(\tilde{s}_2)+u_{-4}
\big),\\
\p_{t_3}u_{-2}=(\p_q^3
u_{-2})+(3)_q\theta(\p_q^2u_{-3})+\tilde{s}_2(\p_q^2
u_{-2})+(3)_q\theta^2(\p_q u_{-4})+(2)_q\tilde{s}_2\theta(\p_q u_{-3})\notag \\
\phantom{\p_{t_3}u_{-2}=}{}+\tilde{s}_1(\p_q
u_{-2})+\theta^3(u_{-5})+\tilde{s}_2\theta^2(u_{-4})+\tilde{s}_1\theta(u_{-3})+\tilde{s}_0u_{-2}\notag
\\
\phantom{\p_{t_3}u_{-2}=}{}-\big(-q^{-6}u_{-1}\theta^{-4}(\p_q^3
\tilde{s}_2)+q^{-3}u_{-1}\theta^{-3}(\p_q^2
\tilde{s}_1)+q^{-5}(3)_qu_{-2}\theta^{-4}(\p_q^2
\tilde{s}_2) \notag \\
\phantom{\p_{t_3}u_{-2}=}{}-q^{-1}u_{-1}\theta^{-2}(\p_q
\tilde{s}_0) -q^{-2}(2)_qu_{-2}\theta^{-3}(\p_q
\tilde{s}_1)-q^{-3}(3)_q
u_{-3}\theta^{-4}(\p_q \tilde{s}_2)\\
\phantom{\p_{t_3}u_{-2}=}{} + u_{-2}\theta^{-2}(\tilde{s}_0)+
u_{-3}\theta^{-3}(\tilde{s}_1)+
u_{-4}\theta^{-4}(\tilde{s}_2)+u_{-5}\big), \\
\p_{t_3}u_{-3}=(\p_q^2 u_{-3})+(3)_q\theta(\p_q^2
u_{-4})+\tilde{s}_2(\p_q^2 u_{-3})+(3)_q\theta^2(\p_q
u_{-5})+(2)_q\tilde{s}_2\theta(\p_q u_{-4})\notag \\
\phantom{\p_{t_3}u_{-3}=}{}+\tilde{s}_1 (\p_q
u_{-3})+\theta^3(u_{-6})+\tilde{s}_2\theta^2(u_{-5})+\tilde{s}_1\theta(u_{-4})+\tilde{s}_0u_{-3}\notag \\
\phantom{\p_{t_3}u_{-3}=}{}-\big( q^{-10}u_{-1}\theta^{-5}(\p_q^4
\tilde{s}_2)-q^{-6}u_{-1}\theta^{-4}(\p_q^3
\tilde{s}_1)-q^{-9}(4)_q\theta^{-5}(\p_q^3 \tilde{s}_2)
 \notag \\
\phantom{\p_{t_3}u_{-3}=}{}+q^{-3}u_{-1}\theta^{-3}(\p_q^2
\tilde{s}_0)+q^{-5}(3)_qu_{-2}\theta^{-4}(\p_q^2
\tilde{s}_1)+q^{-7}\dfrac{(3)_q(4)_4}{(2)_q}u_{-3}\theta^{-5}(\p_q^2
\tilde{s}_2)\\
\phantom{\p_{t_3}u_{-3}=}{} -q^{-2}(2)_qu_{-2}\theta^{-3}(\p_q
\tilde{s}_0)-q^{-3}(3)_qu_{-3}\theta^{-4}(\p_q
\tilde{s}_1)-q^{-4}(4)_qu_{-4}\theta^{-5}(\p_q
\tilde{s}_2)\notag \\
\phantom{\p_{t_3}u_{-3}=}{}+u_{-3}\theta^{-3}(\tilde{s}_0)
+u_{-4}\theta^{-4}(\tilde{s}_1)+u_{-5}\theta^{-5}(\tilde{s}_2)+u_{-6}\big).
\end{gather*}
Obviously, $\p_{t_1}=\p$ and equations of f\/lows here are
reduced to usual KP f\/lows (4.10) and (4.11) in  \cite{ostt} when
$q\rightarrow 1$ and $u_0=0$. If we only consider the f\/irst
three f\/lows, i.e.\ f\/lows of ($t_1,t_2, t_3$), then
$u_{-1}=u_{-1}(t_1,t_2,t_3)$ is a $q$-deformation of the solution
of KP equation \cite{ostt}
\[
\dfrac{\partial}{\partial t_1}\left( 4\dfrac{\partial u}{\partial
t_3}-12u\dfrac{\partial u}{\partial t_1} - \dfrac{\partial^3
u}{\partial t_1^3}
  \right)- 3\dfrac{\partial^2 u}{\partial
t_2^2}=0.
\] In other words, $u_{-1}=u(t_1,t_2,t_3)$ in the above
equation when $q\rightarrow 1$, and hence $u_{-1}$ is called 
a~$q$-soliton if $u(t_1,t_2,t_3)=\lim\limits_{q\rightarrow1} u_{-1}$
is a soliton solution of KP equation.

On the other hand, $L$ in (\ref{qlaxoperator}) can be generated by
dressing operator $S=1+ \sum\limits_{k=1}^{\infty}s_k
\partial_q^{-k} $ in the following way
\begin{gather}\label{qdressing}
L=S \circ \partial_q \circ S^{-1}.
\end{gather}
Further, the dressing operator $S$ satisf\/ies the Sato equation
\begin{gather}\label{qdressingequ}
\dfrac{\partial S}{\partial t_n}=-(L^n)_-S, \qquad n=1,2, 3,
\ldots.
\end{gather}
The $q$-wave function $w_q(x,\overline{t})$ and $q$-adjoint wave
function $w^*_q(x,\overline{t})$ for $q$-KP hierarchy are
def\/ined by
\begin{gather}
w_q(x,\overline{t};z)=\left(S
e_q(xz)\exp\left(\sum\limits_{i=1}^{\infty}t_iz^i\right)\right) \label{q-wave} \\
\intertext{and}
 w^*(x,\overline{t};z)=\left((S^*)^{-1}|_{x/q}
e_{1/q}(-xz)\exp\left(-\sum\limits_{i=1}^{\infty}t_iz^i\right)\right),
\label{q-adjoint wave}
\end{gather}
where $\overline{t}=(t_1,t_2,t_3, \ldots)$. Here, for a $q$-PDO
$P=\sum\limits_i p_i(x)\partial_q^i$, the notation
\[
P|_{x/t}=\sum\limits_i p_i(x/t) t^i \partial_q^i
\]
is used in \eqref{q-adjoint wave}. Note that $w_q(x,\overline{t})$
and $w_q^*(x,\overline{t})$ satisfy following linear
$q$-dif\/ferential equations,
\begin{alignat}{3}
&(Lw_q)=zw_q, && \dfrac{\partial w_q}{\partial t_n}=(B_nw_q), & \nonumber\\ 
&(L^*|_{x/q}w^*_q)=zw^*_q, \qquad && \dfrac{\partial
w^*_q}{\partial t_n}=-((B_n|_{x/q})^*w^*_q). & \label{q-linear
adjoint system}
\end{alignat}
Furthermore, $w_q(x,\overline{t})$ and $w_q^*(x,\overline{t})$ can
be expressed by sole function $\tau_q(x,\overline{t})$ as
\begin{gather}
  \omega_q=\frac{\tau_q(x;\overline{t}-[z^{-1}])}{\tau_q(x;\overline{t})}
  e_q(xz)\exp\left(\sum_{i=1}^{\infty}t_iz^i\right),\label{qwavetau}
\\
  \omega_q^{*}=\frac{\tau_q(x;\overline{t}+[z^{-1}])}{\tau_q(x;\overline{t})}e_{1/q}(-xz)
  \exp\left(-\sum_{i=1}^{\infty}t_iz^i\right),
  \nonumber 
\end{gather}
where
\[
  [z]=\left(z,\frac{z^2}{2},\frac{z^3}{3},\ldots\right).
\]
From comparison of (\ref{q-wave}) and (\ref{qwavetau}), the
dressing operator $S$ has the form of
\begin{gather}\label{qStau}
  S=1-\left(\frac{1}{\tau_q}\frac{\partial}{\partial_{t_1}}\tau_q\right)\pq+
  \left[\frac{1}{2\tau_q}\left(\frac{\p^2}{\p t_1^2}-\frac{\p}{\p
  t_2}\right)\tau_q\right]\p_q^{-2}+\cdots.
\end{gather}
Using (\ref{qStau}) in (\ref{qdressing}), and then comparing with
Lax operator in (\ref{qlaxoperator}), we can show that all
dynamical variables $u_i$ $(i=0, -1, -2, -3, \ldots)$ can be
expressed by $\tau_q(x,\overline{t})$, and the f\/irst two are
\begin{gather}
u_0 =s_1-\t(s_1)=-x(q-1)\p_q s_1=x(q-1)\partial_q\partial_{t_1}\ln \tau_q,\nonumber \\
 u_{-1} =-\p_q
s_1+s_2-\t(s_2)+\t(s_1)s_1-s_1^2,   \label{qkpu}\\
 \cdots\cdots\cdots\cdots\cdots\cdots\cdots\cdots\cdots\cdots\cdots\cdots\cdots\cdots
 \cdots\cdots\cdots \notag
\end{gather}
We can see  $u_0=0$, and $u_{-1}=(\partial_{x}^2 \log \tau) $ as
classical KP hierarchy when $q$ $\rightarrow$ $1$, where
$\tau=\tau_q(x,\overline{t})|_{q\rightarrow 1}$. By considering
 $u_{-1}$ depending only on $(q,x,t_1,t_2,t_3)$, we can
regard $u_{-1}$ as $q$-deformation of solution of classical KP
equation. We shall show the $q$-ef\/fect of this solution for
$q$-KP hierarchy after we get  $\tau_q$ in next section. In order
to guarantee that $e_q(x)$ is convergent, we require the parameter
$0<q<1$ and parameter $x$ to be bounded.

 Beside  existence of the Lax operator, $q$-wave function,  $\tau_q$ for $q$-KP
 hierarchy,  another important property is the $q$-deformed Z-S equation and
 associated linear $q$-dif\/ferential equation. In other words, $q$-KP
 hierarchy also has an alternative expression, i.e.,
\begin{gather}\label{qzs}
  \frac{\partial B_m}{\partial t_n}-\frac{\partial B_n}{\partial
  t_m}+[B_m,B_n]=0, \qquad m,n=1,2,3,\ldots.
  \end{gather}
The ``eigenfunction'' $\phi$ and ``adjoint eigenfunction'' $\psi$
of $q$-KP hierarchy associated to  (\ref{qzs}) are def\/ined by
\begin{gather}
 \dfrac{\partial \phi}{\partial t_n}=(B_n\phi) , \label{qzslax1}\\
 \dfrac{\partial \psi}{\partial t_n}=-(B_n^{*}\psi),
\label{qzslax2}
\end{gather}
where $\phi=\phi(\lambda;x,\stackrel{\_}{t})$ and
$\psi=\psi(\mu;x, \stackrel{\_}{t}  )$. Here
 (\ref{qzslax2}) is dif\/ferent from the second equation in~(\ref{q-linear adjoint system}).
$\phi_i\equiv\phi(\lambda_i;x,\overline{t})$ and
$\psi_i\equiv\psi(\mu_i;x,\overline{t})$ will be  generating
functions of gauge transformations.

\section[Gauge transformations of $q$-KP hierarchy]{Gauge transformations of $\boldsymbol{q}$-KP hierarchy}

The authors in \cite{tsl}  reported two types of elementary gauge
transformation operator only for $q$-Gelfand--Dickey hierarchy. We
extended the elementary gauge transformations given in \cite{tsl},
for the $q$-KP hierarchy. At the same time, we shall add some
vital operator identity concerning to the $q$-dif\/ferential
operator and its inverse. Here we shall prove  two transforming
rules of $\tau$ function, ``eigenfunction'' and ``adjoint
eigenfunction'' of the $q$-KP hierarchy under these
transformations. Majority of the proofs are similar to the
classical case given by \cite{csy,og} and \cite{hlc1}, so we will
omit part of the proofs.

     Suppose $T$ is a pseudo-dif\/ferential operator, and
\begin{gather*}
\La{1}=T\circ  L \circ T^{-1} , \qquad \B{1}{n} \equiv
\big(\La{1}\big)^n_+,
\end{gather*}
so that
\begin{gather*}
\dfrac{\partial}{\partial t_n}\La{1}=\big[ \B{1}{n},\La{1} \big]
\end{gather*}
still holds for the transformed Lax operator $\La{1}$;  then $T$
is called a gauge transformation operator\ of the $q$-KP
hierarchy.
\begin{lemma}\label{lemqgtrequirment}
The operator $T$ is a gauge transformation operator, if
\begin{gather}
\big(T\circ B_{n}\circ T^{-1}\big)_+ = T\circ B_{n}\circ T^{-1}
            + \dfrac{\partial T}{\partial t_n}\circ T^{-1}, \label{requirementqT1} \\
\intertext{or} \big(T\circ B_{n}\circ T^{-1}\big)_- =
-\dfrac{\partial T}{\partial t_n} \circ T^{-1}.
\label{requirementqT2}
\end{gather}
\end{lemma}
If the initial Lax operator of $q$-KP is a ``free'' operator
$L=\partial_q$, then the gauge transformation operator\ is also a
dressing operator for new $q$-KP hierarchy whose Lax operator
$\La{1}=T\circ \partial_q \circ T^{-1}$, because of
\eqref{requirementqT2} becomes
\begin{gather}\label{qgtsato}
T_{t_n}=-\big(T\circ B_n \circ T^{-1}\big)_{-}\circ T=-\big(T\circ
\partial_q^n \circ T^{-1}\big)_-\circ T=-\big(L^{(1)}\big)^n_-\circ T,
\end{gather}
which is the Sato equation  \eqref{qdressingequ}. In order to
prove existence of two types of the gauge transformation operator,
the following operator identities are necessary.
\begin{lemma} Let $f$ be a suitable function, and  $A$ be a $q$-deformed pseudo-differential operator,
then
\begin{gather}
(1)\quad \big(\t(f)\circ \partial_q \circ f^{-1}\circ A \circ
f\circ
\partial^{-1}_q
\circ (\t(f))^{-1}  \big)_+ \notag \\
\phantom{(1)\quad\qquad}{} =\t(f)\circ \partial_q \circ
f^{-1}\circ A_+ \circ f\circ
\partial^{-1}_q \circ  (\t(f))^{-1}\notag \\
 \phantom{(1)\quad\qquad =}{}-\t(f)\circ \big[\partial_q \big(f^{-1}\cdot (A_+\cdot f) \big)  \big]
\circ \partial^{-1}_q \circ (\t(f))^{-1},\label{qopertorid1} \\
(2)\quad \big(\t^{-1}(f^{-1})\circ \partial^{-1}_q \circ f \circ A
\circ f^{-1} \circ \partial_q
\circ  \t^{-1}(f)  \big)_-  \notag\\
\phantom{(2)\quad\qquad}{}= \t^{-1}(f^{-1})\circ
\partial^{-1}_q \circ  f  \circ A_- \circ
f^{-1} \circ
\partial_q \circ \t^{-1}(f) \notag\\
\phantom{(2)\quad\qquad =}{} - \t^{-1}(f^{-1}) \circ
\partial^{-1}_q \circ
 \t^{-1}(f)\circ \partial_q\big(
 \t^{-1}[f^{-1}\cdot \big(A^*_+\cdot f\big)] \big).\label{qoperatorid2}
\end{gather}
\end{lemma}
\begin{remark}This lemma is a $q$-analogue of corresponding
identities of pseudo-dif\/ferential ope\-ra\-tors given by \cite{og}.
\end{remark}

\begin{theorem}\label{qgtexistence}
There exist two kinds of gauge transformation operator\ for the
$q$-KP hierarchy, namely
\begin{alignat}{3}
&\mbox{\rm Type I}: \quad  && T_D(\phi_1)=\t(\phi_1)\circ \partial_q \circ \phi^{-1}_1,& \label{qgtI}\\
&\mbox{\rm Type II}: \quad &&
T_I(\psi_1)=(\t^{-1}(\psi_1))^{-1}\circ \partial_q^{-1} \circ
\psi_1. & \label{qgtII}
\end{alignat}
Here $\phi_1$  and $\psi_1$ are defined by  \eqref{qzslax1} and
 \eqref{qzslax2} that are called the generating functions of
gauge transformation.
\end{theorem}

\begin{proof} First of all, for the Type~I case (see
\eqref{qgtI}),
\begin{gather*}
\B{1}{n}\equiv \big( \La{1}\big)^n_+=\big( T_D\circ (L)^n \circ T_D^{-1} \big)_+\\
\phantom{\B{1}{n}}{} =T_D\circ B_{n} \circ T_D^{-1} -
\t(\phi_1)\cdot \partial_q\big(
\phi_1^{-1} \cdot (B_{n}\cdot \phi_1) \big) \circ \partial_q^{-1}\circ (\t(\phi_1))^{-1}\\
\phantom{\B{1}{n}}{} =T_D\circ \B{0}{n} \circ T_D^{-1}
-\Big(\t(\phi_1)\circ
\partial_q \circ \dfrac{(\phi_1)_{t_n}}{\phi_1}\circ
\partial_q^{-1}\circ (\t(\phi_1))^{-1}
\\
\phantom{\B{1}{n}=}{} -\t(\phi_1)\circ
\t\Big(\dfrac{(\phi_1)_{t_n}}{\phi_1}\Big)
\circ \partial_q\circ \partial^{-1}_q \circ (\t(\phi_1))^{-1} \Big)\\
\phantom{\B{1}{n}}{}=T_D\circ B_n \circ T_D^{-1} +
\t\Big(\dfrac{(\phi_1)_{t_n}}{\phi_1}\Big) -\t(\phi_1)\circ
\partial_q\circ \dfrac{(\phi_1)_{t_n}}{\phi_1}\circ
\partial^{-1}_q\circ (\t(\phi_1))^{-1}.
\end{gather*}
Here the operator identity \eqref{qopertorid1}, $B_{n}=(L)^n_+$,
$(\phi_1)_{t_n}=(B_{n}\cdot \phi_1)$ and
\eqref{qleibnizdifferentialfirst} were used. On the other hand,
\begin{gather*}
\dfrac{\partial T_D}{\partial t_n}\circ T_D^{-1}=\big(
\t(\phi_1)\circ
\partial_q
\circ \phi_1^{-1} \big)_{t_n}\circ T_D^{-1}
 =\t((\phi_1)_{t_n}) \circ
\partial_q \circ
\phi_1^{-1} \circ \phi_1 \circ
\partial^{-1}_q \circ (\t(\phi_1))^{-1}\\
\phantom{\dfrac{\partial T_D}{\partial t_n}\circ T_D^{-1}=}{}
-\t(\phi_1)\circ \partial_q
 \circ
\dfrac{(\phi_1)_{t_n}}{\phi_1^2} \circ \phi_1 \circ
\partial^{-1}_q
 \circ (\t(\phi_1))^{-1} \\
\phantom{\dfrac{\partial T_D}{\partial t_n}\circ T_D^{-1}}{}
=\t\Big(\dfrac{(\phi_1)_{t_n}}{\phi_1}\Big) -\t(\phi_1)\circ
\partial_q \circ \dfrac{(\phi_1)_{t_n}}{\phi_1}\circ
\partial^{-1}_q \circ (\t(\phi_1))^{-1}.
\end{gather*}
Taking this expression back into $\B{1}{n}$, we get
\[
\B{1}{n}\equiv \big(\La{1}\big)^n_+ = T_D \circ B_{n} \circ
T_D^{-1}
  +\dfrac{\partial T_D}{\partial t_n}\circ T^{-1}_D,
\]
and that indicates that T$_D$($\phi_1$) is indeed a gauge
transformation operator via Lemma~\ref{lemqgtrequirment}.
Se\-cond, we want to prove that the equation
\eqref{requirementqT2} holds for Type II case (see \eqref{qgtII}).
By direct calculation the left hand side of \eqref{requirementqT2}
is in the form of
\begin{gather*}
\big(T_I \circ B_{n} \circ T_I^{-1}\big)_-=\big(
(\t^{-1}(\psi_1))^{-1}\circ \partial^{-1}_q \circ \psi_1 \circ
B_{n}\circ \psi_1^{-1}\circ
\partial_q \circ \t^{-1}(\psi_1)\big)_-\\
\phantom{\big(T_I \circ B_{n} \circ T_I^{-1}\big)_-}{} =
(\t^{-1}(\psi_1))^{-1} \circ \partial^{-1}_q \circ \psi_1 \circ
(B_{n})_- \circ \psi_1^{-1}\circ
\partial_q \circ \t^{-1}(\psi_1) \\
\phantom{\big(T_I \circ B_{n} \circ T_I^{-1}\big)_-=}{} -
(\t^{-1}(\psi_1))^{-1} \circ \partial^{-1}_q \circ \t^{-1}(\psi_1)
\circ
 \left(\partial_q \t^{-1}\left(\dfrac{\big( B^*_{n} \cdot \psi_1 \big)}{\psi_1}\right)\right) \\
\phantom{\big(T_I \circ B_{n} \circ T_I^{-1}\big)_-}{}=
(\t^{-1}(\psi_1))^{-1}  \circ \partial^{-1}_q \circ
\t^{-1}(\psi_1) \circ \left(\partial_q
\t^{-1}\left(\dfrac{\big(\psi_1
\big)_{t_n}}{\psi_1}\right)\right).
\end{gather*}
In the above calculation, the operator identity
\eqref{qoperatorid2}, $(B_{n})_-=0$, $(\psi_1)_{t_n}=-(B^*_{n}
\cdot\psi_1) $ were used. Moreover, with the help of
\eqref{qleibnizdifferentialfirst}, we have
\begin{gather*}
- \dfrac{\partial T_I }{\partial t_n}\circ T_I^{-1}
=-\dfrac{\partial }{\partial t_n}\big((\t^{-1}(\psi_1))^{-1}\circ
\partial^{-1}_q \circ \psi_1 \big)
\circ \psi_1^{-1}\circ \partial_q \circ \t^{-1}(\psi_1) \\
\phantom{- \dfrac{\partial T_I }{\partial t_n}\circ T_I^{-1}}{} =
\dfrac{\t^{-1}((\psi_1)_{t_n})}{(\t^{-1}(\psi_1))^{2}} \circ
\partial^{-1}_q\circ \psi_1\circ
\psi_1^{-1} \circ \partial_q \circ
\t^{-1}(\psi_1)\\
\phantom{- \dfrac{\partial T_I }{\partial t_n}\circ T_I^{-1}=}{}
 -(\t^{-1}(\psi_1))^{-1}\circ
\partial^{-1}_q\circ (\psi_1)_{t_n}\circ \psi_1^{-1}\circ
 \partial_q \circ
\t^{-1}(\psi_1)\\
\phantom{- \dfrac{\partial T_I }{\partial t_n}\circ T_I^{-1}}{}
=\t^{-1}\left(\dfrac{(\psi_1)_{t_n}}{\psi_1}\right)
-\dfrac{1}{\t^{-1}(\psi_1)}\circ\partial_q^{-1}\circ\left[\partial_q
\circ \t^{-1}\left(\dfrac{(\psi_1)_{t_n}}{\psi_1}\right)\right.\\
\left.\phantom{- \dfrac{\partial T_I }{\partial t_n}\circ
T_I^{-1}=}{}
 -\left(\partial_q\cdot\t^{-1}\left(\dfrac{(\psi_1)_{t_n}}{\psi_1}\right)\right)
 \right]\circ \t^{-1}(\psi_1)= \t^{-1}\left(\dfrac{(\psi_1)_{t_n}}{\psi_1}\right)
\\
\phantom{- \dfrac{\partial T_I }{\partial t_n}\circ T_I^{-1}=}{}
 -
\t^{-1}\left(\dfrac{(\psi_1)_{t_n}}{\psi_1}\right)+
\dfrac{1}{\t^{-1}(\psi_1)}\circ \partial_q^{-1}\circ
 \left(\partial_q\cdot\t^{-1}\left(\dfrac{(\psi_1)_{t_n}}{\psi_1}\right)\right) \circ \t^{-1}(\psi_1) \\
\phantom{- \dfrac{\partial T_I }{\partial t_n}\circ T_I^{-1}}{}
=\dfrac{1}{\t^{-1}(\psi_1)}\circ \partial_q^{-1}\circ
 \left(\partial_q\cdot\t^{-1}\left(\dfrac{(\psi_1)_{t_n}}{\psi_1}\right)\right) \circ \t^{-1}(\psi_1) .
\end{gather*}
The two equations obtained above show that  T$_I(\psi_1)$
satisf\/ies \eqref{requirementqT2}, so $T_I(\psi_1)$ is also a
gauge transformation operator\ of the $q$-KP hierarchy according
to Lemma \ref{lemqgtrequirment}.
\end{proof}

\begin{remark}
There are two convenient expressions for $T_D$ and $T_I$,
\begin{gather}
T_D=\partial_q-\alpha_1,\qquad
T_D^{-1}=\partial_q^{-1}+\t^{-1}(\alpha_1)\partial_q^{-2}+\cdots ,
\qquad \alpha_1= \dfrac{\partial_q\phi_1}{\phi_1}, \label{covenientgt1} \\
T_I=(\partial_q+\beta_1)^{-1}=\partial_q^{-1}-\t^{-1}(\beta_1)\partial_q^{-2}+\cdots,
\qquad  \beta_1= \dfrac{\partial_q \t^{-1}(\psi_1)}{\psi_1}.
\label{covenientgt2}
\end{gather}
\end{remark}

  In order to get a new solution of $q$-KP hierarchy from the input
solution, we should  know the transformed expressions of
$\fun{u}{1}{i}$, $\tu{1}_q$, $\fun{\phi}{1}{i}$,
$\fun{\psi}{1}{i}$. The following two theorems are related to
this. Before we start to discuss explicit forms of them, we would
like to def\/ine the generalized $q$-Wronskian for a set of
functions \{$\psi_k,
\psi_{k-1},\ldots,\psi_1;\phi_1,\phi_2,\ldots,\phi_n$ \} as
\[
IW_{k,n}^q(\psi_k,\ldots,\psi_1;\phi_1,\ldots,\phi_n)=\begin{vmatrix}
          \pq\psi_k\phi_1    &\pq\psi_k\phi_2    &\cdots  &\pq\psi_k\phi_n\\
          \vdots             &\vdots             &\cdots  &\vdots\\
          \pq\psi_1\phi_1    &\pq\psi_1\phi_2    &\cdots  &\pq\psi_1\phi_n\\
          \phi_1             &\phi_2             &\cdots  &\phi_n\\
          \p_q\phi_1         &\p_q\phi_2         &\cdots  &\p_q\phi_n\\
          \vdots             &\vdots             &\cdots  &\vdots\\
          \p_q^{n-k-1}\phi_1 &\p_q^{n-k-1}\phi_2 &\cdots  &\p_q^{n-k-1}\phi_n
          \end{vmatrix},
\]

which reduce to the $q$-Wronskian when $k=0$,
\[
W_{n}^q(\phi_1,\cdots,\phi_n)=\begin{vmatrix}
          \phi_1             &\phi_2             &\cdots  &\phi_n\\
          \p_q\phi_1         &\p_q\phi_2         &\cdots  &\p_q\phi_n\\
          \vdots             &\vdots             &\cdots  &\vdots\\
          \p_q^{n-1}\phi_1 &\p_q^{n-1}\phi_2 &\cdots  &\p_q^{n-1}\phi_n
          \end{vmatrix}.
\]
 Suppose $\phi_1(\lambda_1;x,\overline{t}) $ is
a known  ``eigenfunction'' of $q$-KP with the initial function
$\tau_q$, which generates gauge transformation operator\
$T_D(\phi_1)$. Then we have

\begin{theorem}\label{thmgtI}
Under the gauge transformation\ $\La{1}=T_D(\phi_1)\circ L \circ
(T_D(\phi_1))^{-1}$, new ``eigenfunction'', ``adjoint
eigenfunction'' and $\tau$ function of the transformed $q$-KP
hierarchy are
\begin{gather*}
\phi\longrightarrow \fun{\phi}{1}{}(\lambda;x,\overline{t})
=(T_D(\phi_1)\cdot\phi)=\dfrac{W_2^{q}(\phi_1,\phi)}{\phi_1},\\
\psi\longrightarrow
\fun{\psi}{1}{}(\lambda;x,\overline{t})=\big(\big(T_D(\phi_1)^{-1}\big)^*\cdot\psi\big)=\dfrac{\t(\partial^{-1}_q
\phi_1\psi)}{\t(\phi_1)}, \\
\tau_q \longrightarrow \tu{1}_q=\phi_1\tau_q.
\end{gather*}
$
\fun{\phi}{1}{k}=\fun{\phi}{1}{}(\lambda=\lambda_k;x,\overline{t})$,
$\fun{\psi}{1}{k}=\fun{\psi}{1}{}(\lambda=\lambda_k;x,\overline{t})$.
Note $\fun{\phi}{1}{1}=0$.
\end{theorem}
\begin{proof}
(1) By direct calculations, we have
\begin{gather*}
\big(\partial_{t_n}\fun{\phi}{1}{}\big)=(\partial_{t_n}(T_D\cdot\phi))=(\partial_{t_n}T_D)\cdot
\phi +( T_D\cdot\partial_{t_n}\phi)  \notag \\
\phantom{\big(\partial_{t_n}\fun{\phi}{1}{}\big)}{}=\big(\partial_{t_n}T_D\circ
T_D^{-1}\big) \cdot (T_D\phi)+ T_D\cdot(B_n\phi)
=\big(\partial_{t_n}T_D\circ T_D^{-1} + T_D\circ B_n \circ
T_D^{-1} \big)\cdot (T_D\phi)\!\\
\phantom{\big(\partial_{t_n}\fun{\phi}{1}{}\big)}{}=\big(\fun{B}{1}{n}\cdot\fun{\phi}{1}{}\big),
\end{gather*}
in which \eqref{qzslax1} and \eqref{requirementqT1} were used.

(2) Similarly, with the help of
$(\fun{B}{1}{n})^*=(T_D^{-1})^*\circ
\partial_{t_n}T_D^*+ (T_D^{-1})^*\circ B_n^*\circ T_D^*$ and \eqref{qzslax2}, we can
obtain
\begin{gather*}
\partial_{t_n}\fun{\psi}{1}{}=\big((T_D^{-1})^*\cdot \psi\big)_{t_n}
=\big(-(T_D^*)^{-1}\circ \partial_{t_n}T_D^*\circ
(T_D^*)^{-1}\big)\cdot\psi+
(T_D^*)^{-1}\cdot\partial_{t_n}\psi\\
\phantom{\partial_{t_n}\fun{\psi}{1}{}}{}=-\big((T_D^{-1})^*\circ
\partial_{t_n}T_D^*+ (T_D^{-1})^*\circ B_n^*\circ T_D^*\big)\cdot\big((T_D^{-1})^*\cdot
\psi\big)=-\big(\fun{B}{1}{n}\big)^*\cdot\fun{\psi}{1}{}.
\end{gather*}

(3) According to the def\/inition of $T_D$ in \eqref{qgtI} and
with the help of \eqref{covenientgt1}, $\La{1}$ can be expressed
as
\begin{gather*}
\fun{L}{1}{q}=\partial_q+ \fun{u}{1}{0}+
\fun{u}{1}{-1}\partial_q^{-1}+\cdots, \qquad \fun{u}{1}{0}
=x(q-1)\partial_q\alpha_1+ \t(u_0).
\end{gather*}
On the other hand, $(\phi_1)_{t_1}=((L)_+\phi_1)$ implies
$\alpha_1=
\partial_{t_1}\ln\phi_1-u_0$, then $\fun{u}{1}{0}$ becomes
\begin{gather*}
\fun{u}{1}{0}=x(q-1)\partial_q\partial_{t_1}\ln \phi_1+ u_0
=x(q-1)\partial_q\partial_{t_1}\ln \phi_1+ x(q-1)\partial_q\partial_{t_1}\ln \tau_q \\
\phantom{\fun{u}{1}{0}}{}
=x(q-1)\partial_q\partial_{t_1}\ln\phi_1\tau_q.
\end{gather*}
Then
\[
\tu{1}_q=\phi_1\tau_q.
\]
This completes the proof of the theorem.
\end{proof}

For the gauge transformation operator of Type II, there exist
similar results. Let $\psi_1(\mu_1;x,\overline{t})$ be a known
``adjoint eigenfunction'' of $q$-KP with the initial function
$\tau_q$, which generates the gauge transformation operator
$T_I(\psi_1)$. Then we have
\begin{theorem}\label{thmgtII}
Under the gauge transformation\ $\La{1}=T_I(\psi_1)\circ
L\circ(T_I(\psi_1))^{-1}$, new ``eigenfunction'', ``adjoint
eigenfunction'' and $\tau$ function of the transformed $q$-KP
hierarchy are
\begin{gather*}
\phi\longrightarrow
\fun{\phi}{1}{}(\lambda;x,\overline{t})=(T_I(\psi_1)\cdot\phi)=
\dfrac{(\partial^{-1}_q \psi_1\phi)}{\t^{-1}(\psi_1)}
,\\
\psi\longrightarrow
\fun{\psi}{1}{}(\lambda;x,\overline{t})=\big(\big(T_I(\psi_1)^{-1}\big)^*\cdot\psi\big)=
\dfrac{\widetilde{W}_2^{q}(\psi_1,\psi)}{\psi_1}, \\
\tau_q \longrightarrow \tu{1}_q=\t^{-1}(\psi_1)\tau_q.
\end{gather*}
$
\fun{\phi}{1}{k}=\fun{\phi}{1}{}(\lambda=\lambda_k;x,\overline{t})$,
$\fun{\psi}{1}{k}=\fun{\psi}{1}{}(\lambda=\lambda_k;x,\overline{t})$.
Note $\fun{\psi}{1}{1}=0$. $\widetilde{W}_n^q$ is obtained from
$W_n^q$ by replacing $\partial_q$ with $\partial^*_q$.
\end{theorem}
\begin{proof}
 The proof is analogous to the proof of the previous theorem. So it
is omitted.
\end{proof}

\section{Successive applications of gauge transformations}

We now discuss  successive applications of the two types of gauge
transformation operators in a general way, which is similar to the
classical case \cite{csy,hlc,hlc1}.  For example, consider the
chain of gauge transformation operators,
\begin{gather}
L\xrightarrow{\T{1}{D}{\phi_1}}
\La{1}\xrightarrow{\T{2}{D}{\fun{\phi}{1}{2}}}
\La{2}\xrightarrow{\T{3}{D}{\fun{\phi}{2}{3}}}
\La{3}\xrightarrow{} \cdots\xrightarrow{}
\La{n-1}\xrightarrow{\T{n}{D}{\fun{\phi}{n-1}{n}}}
\La{n} \nonumber\\
\xrightarrow{\T{n+1}{I}{\psi_1}}
\La{n+1}\xrightarrow{\T{n+2}{I}{\fun{\psi}{n +1}{2}}}
\La{n+2}\xrightarrow{}\cdots\xrightarrow{}
\La{n+k-1}\xrightarrow{\T{n+k}{I}{\fun{\psi}{n+k-1}{k}}}
\La{n+k}.\!\!\label{succcessiveqgt}
\end{gather}
Here the index ``$i$`` in a gauge transformation operator means
the $i$-th gauge transformation, and~$\fun{\phi}{j}{i}$ (or
$\fun{\psi}{j}{i}$) is transformed by $j$-steps gauge
transformations\ from $\phi_i$ (or $\psi_i$), $\La{i}$ is
transformed by $j$-step gauge transformations\ from the initial
Lax operator $L$. Successive applications of gauge transformation
operator in  \eqref{succcessiveqgt} can be represented by
\begin{gather*}
T_{n+k}=T^{(n+k)}_I\big(\fun{\psi}{n+k-1}{k}\big)  \cdots
T^{(n+2)}_I\big(\fun{\psi}{n+1}{2}\big) \circ T^{(n+1)}_I\big(\fun{\psi}{n}{1}\big)\nonumber\\
\phantom{T_{n+k}=}{} \circ T^{(n)}_D\big(\fun{\phi}{n-1}{n}\big)
\cdots
T^{(2)}_D\big(\fun{\phi}{1}{2}\big)\circ T^{(1)}_D(\phi_1).
\end{gather*}
Our goal is to obtain $\fun{\phi}{n+k}{}$, $\fun{\psi}{n+k}{}$,
$\tau^{(n+k)}_q$ of $\La{n+k}$ transformed from $L$ by the
$T_{n+k}$ in the above chain. All of these are based on the
determinant representation of gauge transformation operator\
$T_{n+k}$. As the proof of the determinant representation of
$T_{n+k}$ is similar extremely to the case of classical KP
hierarchy \cite{hlc}, we will omit it.

\begin{lemma}\label{lemgtdetrep1}
The gauge transformation operator $T_{n+k}$ has the following
determinant representation $(n>k)$:
\begin{gather*}
  T_{n+k}=\frac{1}{IW_{k,n}^q(\psi_k,\ldots,\psi_1;\phi_1,\ldots,\phi_n)}
  \begin{vmatrix}
  \p_q^{-1}\psi_k\phi_1 &\cdots  & \p_q^{-1}\psi_k\phi_n & \p_q^{-1}\circ\psi_k\\
  \vdots                &\cdots  & \vdots                & \vdots\\
  \p_q^{-1}\psi_1\phi_1 &\cdots  & \p_q^{-1}\psi_1\phi_n & \p_q^{-1}\circ\psi_1\\
  \phi_1                &\cdots  & \phi_n                & 1\\
  \p_q\phi_1            &\cdots  & \p_q\phi_n            & \p_q\\
  \vdots                &\cdots  & \vdots                & \vdots\\
  \p_q^{n-k}\phi_1      &\cdots  & \p_q^{n-k}\phi_n      & \p_q^{n-k}
  \end{vmatrix}
\end{gather*}
and
\begin{gather*}
   T_{n+k}^{-1}=  \begin{vmatrix}
   \phi_1\circ\pq &\theta(\pq\psi_k\phi_1)&\cdots&\theta(\pq\psi_1\phi_1)&\theta(\phi_1)& \cdots &\theta(\p_q^{n-k-2}\phi_1)\\
   \phi_2\circ\pq &\theta(\pq\psi_k\phi_2)&\cdots&\theta(\pq\psi_1\phi_2)&\theta(\phi_2)& \cdots &\theta(\p_q^{n-k-2}\phi_2)\\
   \vdots         &\vdots                 &\cdots&\vdots                 &\vdots        & \cdots &\vdots\\
   \phi_n\circ\pq &\theta(\pq\psi_k\phi_n)&\cdots&\theta(\pq\psi_1\phi_n)&\theta(\phi_n)& \cdots &\theta(\p_q^{n-k-2}\phi_n)
   \end{vmatrix}\nonumber\\
\phantom{T_{n+k}^{-1}=}{}   \times
\frac{(-1)^{n-1}}{\theta(IW_{k,n}^q(\psi_k,\ldots,\psi_1;\phi_1,\ldots,\phi_n)}.
\end{gather*}

\end{lemma}
\begin{lemma}\label{lemgtdetrep2}
Under the case of $n=k$, $T_{n+k}$ is given by
\begin{gather*}
   T_{n+n}=\frac{1}{IW_{n,n}^q(\psi_n,\ldots,\psi_1;\phi_1,\ldots,\phi_n)}
  \begin{vmatrix}
  \p_q^{-1}\psi_n\phi_1 &\cdots  & \p_q^{-1}\psi_n\phi_n & \p_q^{-1}\circ\psi_n\\
  \vdots                &\cdots  & \vdots                & \vdots\\
  \p_q^{-1}\psi_1\phi_1 &\cdots  & \p_q^{-1}\psi_1\phi_n & \p_q^{-1}\circ\psi_1\\
  \phi_1                &\cdots  & \phi_n                & 1
  \end{vmatrix}
\end{gather*}
but $T_{n+n}^{-1}$ becomes
\begin{gather*}
   T_{n+n}^{-1} = \begin{vmatrix}
   -1             &\psi_n                 &\cdots&\psi_1  \\
   \phi_1\circ\pq &\theta(\pq\psi_n\phi_1)&\cdots&\theta(\pq\psi_1\phi_1)\\
   \vdots         &\vdots                 &\cdots&\vdots                 \\
   \phi_n\circ\pq &\theta(\pq\psi_n\phi_n)&\cdots&\theta(\pq\psi_1\phi_n)
   \end{vmatrix}  \frac{(-1)}{\theta(IW_{n,n}^q(\psi_n,\ldots,\psi_1;\phi_1,\ldots,\phi_n)}.\!\! 
\end{gather*}
\end{lemma}

In the above lemmas, $T_{n+k}$ are  expanded with respect to the
last column collecting  all sub-determinants on the left of the
symbols $\partial_q^i$ ($i=-1,0,1, 2, \ldots, n-k $);
$T_{n+k}^{-1}$  are  expanded  with respect to the f\/irst column
by means of collection of all minors on the right of $\phi_i
\partial_q^{-1}$. Basing on the determinant representation, f\/irst
of all, we would like to consider the case of $k=0$ in
 \eqref{succcessiveqgt}, i.e.
\begin{gather*}
L\xrightarrow{\T{1}{D}{\phi_1}}
\La{1}\xrightarrow{\T{2}{D}{\fun{\phi}{1}{2}}}
\La{2}\xrightarrow{\T{3}{D}{\fun{\phi}{2}{3}}}
\La{3}\xrightarrow{}\cdots\xrightarrow{}
\La{n-1}\xrightarrow{\T{n}{D}{\fun{\phi}{n-1}{n}}} \La{n},
\end{gather*}
whose corresponding equivalent gauge transformation operator\ is
\begin{gather}\label{qgtTn}
T_{n}= T^{(n)}_D\big(\fun{\phi}{n-1}{n}\big) \cdots
T^{(2)}_D\big(\fun{\phi}{1}{2}\big)\circ T^{(1)}_D(\phi_1).
\end{gather}

\begin{theorem}\label{thmqgtn}
Under the gauge transformation $T_{n}$ $(n\geq 1)$,
\begin{gather}
\phi^{(n)}(\lambda;x,\overline{t})=(T_{n}\cdot
\phi)=\dfrac{W_{n+1}^q(\phi_1,\ldots,\phi_n,\phi)}
{W_{n}^q(\phi_1,\ldots,\phi_n)},\label{qgtnphi}\\
\fun{\psi}{n}{}(\mu;x,\overline{t})=
\big(\big(T_{n}^{-1}\big)^*\cdot \psi\big)
=(-1)^n\t\left(\dfrac{IW^q_{1,n}(\psi;\phi_{1},\ldots,\phi_{n})}
{W^q_{n}(\phi_1,\ldots,\phi_n)}\right), \label{qgtnpsi}\\
 \tu{n}_q= W^q_{n}(\phi_1,\ldots,\phi_n)\cdot \tau_q.\nonumber 
\end{gather}
Furthermore, $\fun{\phi}{n}{i}=
\phi^{(n)}(\lambda=\lambda_i;x,\overline{t})$, $\fun{\psi}{n}{i}=
\psi^{(n)}(\mu=\mu_i;x,\overline{t})$. Note $\fun{\phi}{n}{i}=0$
if $i\in \{1,2,\ldots,n\}$.
\end{theorem}

\begin{proof} (1) Successive application of Theorem~\ref{thmgtI} implies
\begin{gather*}
\fun{\phi}{n}{}=\fun{T}{n}{D}\big(\fun{\phi}{n-1}{n}\big)\fun{\phi}{n-1}{}
=\fun{T}{n}{D}\big(\fun{\phi}{n-1}{n}\big)\fun{T}{n-1}{D}\big(\fun{\phi}{n-2}{n-1}\big)\fun{\phi}{n-2}{}=\cdots\\
\phantom{\fun{\phi}{n}{}}{} =T^{(n)}_D\big(\fun{\phi}{n-1}{n}\big)
\cdots T^{(2)}_D\big(\fun{\phi}{1}{2}\big)\circ
T^{(1)}_D(\phi_1)\phi=(T_n\cdot\phi).
\end{gather*}
Using the determinant representation of $T_n$ in it leads to
$\phi^{(n)}$. Here $T^{(1)}_D(\phi_1)= T_D(\phi_1)$.

(2) Similarly, according to Theorem~\ref{thmgtI} we have
\begin{gather*}
\fun{\psi}{n}{}=\big({\fun{T}{n}{D}}^{-1}\big)^{*}\fun{\psi}{n-1}{}
=\big({\fun{T}{n}{D}}^{-1}\big)^{*}\big({\fun{T}{n-1}{D}}^{-1}\big)^{*}\fun{\psi}{n-2}{}=\cdots\\
\phantom{\fun{\psi}{n}{}}{}
=\big(\big({\fun{T}{n}{D}}^{-1}\big)^{*}\big({\fun{T}{n-1}{D}}^{-1}\big)^{*}
\cdots
\big({\fun{T}{3}{D}}^{-1}\big)^{*}\big({\fun{T}{2}{D}}^{-1}\big)^{*}
\big(T^{-1}_D\big)^*\big)\cdot \psi
=\big(\big(T^{-1}_n\big)^*\cdot\psi\big).
\end{gather*}
Then $\psi^{(n)}$ can be deduced by using the determinant
representation of $T_n^{-1}$ in  the Lemma~\ref{lemgtdetrep1} with
$k=0$. Here we omit the generating functions in $T^{(i)}_D$
$(i=1,2,\ldots,n)$, which are the same as~(1).

(3) Meanwhile, we can get $\tu{n}$ by repeated iteration according
to the rule in Theorem~\ref{thmgtI},
\begin{gather*}
\tu{n}_q=\fun{\phi}{n-1}{n}\tu{n-1}_q =
\fun{\phi}{n-1}{n}\fun{\phi}{n-2}{n-1}\tu{n-2}_q
=\fun{\phi}{n-1}{n}\fun{\phi}{n-2}{n-1}\fun{\phi}{n-3}{n-2}\tu{n-3}_q
=\cdots\\
\phantom{\tu{n}_q}{}
=\fun{\phi}{n-1}{n}\fun{\phi}{n-2}{n-1}\fun{\phi}{n-3}{n-2}\cdots\fun{\phi}{3}{4}
\fun{\phi}{2}{3}
\fun{\phi}{1}{2}\phi_1\tau_q\\
\phantom{\tu{n}_q}{} =\dfrac{W_n^q(\phi_1,
\phi_2,\phi_3,\ldots,\phi_n)} {W_{n-1}^q(\phi_1, \phi_2, \phi_3,
\ldots,\phi_{n-1})} \dfrac{W_{n-1}^q(\phi_1, \phi_2, \phi_3,
\ldots,\phi_{n-1})} {W_{n-2}^q(\phi_1, \phi_2,
\phi_3,\ldots,\phi_{n-2})} \dfrac{W_{n-2}^q(\phi_1, \phi_2,
\phi_3, \ldots,\phi_{n-2})}
{W_{n-3}^q(\phi_1, \phi_2, \phi_3,\ldots,\phi_{n-3})}\\
\phantom{\tu{n}_q=}{}\cdots \dfrac{W_4^q(\phi_1, \phi_2,
\phi_3,\phi_4)}{W_3^q(\phi_1, \phi_2,\phi_3)} \dfrac{W_3^q(\phi_1,
\phi_2, \phi_3)}{W_2^q(\phi_1, \phi_2)} \dfrac{W_2^q(\phi_1,
\phi_2)}{W_1^q(\phi_1)}\phi_1\tau_q=W^q_n(\phi_1,\phi_2,\ldots,\phi_n)\tau_q.
\end{gather*}
with the help of the determinant representation of
Lemma~\ref{lemgtdetrep1} with $k=0$. Here $W_1^q(\phi_1)=\phi_1$.
\end{proof}

 It should be noted that there is a $\t$ action in
 \eqref{qgtnpsi}, which is the main dif\/ference between the $q$-KP
and classical KP beside dif\/ferent elements in determinant.
Furthermore, for more complicated chain of gauge transformation
operators in
 \eqref{succcessiveqgt}, $\fun{\phi}{n+k}{}$,
$\fun{\psi}{n+k}{}$, $\tau^{(n+k)}_q$ of $\La{n+k}$ can be
expressed by the generalized $q$-Wronskian.

\begin{theorem}\label{thmqgtnk}
Under the gauge transformation $T_{n+k}$ $(n>k>0)$,
\begin{gather*}
\phi^{(n+k)}(\lambda;x,\overline{t})=(T_{n+k}\cdot
\phi)=\dfrac{IW_{k,n+1}^q(\psi_k,\ldots,\psi_1;\phi_1,\ldots,\phi_n,\phi)}
{IW_{k,n}^q(\psi_k,\ldots,\psi_1;\phi_1,\ldots,\phi_n)}, \\ 
\fun{\psi}{n+k}{}(\mu;x,\overline{t})=
\big(\big(T_{n+k}^{-1}\big)^*\cdot \psi\big)
=(-1)^n\dfrac{IW^q_{k+1,n}(\psi,\psi_k,\psi_{k-1},
\ldots,\psi_{1};\phi_{1},\ldots,\phi_{n})}
{IW^q_{k,n}(\psi_k,\ldots,\psi_1;\phi_1,\ldots,\phi_n)}, \\ 
 \tu{n+k}_q=
IW^q_{k,n}(\psi_k,\ldots,\psi_1;\phi_1,\cdots,\phi_n)\cdot \tau_q.
\end{gather*}
Furthermore,  $ \fun{\phi}{n+k}{i}=
\phi^{(n+k)}(\lambda=\lambda_i;x,\overline{t})$;
$\fun{\psi}{n+k}{i}= \psi^{(n+k)}(\mu=\mu_i;x,\overline{t})$. Note
$\fun{\phi}{n+k}{i}=0$ if  $i\in \{1,2,\ldots,n\}$,
$\fun{\psi}{n+k}{i}=0$ if  $i\in \{1,2,\ldots,k \}$.
\end{theorem}

\begin{proof} (1) The repeated iteration of Theorems~\ref{thmgtI} and
\ref{thmgtII} according to the ordering of $T_I$ and $T_D$ deduces
\begin{gather*}
\fun{\phi}{n+k}{}=\T{n+k}{I}{\fun{\psi}{n+k-1}{n+k}}\cdot
\fun{\phi}{n+k-1}{}\\
\phantom{\fun{\phi}{n+k}{}}{}
=\T{n+k}{I}{\fun{\psi}{n+k-1}{n+k}}\T{n+k-1}{I}{\fun{\psi}{n+k-2}{n+k-1}}\cdot
\fun{\phi}{n+k-2}{}=\cdots\\
\phantom{\fun{\phi}{n+k}{}}{}
=\T{n+k}{I}{\fun{\psi}{n+k-1}{n+k}}\T{n+k-1}{I}{\fun{\psi}{n+k-2}{n+k-1}}\cdots\T{n+2}{I}{\fun{\psi}{n+1}{n+2}}
 \T{n+1}{I}{\fun{\psi}{n}{n+1}}\cdot\fun{\phi}{n}{}.
\end{gather*}
Then taking in it $\fun{\phi}{n}{}=(T_n\cdot\phi$)
from~\eqref{qgtnphi} , we get
\begin{gather*}
\fun{\phi}{n+k}{}=\big(\T{n+k}{I}{\fun{\psi}{n+k-1}{n+k}}\T{n+k-1}{I}
{\fun{\psi}{n+k-2}{n+k-1}}\cdots\T{n+2}{I}{\fun{\psi}{n+1}{n+2}}
 \T{n+1}{I}{\fun{\psi}{n}{n+1}} T_n\big)\cdot \phi\\
\phantom{\fun{\phi}{n+k}{}}{} =(T_{n+k}\cdot \phi).
\end{gather*}
Therefore the determinant form of $\fun{\phi}{n+k}{}$ is given by
Lemma~\ref{lemgtdetrep1}.

(2) Using Theorems \ref{thmgtI} and \ref{thmgtII} iteratively
according to the chain in  \eqref{succcessiveqgt}, similarly to
the step~(1), we can get
\begin{gather*}
\fun{\psi}{n+k}{}=\big({T_{I}^{(n+k)}}^{-1}\big)^*\cdot\fun{\psi}{n+k-1}{}=\big({T_{I}^{(n+k)}}^{-1}\big)^*
\big({T_{I}^{(n+k-1)}}^{-1}\big)^*\cdot\fun{\psi}{n+k-2}{}=\cdots\\
\phantom{\fun{\psi}{n+k}{}}{}=\big({T_{I}^{(n+k)}}^{-1}\big)^*
\big({T_{I}^{(n+k-1)}}^{-1}\big)^*\cdots
\big({T_{I}^{(n+2)}}^{-1}\big)^*
\big({T_{I}^{(n+1)}}^{-1}\big)^*\cdot\fun{\psi}{n}{}.
\end{gather*}
Noting that $\fun{\psi}{n}{}$ is given by  \eqref{qgtnpsi}, we get
$ \fun{\psi}{n+k}{}=((T_{n+k}^{-1})^*\cdot\psi). $ The explicit
form of $\fun{\psi}{n+k}{}$ is given from the determinant
representation of $T_{n+k}^{-1}$.

(3) According to the changing rule under gauge transformation in
Theorems \ref{thmgtI} and \ref{thmgtII}, the new $\tau$ function
of $q$-KP hierarchy $\tu{n+k}_q$ produced by chain of gauge
transformations in \eqref{succcessiveqgt} is
\begin{gather*}
\tu{n+k}_q=\t^{-1}\big(\fun{\psi}{n+k-1}{k}\big)\fun{\tau}{n+k-1}{q}=\t^{-1}\big(\fun{\psi}{n+k-1}{k}\big)\t^{-1}\big(
\fun{\psi}{n+k-2}{k-1}\big)\fun{\tau}{n+k-2}{q}\\
\phantom{\tu{n+k}_q}{}=\t^{-1}\big(\fun{\psi}{n+k-1}{k}\big)\t^{-1}
\big(\fun{\psi}{n+k-2}{k-1}\big)\cdots\t^{-1}\big(\fun{\psi}{n+1}{2}\big)\t^{-1}\big(\fun{\psi}{n}{1}\big)
\fun{\tau}{n}{q}.
\end{gather*}
So the explicit form of $\fun{\psi}{n+i-1}{i}$ $(i=1,2,\ldots,k)$
and $\tu{n}_q$ implies
\begin{gather*}
\tu{n+k}_q= (-1)^n\dfrac{IW^q_{k,n}(\psi_k,
\psi_{k-1},\ldots,\psi_1;\phi_1,\phi_2,\ldots,\phi_n
)}{IW^q_{k-1,n}(\psi_{k-1},\ldots,\psi_1;\phi_1,\phi_2,\ldots,\phi_n
)}\\
\phantom{\tu{n+k}_q=}{} \times
(-1)^n\dfrac{IW^q_{k-1,n}(\psi_{k-1},
\psi_{k-1},\ldots,\psi_1;\phi_1,\phi_2,\ldots,\phi_n
)}{IW^q_{k-2,n}(\psi_{k-2},\ldots,\psi_1;\phi_1,\phi_2,\ldots,\phi_n
)}\\
\phantom{\tu{n+k}_q=}{}
\cdots(-1)^n\dfrac{IW^q_{2,n}(\psi_2,\psi_1;\phi_1,\phi_2,\ldots,\phi_n
)}{IW^q_{1,n}(\psi_1;\phi_1,\phi_2,\ldots,\phi_n )}\\
\phantom{\tu{n+k}_q=}{}\times
(-1)^n\dfrac{IW^q_{1,n}(\psi_1;\phi_1,\phi_2,\ldots,\phi_n
)}{W^q_{n}(\phi_1,\phi_2,\ldots,\phi_n )}
W^q_{n}(\phi_1,\phi_2,\ldots,\phi_n )\tau_q \\
\phantom{\tu{n+k}_q} \thickapprox IW^q_{k,n}(\psi_k,
\psi_{k-1},\ldots,\psi_1;\phi_1,\phi_2,\ldots,\phi_n )\tau_q.
\end{gather*}
We omitted the trivial factor $(-1)^n$ in the last step, because
it will not af\/fect $u_i$ in the $q$-KP hierarchy.
\end{proof}

\begin{remark}
There exists another complicated chain of gauge transformation
operators\ for $q$-KP hierarchy (that may be regarded as motivated
by the classical KP hierarchy)
\begin{gather*}
L\xrightarrow{\T{1}{I}{\psi_1}}
\La{1}\xrightarrow{\T{2}{I}{\fun{\psi}{1}{2}}}
\La{2}\xrightarrow{\T{3}{I}{\fun{\psi}{2}{3}}}
\La{3}\xrightarrow{}\cdots\xrightarrow{}
\La{n-1}\xrightarrow{\T{n}{I}{\fun{\psi}{n-1}{n}}}
\La{n} \\
\xrightarrow{\T{n+1}{D}{\fun{\phi}{n}{1}}}
\La{n+1}\xrightarrow{\T{n+2}{D}{\fun{\phi}{n +1}{2}}}
\La{n+2}\xrightarrow{}\cdots\xrightarrow{}
\La{n+k-1}\xrightarrow{\T{n+k}{D}{\fun{\phi}{n+k-1}{k}}} \La{n+k},
\end{gather*}
that can lead to another form of $\tu{n+k}_q$. This is parallel to
the classical case of~\cite{csy}.
\end{remark}

   If the initial $q$-KP is a ``free'' operator, then $L=\partial_q$ that means the initial $\tau$ function is
$1$. We can write down the explicit form of $q$-KP hierarchy
generated by $T_{n+k}$. Under this situation,  \eqref{qzslax1} and
 \eqref{qzslax2} become
\begin{gather}
 \dfrac{\partial \phi}{\partial t_n}=(\partial_q^n\phi) , \qquad
 \dfrac{\partial \psi}{\partial
t_n}=-(\partial_q^{n*}\psi),\label{zeroqzslax1+zeroqzslax2}
\end{gather}
that possess set of solution  \{$\phi_i, \psi_i$\}  as follows
\begin{gather}
\phi_i(x;\overline{t})=e_q(\lambda_{i_1}x)e^{\sum\limits_{j=1}^{\infty}t_j\lambda_{i_1}^j}+a_ie_q(\mu_{i_1}x)
e^{\sum\limits_{j=1}^{\infty}t_j\mu_{i_1}^j},\label{zeroinitialvalueegienfun}  \\
\psi_i(x;\overline{t})=e_{1/q}(-\lambda_{i_2}qx)e^{-\sum\limits_{j=1}^{\infty}
t_j\lambda_{i_2}^j}
  +b_ie_{1/q}(-\mu_{i_2}qx)e^{-\sum\limits_{j=1}^{\infty}t_j\mu_{i_2}^j}.
  \label{zeroinitialvalueadjointegienfun}
  \end{gather}
After the $(n+k)$-th step gauge transformation $T_{n+k}$, the
f\/inal form of $\tau_q$ can be given in following corollary,
which can be deduced directly from Theorems \ref{thmqgtn} and
\ref{thmqgtnk}.

\begin{corollary}\label{corzeroinitialtauqkp}
The gauge transformation  can generate the following two forms of
$\tau$ function of the $q$-KP hierarchy,
\begin{gather}
\tu{n+k}_q=
IW_{k,n}^q(\psi_k,\ldots,\psi_1;\phi_1,\ldots,\phi_n)=\begin{vmatrix}
          \pq\psi_k\phi_1    &\pq\psi_k\phi_2    &\cdots  &\pq\psi_k\phi_n\\
          \vdots             &\vdots             &\cdots  &\vdots\\
          \pq\psi_1\phi_1    &\pq\psi_1\phi_2    &\cdots  &\pq\psi_1\phi_n\\
          \phi_1             &\phi_2             &\cdots  &\phi_n\\
          \p_q\phi_1         &\p_q\phi_2         &\cdots  &\p_q\phi_n\\
          \vdots             &\vdots             &\cdots  &\vdots\\
          \p_q^{n-k-1}\phi_1 &\p_q^{n-k-1}\phi_2 &\cdots  &\p_q^{n-k-1}\phi_n
          \end{vmatrix},\nonumber\\ 
\tu{n}_q=W_{n}^q(\phi_1,\ldots,\phi_n)=\begin{vmatrix}
          \phi_1             &\phi_2             &\cdots  &\phi_n\\
          \p_q\phi_1         &\p_q\phi_2         &\cdots  &\p_q\phi_n\\
          \vdots             &\vdots             &\cdots  &\vdots\\
          \p_q^{n-1}\phi_1 &\p_q^{n-1}\phi_2 &\cdots  &\p_q^{n-1}\phi_n
          \end{vmatrix}.\label{zerointialtaun}
\end{gather}
Here \{$\phi_i, \psi_i$\} are def\/ined by
 \eqref{zeroinitialvalueegienfun} and
 \eqref{zeroinitialvalueadjointegienfun}.
\end{corollary}

On the other hand, we know from  \eqref{qgtsato} that $T_n$
def\/ined by  \eqref{qgtTn} is a dressing operator if its
generating functions are given by
\eqref{zeroinitialvalueegienfun}. Therefore we can def\/ine one
$q$-wave function
\begin{gather}\label{zeroinitialqwave}
\omega_q=T_n\p_q^{-n}e_q(xz)e^{\sum\limits_{i=1}^{\infty}z^it_i}=\frac{1}{W_n^q}\begin{vmatrix}
     \phi_1            &\cdots&\phi_n            & z^{-n} \\
     \partial_q \phi_1 &\cdots&\partial_q \phi_n & z^{-n+1}\\
     \vdots &\cdots & \vdots &\vdots\\
     \partial_q^n\phi_1&\cdots &\partial_q^n \phi_n   &1
     \end{vmatrix}e_q(xz)e^{\sum\limits_{i=1}^{\infty}z^it_i}.
\end{gather}

\begin{corollary}
The relationship in  \eqref{qwavetau} between the $q$-wave
function and $\tau_q$ is satisf\/ied by $\tu{n}_q$
in~\eqref{zerointialtaun} and $q$-wave function in
\eqref{zeroinitialqwave}, i.e.,
\begin{gather} \label{qrelationwavetau}
  \omega_q=\frac{\tu{n}_q(x;\overline{t}-[z^{-1}])}{\tu{n}_q(x;\overline{t})}e_q(xz)
  \exp\left(\sum_{i=1}^{\infty}t_iz^i\right).
\end{gather}
\end{corollary}

\begin{proof}
We follow the Dickey's method on page 100 of \cite{dl1} to prove
the corollary. By direct computations,
\begin{gather*}
\phi_k(x;t-[z^{-1}])=e_q(\lambda_kx)e^{\sum\limits_{i=1}^{\infty}z^it_i-\left(\frac{\lambda_k}{z}
+\frac{\lambda_k^2}{2z^2}+\cdots\right)}
+a_ke_q(\mu_kx)e^{\sum\limits_{i=1}^{\infty}z^it_i-
\left(\frac{\mu_k}{z}+\frac{\mu_k^2}{2z^2}+\cdots\right)}\\
\phantom{\phi_k(x;t-[z^{-1}])}{}
=\phi_k-\frac{1}{z}\partial_q\phi_k
\end{gather*}
whence
\[
\frac{\tu{n}_q(x;\overline{t}-[z^{-1}])}{\tu{n}_q(x;\overline{t})}=\frac{1}{W_q}\begin{vmatrix}
     \phi_1-\frac{1}{z}\partial_q\phi_1     & \phi_2-\frac{1}{z}\partial_q\phi_2      &\cdots&\phi_n-\frac{1}{z}\partial_q\phi_n \\
     \partial_q \phi_1-\frac{1}{z}\partial_q^2\phi_1& \partial_q \phi_2-\frac{1}{z}\partial_q^2\phi_2  &\cdots&\partial_q \phi_n-\frac{1}{z}\partial_q^2\phi_n\\
     \vdots & \vdots& \cdots & \vdots \\
     \partial_q^{n-1}\phi_1-\frac{1}{z}\partial_q^n\phi_1&\partial_q^{n-1}\phi_2-\frac{1}{z}\partial_q^n\phi_2 &\cdots &\partial_q^n
     \phi_n-\frac{1}{z}\partial_q^n\phi_n
     \end{vmatrix}.
\]
Comparing the fraction above of  the determinant term with
 \eqref{zeroinitialqwave}, we can see that they are similar, although the form of the
determinant in  the numerator is dif\/ferent. The determinant in
the numerator of  \eqref{zeroinitialqwave} can be reduced to the
same form of
 \eqref{qrelationwavetau} if the second row, divided by $z$, is
subtracted from the f\/irst one, the third from the second etc.
\end{proof}

At the end of this section, we would like to discuss $q$-ef\/fects
in the solution of $q$-KP hierarchy. By direct calculation, we get
that the f\/irst f\/low of $q$-KP is
\begin{gather*}
\partial_{t_1}u_0=x(q-1)(\partial_qu_1),\\
\partial_{t_1}u_{-1}=(\partial_qu_{-1})+u_0u_{-1}+\t(u_{-2})-u_{-2}-u_{-1}\t^{-1}(u_0)),\\
\partial_{t_1}u_{-2}=(\partial_qu_{-2})+u_0u_{-2}+\t(u_{-3})
+\big[-u_{-3}+q^{-1}u_{-1}\t^{-2}(\partial_qu_0)-u_{-2}\t^{-2}(u_0)\big],\\
\partial_{t_1}u_{-3}=(\partial_qu_{-3})+u_0u_{-3}+\t(u_{-4})+\big[-u_{-4}-q^{-3}u_{-1}\t^{-3}(\partial_q^2u_0) \notag \\
\phantom{\partial_{t_1}u_{-3}=}{} +(q^{-1}+q^{-2})u_{-2}\t^{-3}(\partial_qu_0)-u_{-3}\t^{-3}(u_0)\big],\\
\partial_{t_1}u_{-i}=(\partial_qu_{-i})+u_0u_{-i}+\t(u_{-i-1})+\big[-u_{-i-1}+(\cdots)-u_{-i}\t^{-i}(u_0)\big],\\
\cdots\cdots\cdots\cdots\cdots\cdots\cdots\cdots\cdots\cdots\cdots\cdots\cdots\cdots\cdots\cdots\cdots\cdots\cdots\cdots\cdots
\end{gather*}
in which
$(\cdots)=\sum\limits_{k=1}^{i-1}a_{-k}u_{-k}\t^{-i}(\partial_q^{i-k}u_0)$
$(i=2,3,\ldots)$, and $a_{-k}$  depends on $q$ only. We can see
that
\begin{gather*}
\partial_{t_1}u_0=0,\qquad
\partial_{t_1}u_{-i}=\partial_qu_{-i}=\partial_xu_{-i}, \qquad i\geq 1.
\end{gather*}
when $q\rightarrow 1$. This result shows that the variable $t_1$
in $q$-KP hierarchy is corresponding to the variable $x$ in KP
hierarchy. So we have two global parameters in $q$-KP hierarchy,
namely~$x$ and~$q$. In order to show $q$-ef\/fect, we will write
out the concrete form of single $q$-soliton  of $q$-KP equation,
namely, we let $u_{-1}$ depend on three variable $(t_1,t_2, t_3)$
beside two parameters~$(x,q)$. We consider $\La{1}$ generated by
one step of $T_D(\phi_1)$ from $L=\partial_q$, and the generating
function is given by
\begin{gather}\label{oneqsolitongeneratingfun}
\phi_1=e_q(\lambda_1x)e^{\xi_1}+ B_1e_q(\lambda_2x)e^{\xi_2}
\end{gather}
from \eqref{zeroinitialvalueegienfun},
 then the Corollary \ref{corzeroinitialtauqkp} shows that the
$\tau$ function of $\La{1}$ is $\tu{1}_q=\phi_1$ in
 \eqref{oneqsolitongeneratingfun}. Here
$\xi_k=\lambda_kt_1+\lambda^2_kt_2+\lambda_k^3t_3$ $(k=1,2)$,
$B_1$ is real constant. Taking this $\tu{1}_q$ back into
 \eqref{qStau}, then~\eqref{qkpu}, we get $q$-soliton of $q$-KP as
\begin{gather*}
u_{-1}=\left[ 1+x(q-1)\left(
\dfrac{\lambda_1e_q(\lambda_1x)e^{\xi_1}+\lambda_2B_1e_q(\lambda_2x)e^{\xi_2}}
{e_q(\lambda_1x)e^{\xi_1}+ B_1e_q(\lambda_2x)e^{\xi_2}} \right)
\right] \notag\\
\phantom{u_{-1}=}{} \times\Bigg\{\dfrac{(\lambda_1^2
e_q(\lambda_1x)e^{\xi_1}+
B_1\lambda_2^2e_q(\lambda_2x)e^{\xi_2})(e_q(\lambda_1qx)e^{\xi_1}+
B_1e_q(\lambda_2qx)e^{\xi_2})} {(e_q(\lambda_1qx)e^{\xi_1}+
B_1e_q(\lambda_2qx)e^{\xi_2})(e_q(\lambda_1x)e^{\xi_1}+
B_1e_q(\lambda_2x)e^{\xi_2})} \notag \\
\phantom{u_{-1}=}{} - \dfrac{ (\lambda_1e_q(\lambda_1x)e^{\xi_1}+
B_1\lambda_2e_q(\lambda_2x)e^{\xi_2})(\lambda_1e_q(\lambda_1qx)e^{\xi_1}+
B_1\lambda_2e_q(\lambda_2qx)e^{\xi_2})}
{(e_q(\lambda_1qx)e^{\xi_1}+
B_1e_q(\lambda_2qx)e^{\xi_2})(e_q(\lambda_1x)e^{\xi_1}+
B_1e_q(\lambda_2x)e^{\xi_2})} \Bigg\}.
\end{gather*}
In particular,  if $q\rightarrow 1$, we have
\begin{gather*}
u_{-1}=\dfrac{B_1(\lambda_1-\lambda_2)^2}{e^{\hat{\xi}_1-\hat{\xi}_2}+B_1^2e^{\hat{\xi}_2-\hat{\xi}_1}+2B_1},
\end{gather*}
which is a single soliton of the classical KP when $x \rightarrow
0$. Here $\hat{\xi}_k=\lambda_kx+\xi_k$ $(k=1,2)$. In order to
plot a f\/igure for $u_{-1}$, we f\/ix $\lambda_1=2$,
$\lambda_2=-1.5$ and $B_1=1$, so $u_{-1}=u_{-1}(x,t_1,t_2,t_3,q)$.
The single $q$-soliton $u_{-1}(0.001,t_1,t_2,t_3,0.999)$ is
plotted in Fig.~1, which is close to classical soliton of KP
equation as we analysed above. From Figs.~2--5\footnote{For
Figs.~2--5, $q$-ef\/fect $\mbox{Du}_{-1}\equiv \triangle u_{-1}
\triangleq u_{-1}(q=0.999)- u_{-1}(q=i)$ with $x=0.5$ and $t_3=0$,
where $i=0.7,0.5,0.3,0.1$. Figs.~6--9, are projection of
Figs.~2--5, by f\/ixing $t_2=-5$.} we can see the varying trends
of $\vartriangle\!\!
u_{-1}=u_{-1}(0.5,t_1,t_2,0,0.999)-u_{-1}(0.5,t_1,t_2,0,q)\triangleq
u_{-1}(q=0.999)- u_{-1}(q)$ for certain values of $q$,  where
$q=0.7,0.5,0.3,0.1$ respectively. Furthermore, in order to see the
$q$-ef\/fects more clearly, we further f\/ixed $t_2=-5$ in
$\vartriangle\!\! u_{-1}$, which are plotted in Figs.~6--9.
Dependence of
$\vartriangle\!\!u_{-1}=u_{-1}(x,t_1,-5,0,0.999)-u_{-1}(x,t_1,-5,0,0.1)
\triangleq u_{-1}(x,q=0.999)- u_{-1}(x,q=0.1) $ on  $x$ is shown
in Figs.~10--14,  and $x=0.3,0.4,0.52,0.54,0.55 $ respectively. It
is obvious from f\/igures that $\vartriangle\!\!u_{-1}$ goes to
zero when $q\rightarrow 1$ and $x\rightarrow 0$, $q$-soliton
($u_{-1}$) of $q$-KP goes to a usual soliton of KP, which
reproduces the process of $q$-deformation.  On the other hand,
Figs.~10--14\footnote{For Figs.~10--14, the variable $x$, varies
as follows: $0.3,0.4,0.52,0.54,0.55$, while $q=i=0.1$ in $\mbox{Du}_{-1}$ is f\/ixed.}
show parameter $x$ amplif\/ies $q$-ef\/fects. In other word, for a
given $\vartriangle\!\! q$, $\vartriangle\!\! u_{-1}$ will
increase along~$x$. However, $x$ is bounded so that $e_q(\lambda_k
x)$ and $e_q(\lambda_k q x)$ $(k=1,2)$ are convergent. This is the
reason for plotting $u_{-1}$ with $x\leq 0.55$. Obviously, the
convergent interval depends on $q$ and $\lambda_k$.  We would like
to emphasize that from  Figs.~6--14\footnote{For Figs.~6--14,
$\mbox{Du}_{-1}$, $u_{-1}(q=0.999)$, are represented by continuous
line and dashed line (long), respectively, while dashed line
(short) represent $u_{-1}(q=i)$, $i=0.7,0.5,0.3,0.1$ for
Figs.~6--9.} the
$q$-deformation does not destroy the prof\/ile of soliton; it just
similar to an ``impulse'' to soliton.

\begin{figure}[t]
  \centering
\includegraphics[width=7cm]{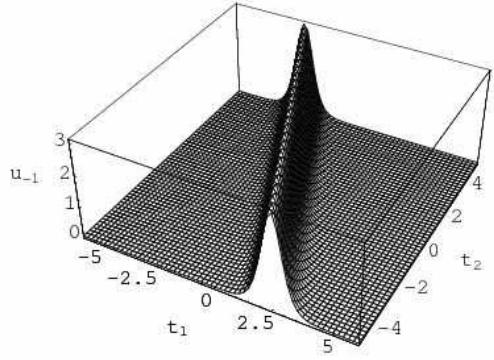}
  \vspace{-2mm}
  \caption{$u_{-1}(x=0.001,q=0.999)$ with $t_3=0$.}
  \label{fig1}
\end{figure}

\begin{figure}[t]
\begin{minipage}[b]{7.5cm}
\centering \includegraphics[width=7cm]{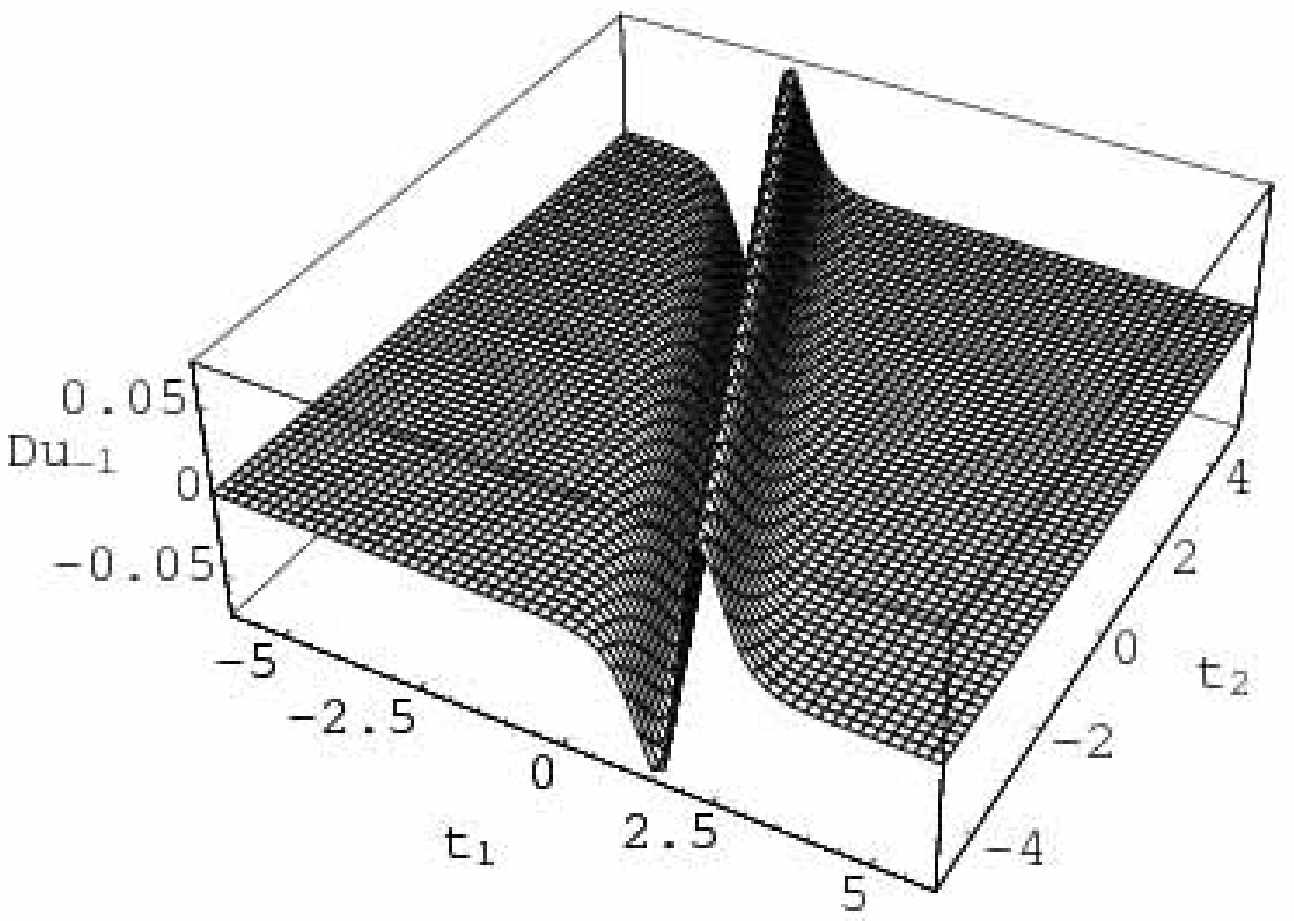}\vspace{-2mm}
\caption{}  \label{fig2}
\end{minipage}\hfill
\begin{minipage}[b]{7.5cm}
  \centering
  \includegraphics[width=7cm]{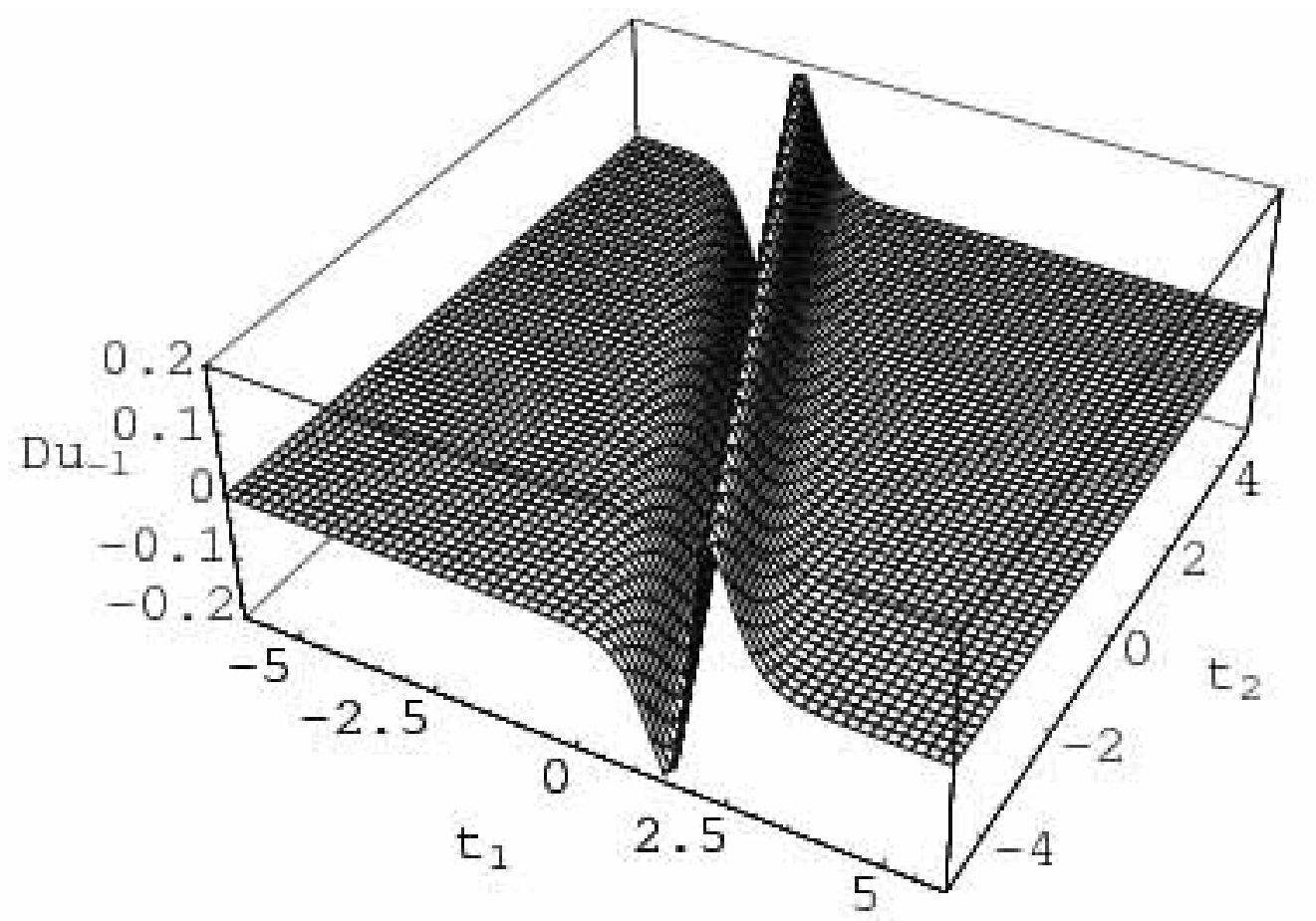}\vspace{-2mm}
  \caption{}
  \label{fig3}
\end{minipage}
\end{figure}

\begin{figure}[t]
\begin{minipage}[b]{7.5cm}
\centering \includegraphics[width=7cm]{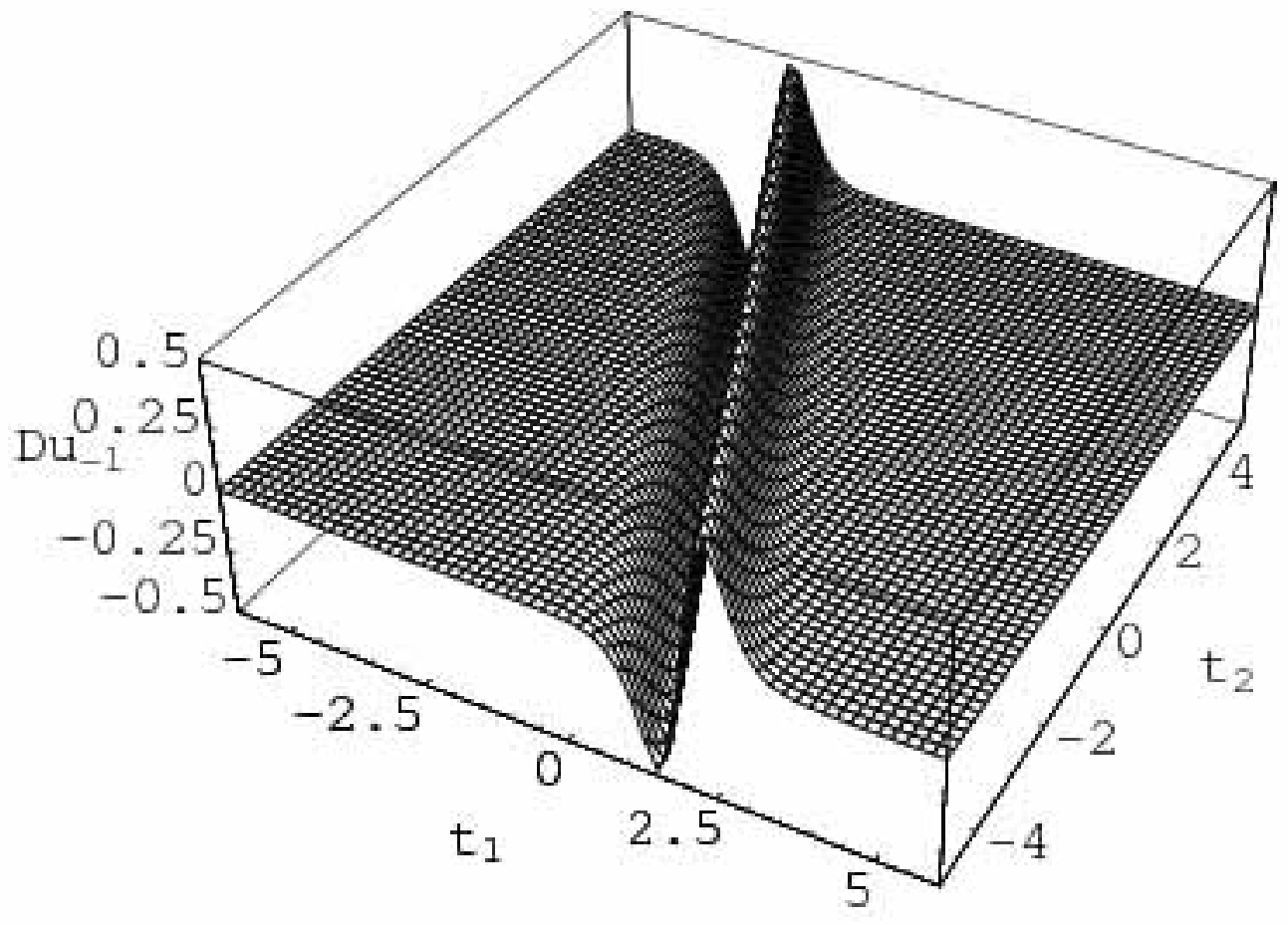}\vspace{-2mm}
\caption{}  \label{fig4}
\end{minipage}\hfill
\begin{minipage}[b]{7.5cm}
  \centering
  \includegraphics[width=7cm]{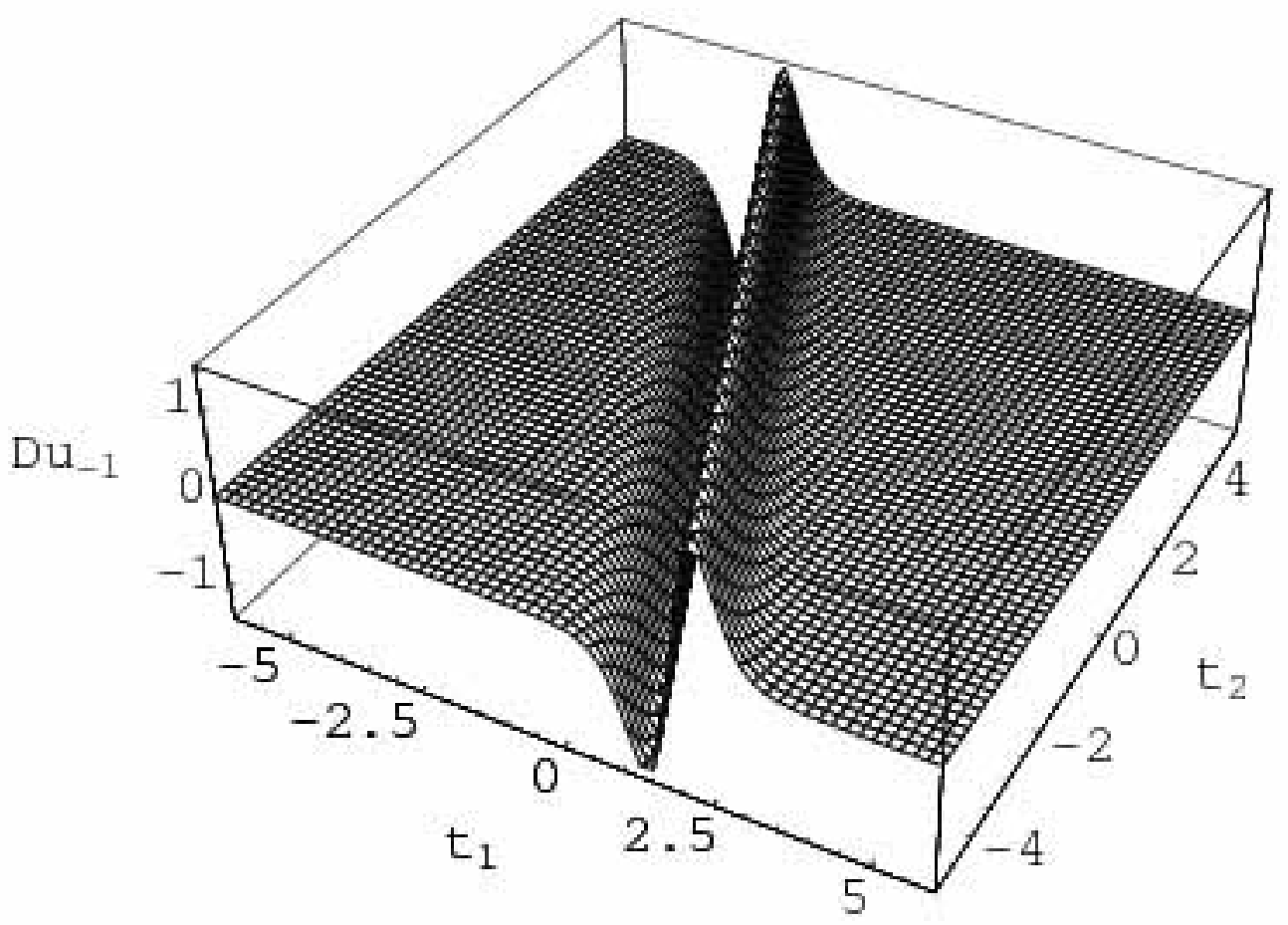}\vspace{-2mm}
  \caption{}
  \label{fig5}
\end{minipage}
\end{figure}

\begin{figure}[t]
\begin{minipage}[b]{7.5cm}
\centering \includegraphics[width=7cm]{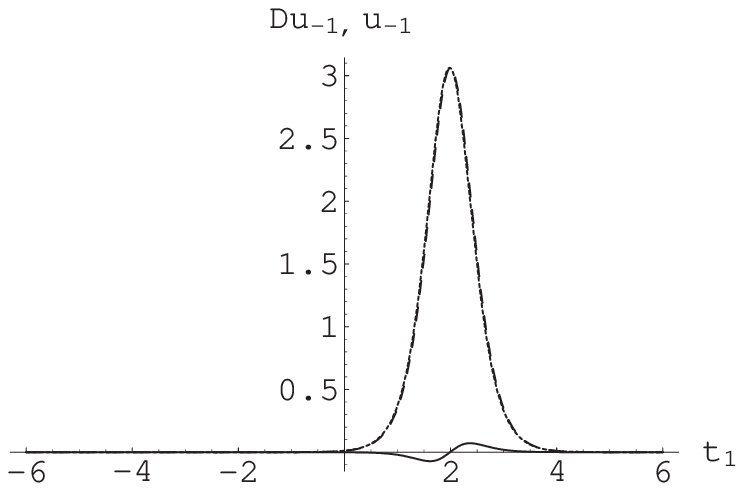}\vspace{-2mm}
\caption{}  \label{fig6}
\end{minipage}\hfill
\begin{minipage}[b]{7.5cm}
  \centering
  \includegraphics[width=7cm]{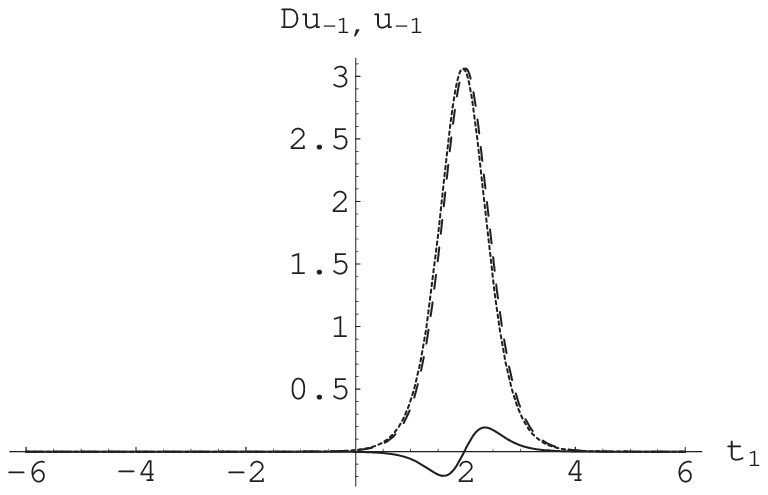}\vspace{-2mm}
  \caption{}
  \label{fig7}
\end{minipage}
\end{figure}

\begin{figure}[t]
\begin{minipage}[b]{7.5cm}
\centering \includegraphics[width=7cm]{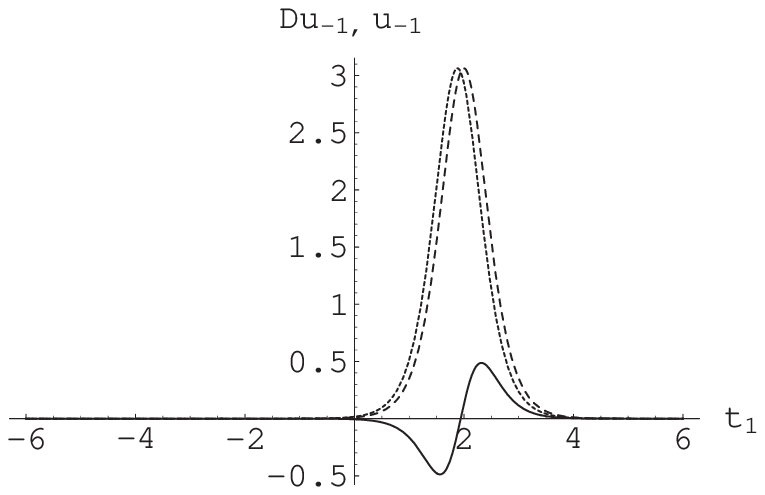}\vspace{-2mm}
\caption{}  \label{fig8}
\end{minipage}\hfill
\begin{minipage}[b]{7.5cm}
  \centering
  \includegraphics[width=7cm]{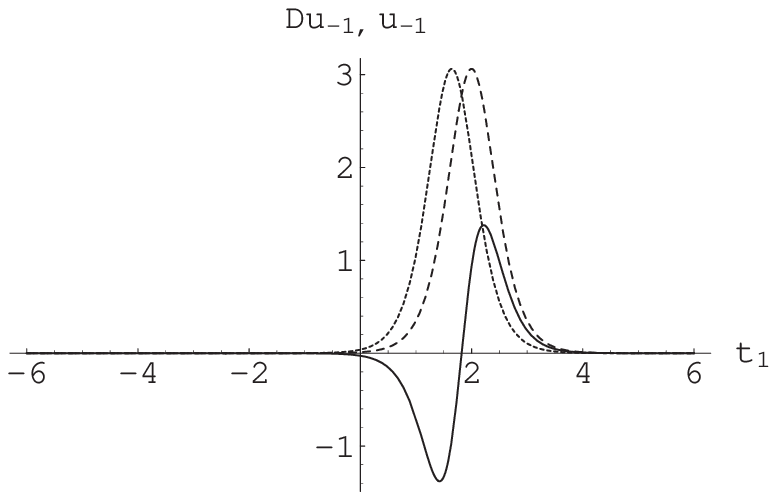}\vspace{-2mm}
  \caption{}
  \label{fig9}
\end{minipage}
\end{figure}

\begin{figure}[t]
\begin{minipage}[b]{7.5cm}
\centering \includegraphics[width=7cm]{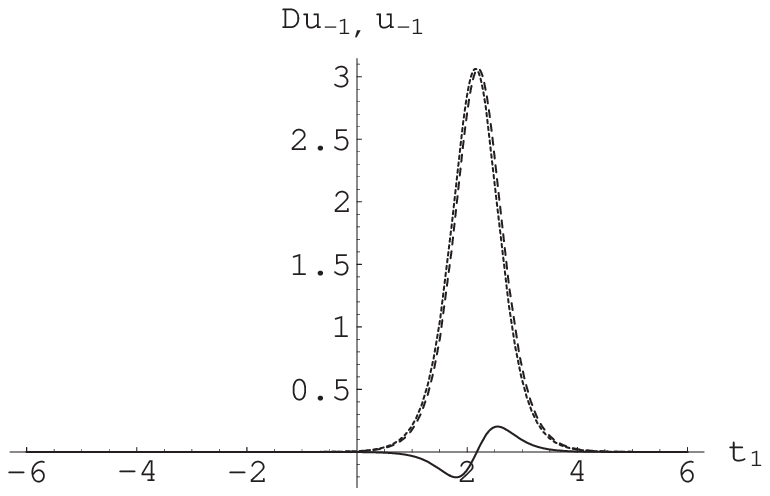}\vspace{-2mm}
\caption{}  \label{fig10}
\end{minipage}\hfill
\begin{minipage}[b]{7.5cm}
  \centering
  \includegraphics[width=7cm]{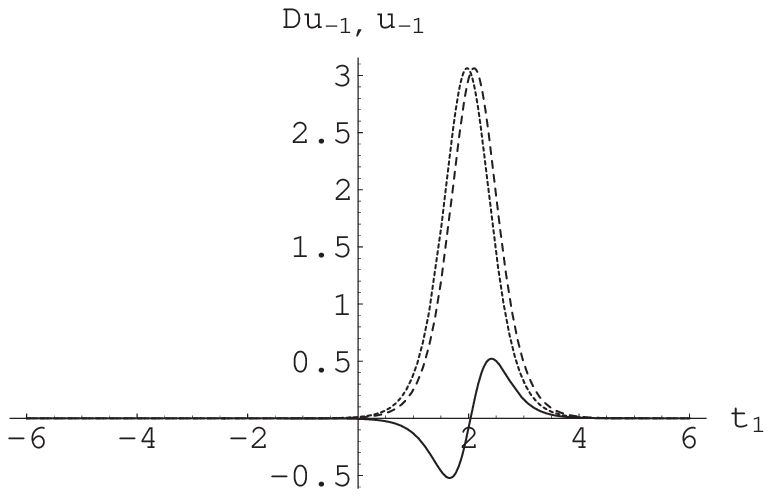}\vspace{-2mm}
  \caption{}
  \label{fig11}
\end{minipage}
\end{figure}

\begin{figure}[t]
\begin{minipage}[b]{7.5cm}
\centering \includegraphics[width=7cm]{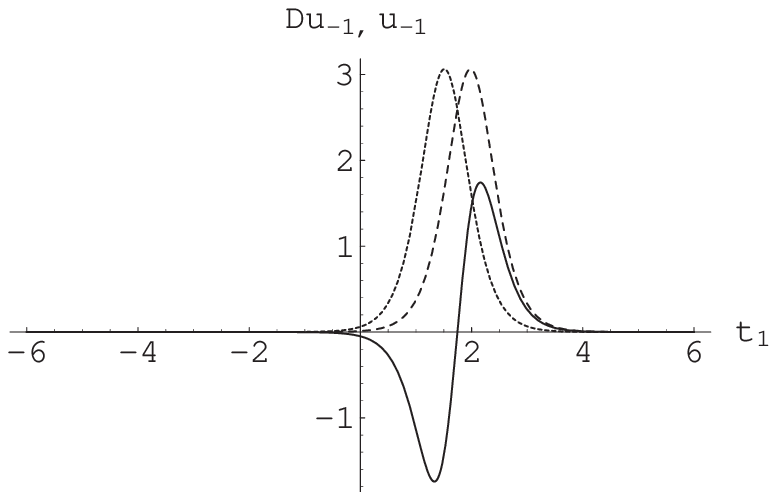}\vspace{-2mm}
\caption{}  \label{fig12}
\end{minipage}\hfill
\begin{minipage}[b]{7.5cm}
  \centering
  \includegraphics[width=7cm]{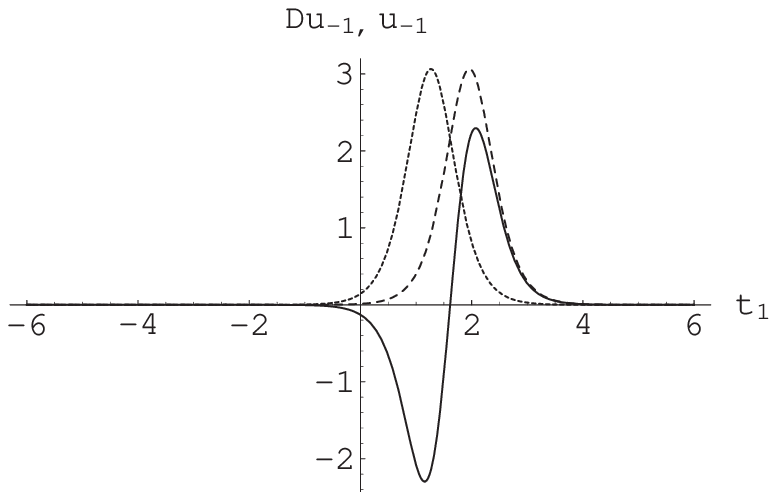}\vspace{-2mm}
  \caption{}
  \label{fig13}
\end{minipage}
\end{figure}

\begin{figure}[t]
\centering \includegraphics[width=7cm]{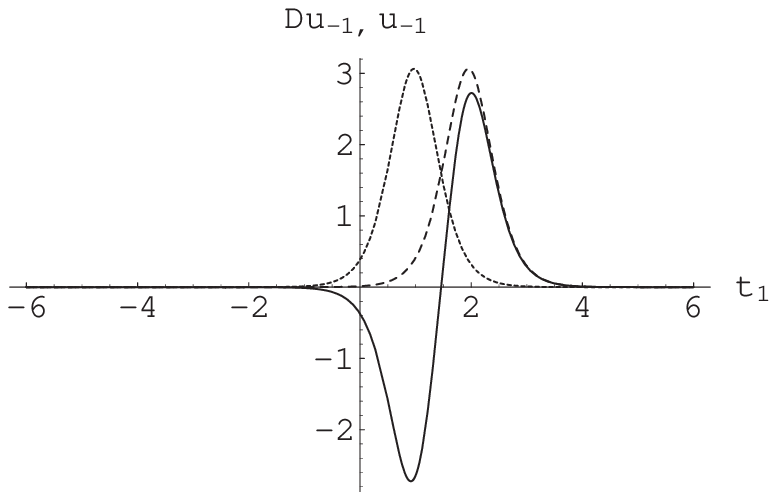}\vspace{-2mm}
\caption{}  \label{fig14}
\end{figure}

\section[Symmetry constraint of $q$-KP: $q$-cKP hierarchy]{Symmetry constraint
of $\boldsymbol{q}$-KP: $\boldsymbol{q}$-cKP hierarchy}

 We know that there exists a constrained
version of KP hierarchy, i.e.\ the constrained KP hierarchy (cKP)
\cite{kss,aratyn2}, introduced by means of the symmetry constraint
from KP hierarchy. With inspiration from it, the symmetry of
$q$-KP was established in \cite{tu}. In the same article the
authors def\/ined one kind of constrained $q$-KP ($q$-cKP)
hierarchy by using the linear combination of generators of
additional symmetry. In this section, we shall brief\/ly introduce
the symmetry and $q$-cKP hierarchy~\cite{tu}.

The linearization of  \eqref{qlaxequation} is given by
\begin{gather}
  \partial_{t_m}(\delta L)=[\delta B_m,L]+[B_m,\delta L], \label{linearizationqlaxeq}
  \end{gather}
  where
\begin{gather*}
  \delta B_m=\left(\sum_{r=1}^mL^{m-r}\delta LL^{r-1}\right)_+.  
\end{gather*}
We call $\delta L=\delta u_0+\delta u_1\partial_q^{-1}+\cdots$ the
symmetry of the $q$-KP hierarchy, if it satisf\/ies
\eqref{linearizationqlaxeq}. Let~$L$ be a ``dressed'' operator
from $\partial_q$, we f\/ind
 \begin{gather}
   \delta L=\delta S\partial_qS^{-1}-S\partial_qS^{-1}\delta
   SS^{-1}=[\delta SS^{-1},L]=[K,L],\label{variationqL}
 \end{gather}
 where $\delta S=\delta s_1\partial_q^{-1}+\delta
 s_2\partial_q^{-2}+\cdots$, and $K=\delta SS^{-1}$. Therefore
\[ \delta B_m=[K,L^m]_+=[K,B_m]_+,
\]
the last identity is resulted by $K=K_-$ and $[K,L^m_-]_+=0$. Then
the linearized equation (\ref{linearizationqlaxeq}) is equivalent
to
\begin{gather}\label{equivlinearizationqlaxeq}
  \partial_{t_m}K=[B_m,K]_-,\qquad  \delta S=KS.
\end{gather}

Let $K_n=-(L^{n})_-$ $(n=1,2,\ldots)$, then it can easily be
checked that $K_n$ satisfies (\ref{equivlinearizationqlaxeq}). For
each $K_n$, $\delta L$ is given by $\delta L=-[(L^n)_-,L]=[B_n,L]$
from
 \eqref{variationqL}. So the $q$-KP hierarchy admits a~reduction
def\/ined by $(L^n)_-=0$, which is called $q$-deformed $n$-th KdV
hierarchy. For example, $n=2$, it leads to $q$-KdV hierarchy,
whose $q$-Lax operator is
\begin{gather*}
L_{q{\rm KdV}}=L^2=L^2_+=\partial_q^2 +x(q-1)u\partial_q +u.
\end{gather*}
There is also another symmetry called additional symmetry, which
is $K=(M^mL^l)_-$~\cite{tu}, and  it also satisf\/ies
(\ref{equivlinearizationqlaxeq}). Here the operator $M$ is
def\/ined by
\begin{gather*}
   \partial_{t_k}M=[L_k^+,M],\qquad M=S\Gamma_qS^{-1},
    \end{gather*}
  and $\Gamma_q$ is def\/ined as
 \[
    \Gamma_q=\sum_{i=1}^{\infty}\left[it_i+\frac{(1-q)^i}{1-q^i}x^i\right]\partial_q^{i-1}.
 \]
The more general generators of additional symmetry are in form of
\begin{gather*}
Y_q(\mu,\lambda)=\sum\limits_{m=0}^{\infty}
 \dfrac{(\mu-\lambda)^m}{m!}\sum\limits_{l=-\infty}^{\infty}\lambda^{-m-l-1}\big(M^mL^{m+l}\big)_-,
\end{gather*}
 which are constructed by combination of $K=(M^mL^l)_-$. The
 operator $Y_q(\mu,\lambda)$ can be expressed as
 \begin{gather*}
Y_q(\mu,\lambda)=\omega_q(x,\overline{t};\mu)\circ
\partial_q^{-1}\circ \t(\omega^*_q(x,\overline{t};\lambda)).
 \end{gather*}
In order to def\/ine the $q$-analogue of the constrained KP
hierarchy, we need to establish one special generator of symmetry
$ Y(t)=\phi(t)\circ
\partial_q^{-1}\circ \psi(t) $
based on $Y_q(\mu,\lambda)$, where
\begin{gather*}
\phi(t)=\int\rho(\mu)\omega_q(x,\overline{t};\mu){\rm d}\mu,
\qquad
\psi(t)=\int\chi(\lambda)\t(\omega^*_q(x,\overline{t};\lambda)){\rm
d}\lambda,
\end{gather*}
further  $\phi(t)$ and $\psi(t)$ satisfy  \eqref{qzslax1} and
 \eqref{qzslax2}.  In other words, we get
a new symmetry of $q$-KP hierarchy,
\begin{gather}\label{generatorqckP}
K=\phi(\lambda;x,\overline{t})\circ \partial_q^{-1}\circ
\psi(\mu;x,\overline{t}),
\end{gather}
where $\phi(\lambda;x,\overline{t})$ and
$\psi(\mu;x,\overline{t})$ is an ``eigenfunction'' and an
``adjoint eigenfunction'', respectively. We can regard from the
process above that $K=\phi(\lambda;x,\overline{t})\circ
\partial_q^{-1}\circ \psi(\mu;x,\overline{t})$ is a special linear
combination of the additional symmetry generator $(M^mL^l)_-$. It
is obvious that generator $K$ in~(\ref{generatorqckP}) satisf\/ies
 \eqref{equivlinearizationqlaxeq}, because of the following two
operator identities,
\begin{gather}\label{qopertoridenitiesII}
   (A\circ a\circ \partial_q^{-1}\circ b)_-
   =(A\cdot a)\circ \partial_q^{-1}\circ b,(a\circ \partial_q^{-1}\circ b\circ A)_-
   = a\circ \partial_q^{-1}\circ (A^*\cdot b).
\end{gather}
Here $A$ is a $q$-PDO, and $a$ and $b$ are two functions.
Naturally, $q$-KP hierarchy  also has a~multi-component symmetry,
i.e.
\begin{gather*}
K=\sum\limits_i^n\phi_i\circ
\partial_q^{-1}\circ \psi_i.
\end{gather*}

It is well known that the integrable KP hierarchy is compatible
with generalized $l$-constraints of this type
$(L^l)_-=\sum\limits_iq_i\circ\partial_x^{-1}\circ r_i$.
Similarly, the $l$-constraints of $q$-KP hierarchy
\begin{gather*}
    (L^l)_-=K=\sum_{i=1}^m \phi_i\circ\partial_q^{-1} \circ \psi_i
\end{gather*}
also lead to $q$-cKP hierarchy. The f\/low equations of this
$q$-cKP hierarchy
 \begin{gather} \label{qcKP}
   \partial_{t_k}L^l =[L_+^k,L^l],\qquad
   L^l=(L^l)_++\sum_{i=1}^m \phi_i\circ\partial_q^{-1}\circ \psi_i
 \end{gather}
are compatible with
\begin{gather*}
   (\phi_i)_{t_k}=((L^k)_+\phi_i),\qquad (\psi_i)_{t_k}=-((L^{*k})_+\psi_i).
    \end{gather*}
It can be obtained directly by using the operator identities in
 \eqref{qopertoridenitiesII}. An important fact is that there
exist two $m$-th order $q$-dif\/ferential operators
\begin{gather*}
   A=\partial_q^m+a_{m-1}\partial_q^{m-1}+\cdots+a_0,
   \qquad
   B=\partial_q^m+b_{m-1}\partial_q^{m-1}+\cdots+b_0,
   \end{gather*}
such that $AL^l$ and $L^lB$ are dif\/ferential operators. From
$(AL^l)_-=0$ and $(L^lB)_-=0$, we get that $A$ and $B$ annihilate
the functions $\phi_i$ and $\psi_i$, i.e.,
$A(\phi_1)=\cdots=A(\phi_m)=0$,
$B^*(\psi_1)=\cdots=B^*(\psi_m)=0$, that implies  $\phi_i \in {\rm
Ker}\, (A)$. It should be noted that ${\rm Ker}\,(A)$ has
dimension $m$. We will use this fact to reduce the number of
components of the $q$-cKP hierarchy in the next section.

\section[$q$-Wronskian solutions of the $q$-cKP hierarchy]{$\boldsymbol{q}$-Wronskian
solutions of $\boldsymbol{q}$-cKP hierarchy}

We know from Corollary~\ref{corzeroinitialtauqkp} that
$q$-Wronskian
\begin{gather}
\tu{N}_q=W_{N}^q(\phi_1,\ldots,\phi_N)=\begin{vmatrix}
          \phi_1             &\phi_2             &\cdots  &\phi_N\\
          \p_q\phi_1         &\p_q\phi_2         &\cdots  &\p_q\phi_N\\
          \vdots             &\vdots             &\cdots  &\vdots\\
          \p_q^{N-1}\phi_1 &\p_q^{N-1}\phi_2 &\cdots
          &\p_q^{N-1}\phi_N
          \end{vmatrix},\label{zerointialtaunb}
\end{gather}
is a $\tau$ function of $q$-KP hierarchy. Here $\phi_i$
$(i=1,2,\ldots,N)$ satisfy linear $q$-partial dif\/ferential
equations,
\begin{gather}\label{zeroinitiallinearfunforqckp}
\dfrac{\partial \phi_i}{\partial t_n}= (\partial_q^n\phi_i),\qquad
n=1,2,3,\ldots.
\end{gather}
In this section, we will reduce $\tu{N}_q$ in
 \eqref{zerointialtaunb} to a $\tau$ function of $q$-cKP
hierarchy. To this end, we will f\/ind the additional conditions
satisf\/ied by $\phi_i$ except the linear $q$-dif\/ferential
equation
 \eqref{zeroinitiallinearfunforqckp}.

  Corollary \ref{corzeroinitialtauqkp} also shows that the $q$-KP hierarchy
with Lax operator $\La{N}=T_N\circ \partial_q \circ T_N^{-1}$ is
generated from the ``free'' Lax operator $L=\partial_q$, which has
the $\tau$ function $\tu{N}_q$ in
 \eqref{zerointialtaunb}. In order to get the explicit form of
such Lax operator $\La{N}$, the  following lemma is necessary.
\begin{lemma}\label{lemTNandInverse}
\begin{gather*}
  T_{N}=\frac{1}{W_{N}^q(\phi_1,\ldots,\phi_N)}
  \begin{vmatrix}
  \phi_1                &\cdots  & \phi_N                & 1\\
  \p_q\phi_1            &\cdots  & \p_q\phi_N            & \p_q\\
  \vdots                &\cdots  & \vdots                & \vdots\\
  \p_q^{N}\phi_1      &\cdots  & \p_q^{N}\phi_N      & \p_q^{N}
  \end{vmatrix}
\end{gather*}
and
\begin{gather*}
T_{N}^{-1}=  \begin{vmatrix}
   \phi_1\circ\pq &\theta(\phi_1)& \cdots &\theta(\p_q^{N-2}\phi_1)\\
   \phi_2\circ\pq &\theta(\phi_2)& \cdots &\theta(\p_q^{N-2}\phi_2)\\
   \vdots         &\vdots        & \cdots &\vdots\\
   \phi_N\circ\pq &\theta(\phi_N)& \cdots &\theta(\p_q^{N-2}\phi_N)
   \end{vmatrix}
   \cdot \frac{(-1)^{N-1}}{\theta(W_{N}^q(\phi_1,\ldots,\phi_N))}
    =\sum\limits_{i=1}^N\phi_i\circ \partial_q^{-1}\circ g_i \!\!\! 
\end{gather*}
with
\begin{gather}\label{gidetrepqTNinverse}
g_i= (-1)^{N-i}\t\big(\dfrac{W_{N}^q(\phi_1,\ldots,
\phi_{i-1},\hat{i},\phi_{i+1},\ldots,\phi_N)}{W_{N}^q(\phi_1,\ldots,
\phi_{i-1},\phi_i, \phi_{i+1},\ldots,\phi_N)}\big).
\end{gather}
Here $\hat{i}$ means that the column containing $\phi_i$ is
deleted from $W_N^q(\phi_1,\ldots, \phi_{i-1},\phi_i,
\phi_{i+1},\ldots,\phi_N)$, and the last row is also deleted.
\end{lemma}

\begin{proof} The proof is a direct consequence of Lemma~\ref{lemgtdetrep1} and Theorem~\ref{thmqgtn}
from the initial ``free'' Lax operator $L=\partial_q$. The
generating functions \{$\phi_i,\, i=1,2,\ldots,N$\} of $T_N$
satisf\/ies equa\-tions~\eqref{zeroinitiallinearfunforqckp}, which is obtained from
def\/inition of ``eigenfunction''  \eqref{qzslax1} of the KP
hierarchy under \mbox{$B_n=\partial_q^n$}.
\end{proof}

In particular, $(T_N \cdot \phi_1)= (T_N
\cdot\phi_2)=\cdots=(T_N\cdot\phi_N)=0$.

Now we can give one theorem reducing the $q$-Wronskian $\tau$
function $\tu{N}_q$ in  \eqref{zerointialtaunb} of $q$-KP
hierarchy to the $q$-cKP hierarchy def\/ined by~\eqref{qcKP}.

\begin{theorem}\label{thmreducingqkptockp}
$\tu{N}_q$ is also a $\tau$ function of the $q$-cKP hierarchy
whose Lax operator $L^l=(L^l)_+
+\sum\limits_{i=1}^{M}q_i\circ\partial_q^{-1}\circ r_i$ with some
suitable functions $\{q_i,\, i=1,2,\ldots,M\}$ and $\{r_i,\,
i=1,2,\ldots,M\}$ if and only if
\begin{gather}
  W_{N+M+1}^q(\phi_1,\ldots,\phi_N,\partial_q^l\phi_{i_1},\ldots,\partial_q^l\phi_{i_{M+1}})=0
  \label{additionalconstraintsa}
\end{gather}
for any choice of $(M+1)$-indices $(i_1,i_2,\ldots,i_{M+1})$ $1
\leqslant i_1 < \cdots < i_{M+1} \leq N $, which can be expressed
equivalently as
 \begin{gather}
   W_{M+1}^q\Bigg(\frac{W_{N+1}^q(\phi_1,\ldots,\phi_N,\partial_q^l\phi_{i_1})}{W_N^q(\phi_1,\ldots,\phi_N)},
   \frac{W_{N+1}^q(\phi_1,\ldots,\phi_N,\partial_q^l\phi_{i_2})}{W_N^q(\phi_1,\ldots,\phi_N)},\ldots,
   \nonumber\\ 
\phantom{W_{M+1}^q\Bigg(}{}
\frac{W_{N+1}^q(\phi_1,\ldots,\phi_N,\partial_q^l\phi_{i_{M+1}})}{W_N^q(\phi_1,\ldots,\phi_N)}\Bigg)
   = 0
   \label{additionalconstraintsb}
   \end{gather}
   for all indices. Here $\{\phi_i,\, i=1,2,\ldots,N \}$
satisfy  \eqref{zeroinitiallinearfunforqckp}.
\end{theorem}

\begin{remark} This theorem is a $q$-analogue of the classical theorem on
cKP hierarchy given by~\cite{os}.
\end{remark}

\begin{proof} The  $q$-Wronskian identity proven in Appendix~C
\begin{gather*}
   W_{M+1}^q\Bigg(\frac{W_{N+1}^q(\phi_1,\ldots,
      \phi_N,f_1)}{W_N^q(\phi_1,\ldots,\phi_N)},\ldots,
   \frac{W_{N+1}^q(\phi_1,\ldots,\phi_N,
   f_{M+1})}{W_N^q(\phi_1,\ldots,\phi_N)}\Bigg)\nonumber\\
\phantom{W_{M+1}^q\Bigg(}{}
   =\frac{W_{N+M+1}^q(\phi_1,\ldots,
   \phi_N,f_1,   \ldots,f_{M+1})}{W_N^q(\phi_1,\ldots,\phi_N)}
\end{gather*}
implies  equivalence between  (\ref{additionalconstraintsa}) and
 (\ref{additionalconstraintsb}). Using $T_N$ and $T_N^{-1}$ in
Lemma~\ref{lemTNandInverse} and the operator identity in
 \eqref{qopertoridenitiesII} we have
\begin{gather}
   (L^l)_-=(T_N\circ \partial_q^l\circ
   T_N^{-1})_-=\sum_{i=1}^N(T_N(\partial_q^l\phi_i))\circ\partial_q^{-1}\circ
   g_i, \label{negativepartofqcKP}
   \end{gather}
where $g_i$ is given by  \eqref{gidetrepqTNinverse} and $T_N$
acting on $(\partial_q^l\phi_i)$ is
 $T_N(\partial_q^l\phi_i)
=\dfrac{W_{N+1}^q(\phi_1,\phi_2,\ldots,\phi_N,\partial_q^l
\phi_i)}{W_N^q(\phi_1,\phi_2,\ldots,\phi_N)}$. So $\tu{N}_q$ is
automatically a tau function of $N$-component $q$-cKP hierarchy
with the form  \eqref{negativepartofqcKP}. Next, we can reduce the
$N$-component to the $M$-component ($M<N$) by a suitable
constraint of~$\phi_i$.

Suppose that the $M$-component ($M< N$) $q$-cKP hierarchy is
obtained by constraint of qKP hierarchy generated by $T_N$, i.e.,
there exist suitable functions \{$q_i,r_i$\} such that
\[
(L^l)_-=\sum\limits_{i=1}^M q_i\circ\partial_q^{-1}\circ r_i=
\sum_{i=1}^N(T_N(\partial_q^l\phi_i))\circ\partial_q^{-1}\circ
   g_i.
\]
As we pointed out in previous section, for a Lax operator whose
negative part is in the form of $(L^l)_-=\sum\limits_{i=1}^M
q_i\circ\partial_q^{-1}\circ r_i$, there exists an $M$-th order
$q$-dif\/ferential operator A such that $AL^l$ is
a~$q$-dif\/ferential operator, then we have
\[
  0=AL^l(T_N(\phi_i))=AT_N\partial_q^l(\phi_i)=A(T_N(\partial_q^l\phi_i))
  \]
from $T_N(\phi_i)=0$that implies $T_N(\partial_q^l\phi_i) \in {\rm
Ker}\,(A)$. Therefore,  at most $M$ of these functions
$T_N(\partial_q^l\phi_i)$ can be linearly independent because  the
Kernel of $A$ has dimension $M$. So
 (\ref{additionalconstraintsb})  is  deduced.

Conversely, suppose   (\ref{additionalconstraintsb})  is true, we
will show that there  exists  one  M-component $q$-ckP ($M<N$)
constrained from  \eqref{negativepartofqcKP}. The equation
 (\ref{additionalconstraintsb}) implies that  at most $M$ of
functions $T_N(\partial_q^l\phi_i)$ $(i=1,2,\ldots,N)$ are
linearly independent.  Then we can f\/ind  suitable $M$ functions
$\{q_1,q_2,\ldots,q_M\}$, which are linearly independent, to
express functions $T_N(\partial_q^l\phi_i)$ as
\[
T_N(\partial_q^l\phi_i)=\frac{W_{N+1}^q(\phi_1,\phi_2\cdots,\phi_N,\partial_q^l\phi_i)}
{W_N^q(\phi_1,\phi_2,\ldots,\phi_N)} =\sum\limits_{j=1}^M
c_{ij}q_j, \qquad  i=1,\cdots,N
\]
with some constants $c_{ij}$. Taking this back
into~\eqref{negativepartofqcKP},  it becomes
\begin{gather*}
  (L^l)_-=\sum\limits_{i=1}^N\left(\sum_{j=1}^M c_{ij}q_j\right)\circ
  \partial_q^{-1}\circ g_i=\sum\limits_{j=1}^M q_j\circ\partial_q^{-1}
  \circ\left(\sum_{i=1}^N c_{ij}g_i\right)=\sum\limits_{j=1}^M q_j\circ\partial_q^{-1} \circ
  r_j,
   \end{gather*}
which is an $M$-component $q$-cKP hierarchy as we expected.
\end{proof}

\section[Example reducing $q$-KP to $q$-cKP hierarchy]{Example reducing $\boldsymbol{q}$-KP
to $\boldsymbol{q}$-cKP hierarchy}

 To illustrate the method in Theorem
\ref{thmreducingqkptockp} reducing the $q$-KP to multi-component a
$q$-cKP hierarchy, we discuss  the $q$-KP generated by
$T_N|_{N=2}$. In order to obtain the concrete solution, we only
consider the three variables ($t_1,t_2,t_3$) in $\overline{t}$.
Furthermore, the $q_1$, $r_1$ and $u_{-1}$ are constructed in this
section.

According to Theorem \ref{thmreducingqkptockp},  the  $q$-KP
hierarchy generated by $T_N|_{N=2}$ possesses a  tau function
\begin{gather}
\tu{2}_q=W_2^q(\phi_1,\phi_2)=\phi_1 (\partial_q \phi_2)-\phi_2(\partial_q \phi_1)\notag\\
\phantom{\tu{2}_q}{}=(\lambda_2-\lambda_1)e_q(\lambda_1x)e_q(\lambda_2x)e^{\xi_1+\xi_2}+
(\lambda_3-\lambda_1)e_q(\lambda_1x)e_q(\lambda_3x)e^{\xi_1+\xi_3}
\notag\\
\phantom{\tu{2}_q=}{}+(\lambda_3-\mu)e_q(\mu
x)e_q(\lambda_3x)e^{\xi+\xi_3} +(\lambda_2-\mu)e_q(\mu
x)e_q(\lambda_2x)e^{\xi+\xi_2} \label{tau2compoentqckp}
\end{gather}
with
\begin{gather*}
  \phi_1=e_q(\lambda_1 x)e^{\xi_1}+e_q(\mu x)e^{\xi},\qquad
  \phi_2=e_q(\lambda_2 x)e^{\xi_2}+e_q(\lambda_3x)e^{\xi_3}.
\end{gather*}
Here  $ \xi_i=c_i+\lambda_i t_1+\lambda_i^2t_2+\lambda_i^3t_3 $
$(i=1,2,3)$,  and $\xi=d+\mu t_1+\mu^2t_2+\mu^3t_3$, $c_i$ and $d$
are arbitrary constants. These functions satisfy the linear
equations
\begin{gather*}
   \frac{\partial\phi_i}{\partial
   t_n}=\partial_q^n\phi_i,\qquad n=1,2,3,\quad i=1,2,
\end{gather*}
as a special case of  \eqref{zeroinitiallinearfunforqckp}. From
 \eqref{negativepartofqcKP},  the  $q$-KP hierarchy generated by
$T_N|_{N=2}$ is in the form of
\begin{gather}
L^l=(L^l)_+ +(T_2(\partial_q^l\phi_1))\circ\partial_q^{-1}\circ
   g_1+(T_2(\partial_q^l\phi_2))\circ\partial_q^{-1}\circ
   g_2, \label{2componentckP}\\
\stackrel{{\rm constraint}}{=====}(L^l)_+ + q_1 \circ
   \partial_q\circ r_1. \label{singlecomponentckP}
\end{gather}
Here $q_1$ and $r_1$ are undetermined, which can be expressed by
$\phi_1$ and $\phi_2$ as follows.

According to \eqref{additionalconstraintsa}, the restriction for
$\phi_1$ and $\phi_2$ to reduce  \eqref{2componentckP} to
 \eqref{singlecomponentckP} is given by
\begin{gather}
0=W_2^q(\phi_1,\phi_2,\partial_q^l\phi_1,\partial_q^l\phi_2)
     =(\mu^l-\lambda_1^l)(\lambda_2^l-\lambda_3^l)V(\lambda_1,\lambda_2,\lambda_3,\mu)e^{c_1+c_2+c_3+d}
     e^{(\lambda_1+\lambda_2+\lambda_3+\mu)t_1} \notag \\
\phantom{0=}{} \times
e^{(\lambda_1^2+\lambda_2^2+\lambda_3^2+\mu^2)t_2}
     e^{(\lambda_1^3+\lambda_2^3+\lambda_3^3+\mu^3)t_3}
     e_q(\lambda_1 x)e_q(\lambda_2 x)e_q(\lambda_3 x)e_q(\mu x)
     \label{qexamplecosntraintsn=2}
     \end{gather}
with
\[
  V(\lambda_1,\lambda_2,\lambda_3,\mu) =     \begin{vmatrix}
     1&\lambda_1 & \lambda_1^2&\lambda_1^3 \\
     1&\lambda_2 &\lambda_2^2 &\lambda_2^3 \\
     1&\lambda_3 & \lambda_3^2&\lambda_3^3 \\
     1&\mu       &\mu^2       &\mu^3
     \end{vmatrix}.
     \]
Obviously, we can let $\mu=\lambda_2$ and  $d=c_2$  such that
 \eqref{qexamplecosntraintsn=2}  holds for $\phi_1$ and
$\phi_2$. Then the $\tau$ function of a single component $q$-cKP
def\/ined by  \eqref{singlecomponentckP} is
  \begin{gather*}
  \tau_{q{\rm cKP}}=(\lambda_2-\lambda_1)e_q(\lambda_1x)e_q(\lambda_2x)e^{\xi_1+\xi_2}+
(\lambda_3-\lambda_1)e_q(\lambda_1x)e_q(\lambda_3x)e^{\xi_1+\xi_3}
\notag\\
\phantom{\tau_{q{\rm cKP}}=}{}
+(\lambda_3-\lambda_2)e_q(\lambda_2x)e_q(\lambda_3x)e^{\xi_2+\xi_3},
  \end{gather*}
which is deduced from \eqref{tau2compoentqckp}. That means we
indeed reduce the $\tau$ function $\tu{2}_q$ in
 \eqref{tau2compoentqckp} of the $q$-KP hierarchy generated by
$T_N|_{N=2}$ to the $\tau$ function $\tau_{q{\rm cKP}}$ of the
one-component $q$-cKP hierarchy. Furthermore, we would like to get
the explicit expression of $(q_1,r_1)$ of $q$-cKP in
\eqref{singlecomponentckP}. Using the determinant representation
of $T_N|_{N=2}$ and $T^{-1}_N|_{N=2}$, we have
\begin{gather*}
f_1\triangleq(T_2(\partial_q^l\phi_1)) =
\dfrac{(\lambda_1^l-\lambda_2^l)(\lambda_3-\lambda_2)(\lambda_2-\lambda_1)(\lambda_3-\lambda_1)e_q(\lambda_1x)
e_q(\lambda_2x)e_q(\lambda_3x)e^{\xi_1+\xi_2+\xi_3}}{\tau_{q{\rm cKP}}},\\
f_2\triangleq(T_2(\partial_q^l\phi_2) =
\dfrac{(\lambda_3^l-\lambda_2^l)(\lambda_3-\lambda_2)(\lambda_2-\lambda_1)(\lambda_3-\lambda_1)e_q(\lambda_1x)
e_q(\lambda_2x)e_q(\lambda_3x)e^{\xi_1+\xi_2+\xi_3}}{\tau_{q{\rm cKP}}},\\
g_1= -\t\left(\dfrac{\phi_2}{\tau_{q{\rm cKP}}} \right), \qquad
g_2= \t\left(\dfrac{\phi_1}{\tau_{q{\rm cKP}}} \right),
\end{gather*}
under the restriction $\mu=\lambda_2$ and  $d=c_2$. One can f\/ind
that $f_1$ and $f_2$ are linearly dependent, and
$(\lambda_3^l-\lambda_2^l) f_1=(\lambda_1^l-\lambda_2^l) f_2$.  So
 \eqref{2componentckP} and  \eqref{singlecomponentckP} reduce to
\begin{gather*}
L^l_-= f_1\circ \partial^{-1}_q \circ g_1 +f_2\circ
\partial^{-1}_q \circ
g_2 \\
\phantom{L^l_-}{} =(\lambda_3^l-\lambda_2^l) f_1\circ
\partial^{-1}_q \circ \dfrac{g_1}{(\lambda_3^l-\lambda_2^l)}
+(\lambda_1^l-\lambda_2^l) f_2\circ
\partial^{-1}_q \circ
\dfrac{g_2}{(\lambda_1^l-\lambda_2^l)} = q_1 \circ \partial_q^{-1}
\circ r_1,
\end{gather*}
in which
 \begin{gather*}
 q_1\triangleq (\lambda_3^l-\lambda_2^l) f_1=(\lambda_1^l-\lambda_2^l)
 f_2\notag \\
\phantom{q_1}{} =
\dfrac{(\lambda_1^l-\lambda_2^l)\lambda_3^l-\lambda_2^l)
(\lambda_3-\lambda_2)(\lambda_2-\lambda_1)(\lambda_3-\lambda_1)e_q(\lambda_1x)
e_q(\lambda_2x)e_q(\lambda_3x)e^{\xi_1+\xi_2+\xi_3}}{\tau_{q{\rm cKP}}},\\
r_1\triangleq\left(\dfrac{g_1}{(\lambda_3^l-\lambda_2^l)}+
\dfrac{g_2}{(\lambda_1^l-\lambda_2^l)} \right)
=\dfrac{1}{(\lambda_1^l-\lambda_2^l)(\lambda_3^l-\lambda_2^l)} \notag \\
{} \times\t\left( \dfrac{e^{-(\xi_1+\xi_2+\xi_3)}\big(
(\lambda_3^l-\lambda_2^l)e_q(\lambda_1x)e^{\xi_1}+(\lambda_3^l-\lambda_1^l)e_q(\lambda_2x)e^{\xi_2}+
(\lambda_2^l-\lambda_1^l)e_q(\lambda_3x)e^{\xi_3} \big)}
 {\tau_{q{\rm cKP}}} \right).
 \end{gather*}
In particular,  we can let $\lambda_1=\lambda$, $\lambda_2=0$,
$\lambda_3=-\lambda$, $c_1=c$, $c_2=-0$, $c_3=-c,$ then
\begin{gather*}
q_1=\dfrac{(-1)^l\lambda^{2l+2}e_q(\lambda x)e_q(-\lambda
x)}{e_q(\lambda x)e_q(-\lambda x)+ \frac{e_q(\lambda
x)e^\eta+e_q(-\lambda x)e^{-\eta}}{2}e^{-\lambda^2 t_2} }
\end{gather*}
and
\begin{gather*}
r_1=\left\{
\begin{array}{ll}
-\dfrac{1}{\lambda^{l+1}}\t\left[ \dfrac{e^{-\lambda^2t_2}+
\frac{e_q(\lambda x)e^\eta +e_q(-\lambda x)e^{-\eta} }{2}}
{e^{\lambda^2t_2}e_q(\lambda x)e_q(-\lambda x)+ \frac{e_q(\lambda
x)e^\eta +e_q(-\lambda x)e^{-\eta} }{2}} \right] & \mbox{if}\ l\
\mbox{is odd},
\vspace{2mm}\\
-\dfrac{1}{\lambda^{l+1}}\t\left[ \dfrac{ \frac{e_q(\lambda
x)e^\eta -e_q(-\lambda x)e^{-\eta} }{2}}
{e^{\lambda^2t_2}e_q(\lambda x)e_q(-\lambda x)+ \frac{e_q(\lambda
x)e^\eta +e_q(-\lambda x)e^{-\eta} }{2}} \right] & \mbox{if}\  l\
\mbox{is even},
\end{array}
\right.
\end{gather*}
where $\eta=c+ \lambda t_1 +\lambda^3 t_3$.

In general, the $l$-constrained one-component $q$-KP hierarchy has
the Lax operator $L=\partial_q+u_0+q_1\circ\partial_q^{-1}\circ
r_1$ when $l=1$. On the other hand, its Lax operator can also be
expressed as
$L=\partial_q+u_0+u_{-1}\partial_q^{-1}+u_{-2}\partial_q^{-2}+\cdots$.
So all of the dynamical variables  \{$u_{-i}, \, i=1,2,3,
\ldots$\} of $q$-KP hierarchy are given by
\[
  u_{-i-1}=(-1)^iq^{-i(i+1)/2}q_1\theta^{-i-1}(\partial_q^ir_1),\qquad i\ge 0.
  \]
For the present situation,
$u_{-1}=u_{-1}(t_1,t_2,t_3)=q_1\t^{-1}(r_1)$ represents the
$q$-deformed solution of the classical KP eqution, which is
constructed from the components of $q$-cKP hierarchy, and is of
the form
\begin{gather}
q_1= \dfrac{-\lambda^4e_q(\lambda x)e_q(-\lambda x)}{e_q(\lambda
x)e_q(-\lambda x)+ \frac{e_q(\lambda x)e^{\eta}
+e_q(-\lambda x)e^{-\eta} }{2}e^{-\lambda^2 t_2}},\label{finnalformq1} \\
r_1=-\dfrac{1}{\lambda^{2}}\t\left[ \dfrac{e^{-\lambda^2t_2}+
\frac{e_q(\lambda x)e^\eta +e_q(-\lambda x)e^{-\eta} }{2}}
{e^{\lambda^2t_2}e_q(\lambda x)e_q(-\lambda x)+ \frac{e_q(\lambda
x)e^\eta +e_q(-\lambda x)e^{-\eta} }{2}} \right],
\label{finnalformr1}\\
u_{-1}=\dfrac{\lambda^2e_q(\lambda x)e_q(-\lambda x)}{e_q(\lambda
x)e_q(-\lambda x)+ \frac{e_q(\lambda x)e^{\eta}
+e_q(-\lambda x)e^{-\eta} }{2}e^{-\lambda^2 t_2}} \notag \\
\phantom{u_{-1}=}{} \times\dfrac{e^{-\lambda^2t_2}+
\frac{e_q(\lambda x)e^\eta +e_q(-\lambda x)e^{-\eta} }{2}}
{e^{\lambda^2t_2}e_q(\lambda x)e_q(-\lambda x)+ \frac{e_q(\lambda
x)e^\eta +e_q(-\lambda x)e^{-\eta} }{2}}. \label{finnalformu1}
\end{gather}

\begin{figure}[t]
  \centering
  \includegraphics[width=7cm]{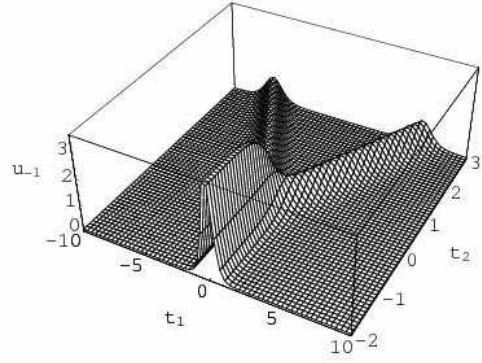}\vspace{-2mm}
  \caption{$u_{-1}(x=0.001, q=0.999)$ from $q$-cKP  and $t_3=0$.}
  \label{fig15}
\end{figure}

Obviously, they will approach to the classical results on the cKP
hierarchy in  \cite{os} when $x\rightarrow 0$ and $q\rightarrow
1$. We will f\/ix $\lambda=2$, $t_3=0$ and $c=0$ to plot their
f\/igures, then get $q_1=q_1(x,t_1,t_2,q)$, $r_1=r_1(x,t_1,t_2,q)$
and $ u_{-1}=u_{-1}(x,t_1,t_2,q)$
from~\eqref{finnalformq1}--\eqref{finnalformu1}. To save space, we
plot the f\/igures for $u_{-1}$ and $q_1$ in $(t_1,t_2,t_3)$
dimension spaces. It can be seen that Fig.~15 of $u_{-1}(0.001,
t_1,t_2,0.999)$ and Fig.~20 of $q_{1}(0.001,t_1,t_2,0.999)$
 match with the prof\/ile of $u_1$ and $q$ in~\cite{os} with the same parameters.
So we def\/ine $q$-ef\/fects quantity $\triangle
u_{-1}=u_{-1}(0.5,t_1,t_2, 0.999)-u_{-1}(0.5,t_1,t_2, q)=
u_{-1}(q=0.999)-u_{-1}(q)$, $\triangle
q_{1}=q_{1}(0.5,t_1,t_2,0.999)-q_{1}(.5,,t_1,t_2,
q)=q_{1}(q=0.999)-q_1(q)$,
 to show their dependence on $q$. Figs.~16--19\footnote{For Figs.~16--19, $q$-ef\/fect $\mbox{Du}_{-1}\equiv \triangle
u_{-1}\triangleq u_{-1}(q=0.999)- u_{-1}(q=i)$, where
$i=0.7,0.5,0.3,0.1$, from $q$-cKP with $x=0.5$ and $t_3=0$.} and
Figs.~21--24\footnote{For Figs.~21--24, $q$-ef\/fect
$\mbox{Dq}_{1}\equiv\triangle q_{1}\triangleq q_{1}(q=0.999)-
q_{1}(q=i)$, where $i=0.7,0.5,0.3,0.1$, from $q$-cKP with $x=0.5$
and $t_3=0$.} are plotted for $\triangle u_{-1}$ and $\triangle
q_{1}$, respectively, where $q=0.7,0.5,0.3,0.1$. Obviously, they
 are decreasing to almost zero when  $q$ goes from $0.1$ to $1$
with f\/ixed $x=0.5$. Furthermore, Figs.~25--29\footnote{For
Figs.~25--29, $q$-ef\/fect $\mbox{Du}_{-1}\equiv \triangle
u_{-1}\triangleq u_{-1}(x=i,q=0.999)- u_{-1}(x=i,q=0.1)$ from
$q$-cKP with $t_3=0$, where $i=0.2,0.4,0.51,0.53,0.55$.} show that the dependence of the
$q$-ef\/fects $\triangle u_{-1}=u_{-1}(x,t_1,t_2,
0.999)-u_{-1}(x,t_1,t_2, 0.1)=u_{-1}(x,q=0.999)-u_{-1}(x,q=0.1)$
on $x$,where $x=0.2,0.4,0.51,0.53,0.55$ in order.  These f\/igures
give us again an opportunity to observe the process of
$q$-deformation in $q$-soliton solution of $q$-KP equation. They
also demonstrate that $q$-deformation keep the prof\/ile of the
soliton, although there exists deformation in some degree. On the
other hand, in fact, $(q_1, r_1)$ can be regarded as a
$q$-deformation of dynamical variables $(q,r)$ of AKNS hierarchy,
because cKP possessing Lax operator $L=\partial +q \circ
\partial^{-1}\circ r$ is equivalent to the AKNS hierarchy.

\begin{figure}[t]
\begin{minipage}[b]{7.5cm}
\centering \includegraphics[width=7cm]{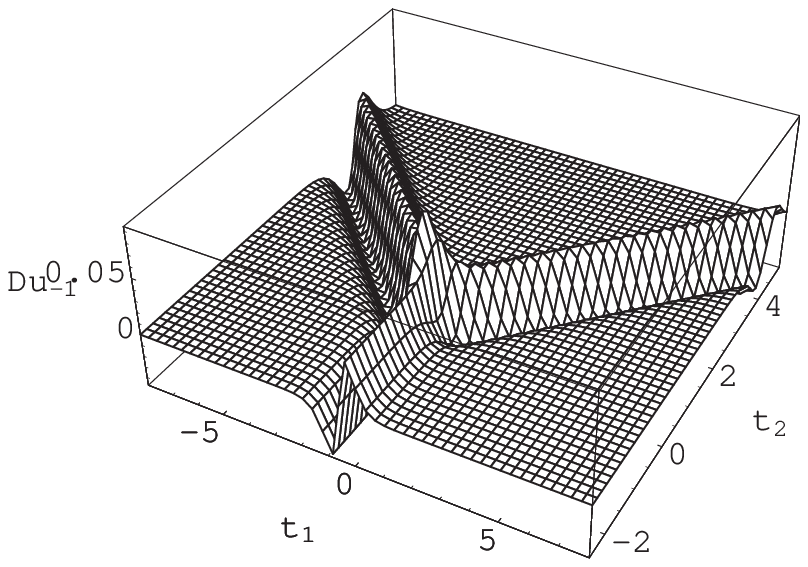}\vspace{-2mm}
\caption{}  \label{fig16}
\end{minipage}\hfill
\begin{minipage}[b]{7.5cm}
  \centering
  \includegraphics[width=7cm]{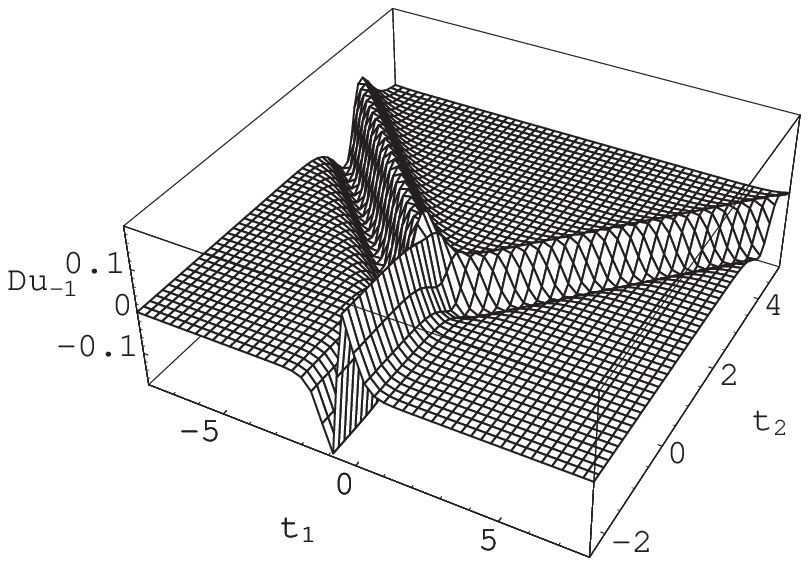}\vspace{-2mm}
  \caption{}
  \label{fig17}
\end{minipage}
\end{figure}

\begin{figure}[t]
\begin{minipage}[b]{7.5cm}
\centering \includegraphics[width=7cm]{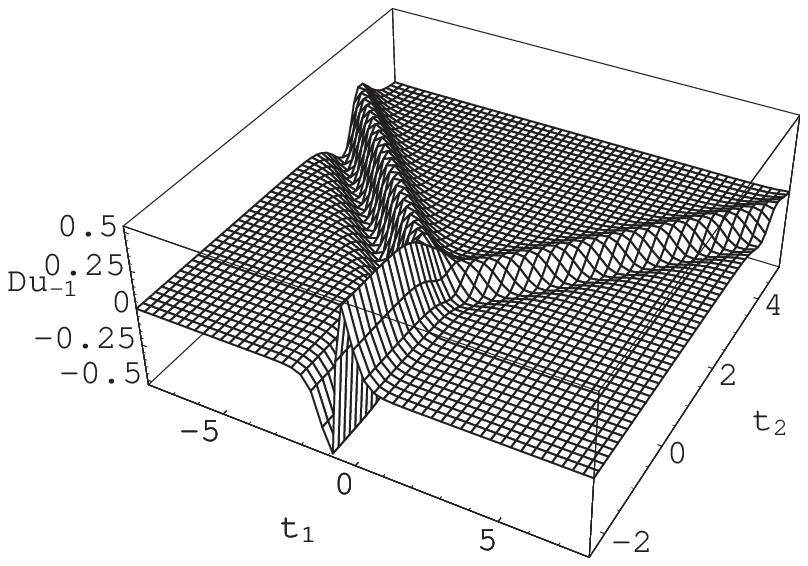}\vspace{-2mm}
\caption{}  \label{fig18}
\end{minipage}\hfill
\begin{minipage}[b]{7.5cm}
  \centering
  \includegraphics[width=7cm]{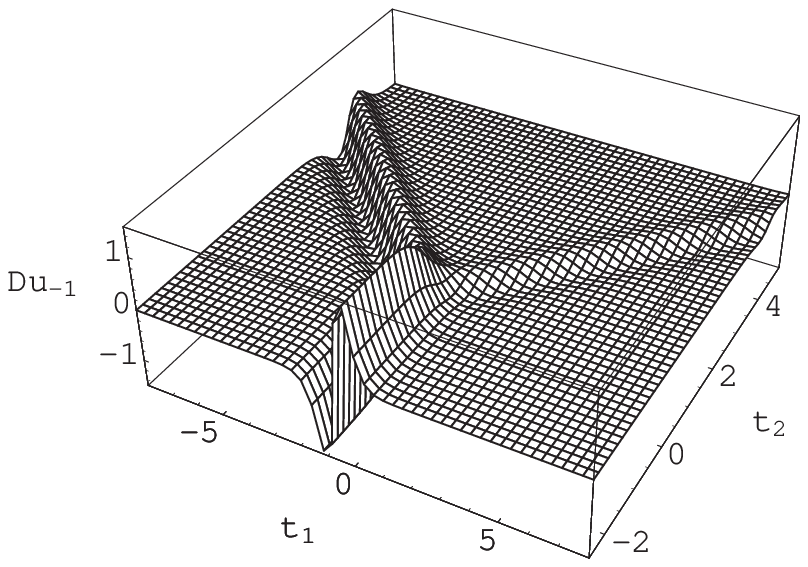}\vspace{-2mm}
  \caption{}
  \label{fig19}
\end{minipage}
\end{figure}

\begin{figure}[t]
  \centering
  \includegraphics[width=7cm]{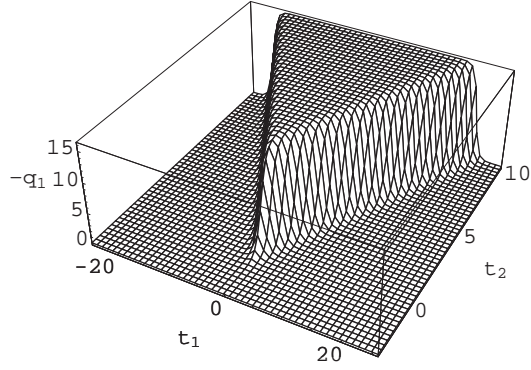}\vspace{-2mm}
  \caption{$q_{1}(x=0.001,q=0.999)$ and $t_3=0$.}
  \label{fig20}
\end{figure}

\begin{figure}[t]
\begin{minipage}[b]{7.5cm}
\centering \includegraphics[width=7cm]{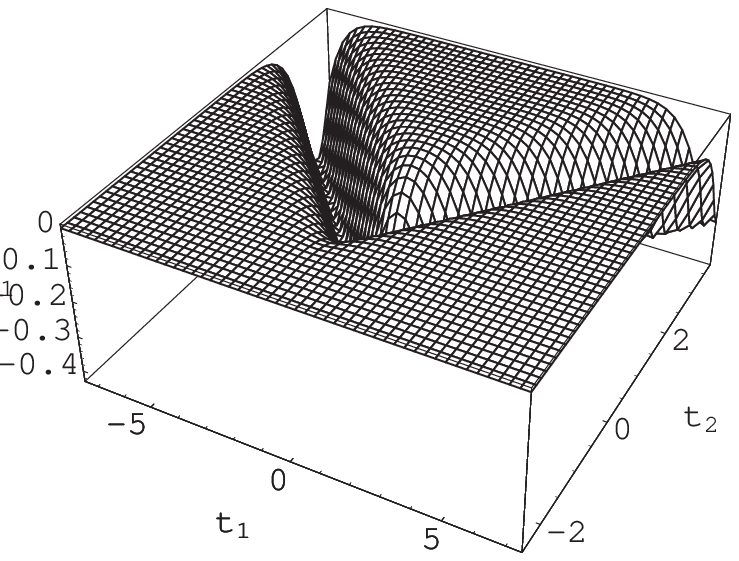}\vspace{-2mm}
\caption{}  \label{fig21}
\end{minipage}\hfill
\begin{minipage}[b]{7.5cm}
  \centering
  \includegraphics[width=7cm]{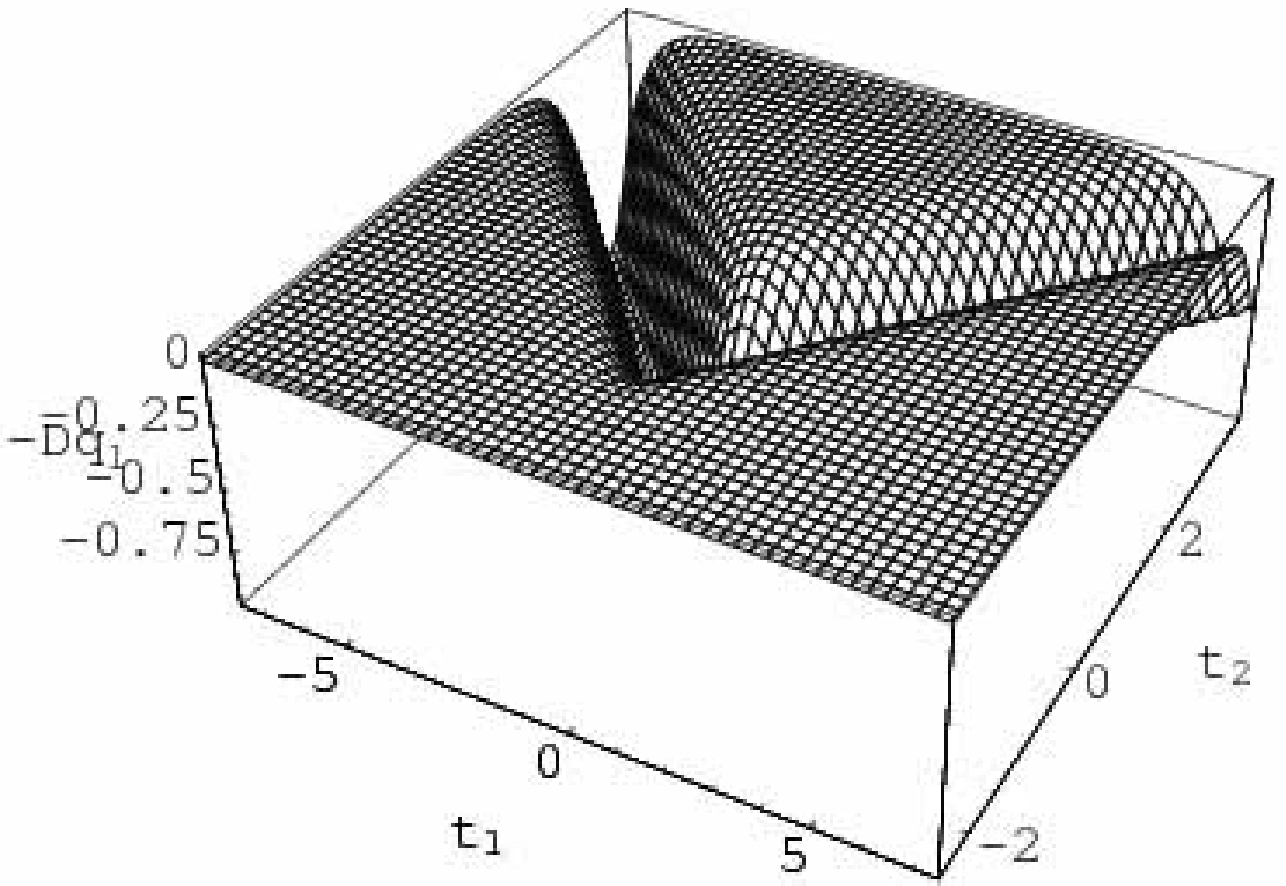}\vspace{-2mm}
  \caption{}
  \label{fig22}
\end{minipage}
\end{figure}

\begin{figure}[t]
\begin{minipage}[b]{7.5cm}
\centering \includegraphics[width=7cm]{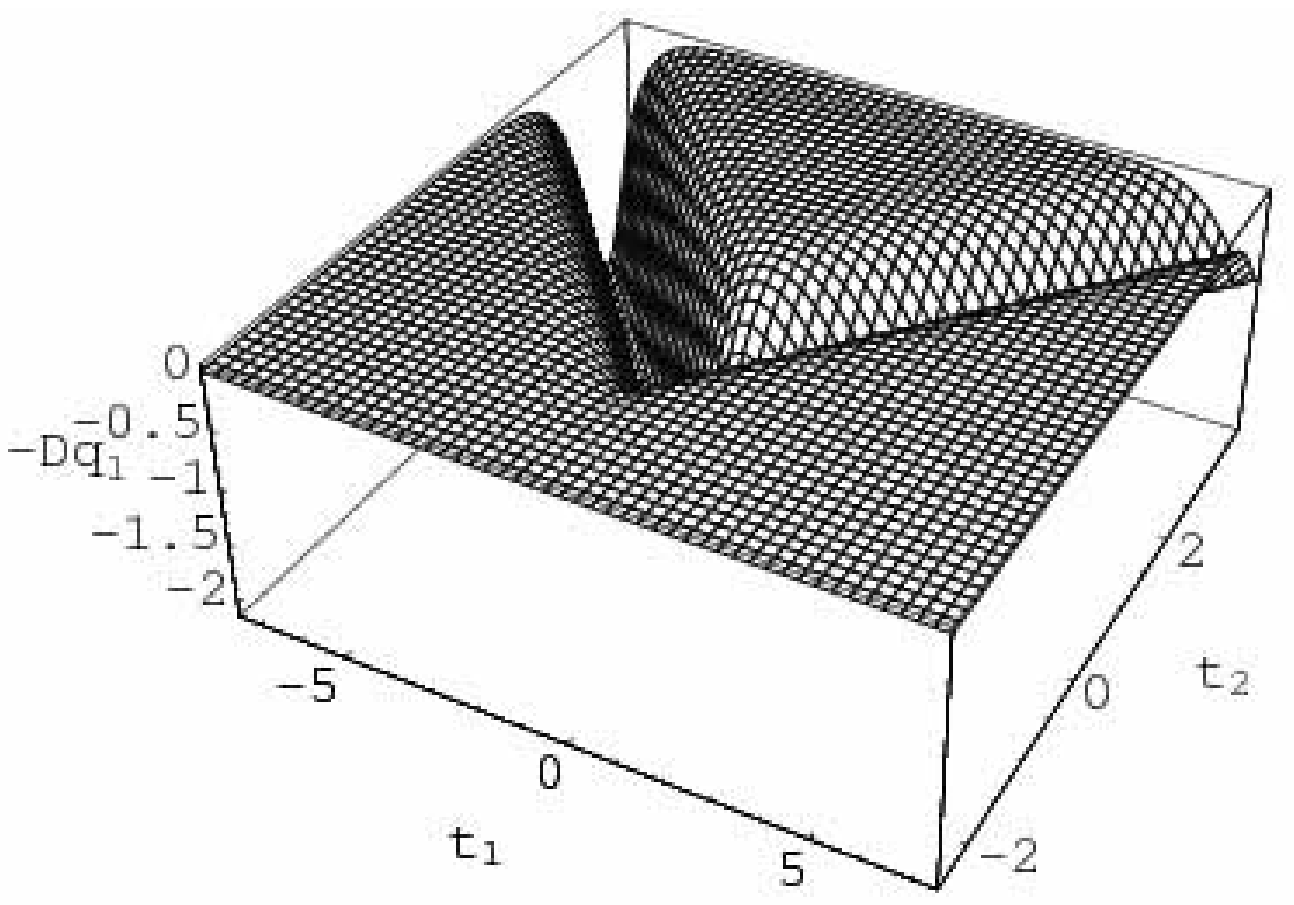}\vspace{-2mm}
\caption{}  \label{fig23}
\end{minipage}\hfill
\begin{minipage}[b]{7.5cm}
  \centering
  \includegraphics[width=7cm]{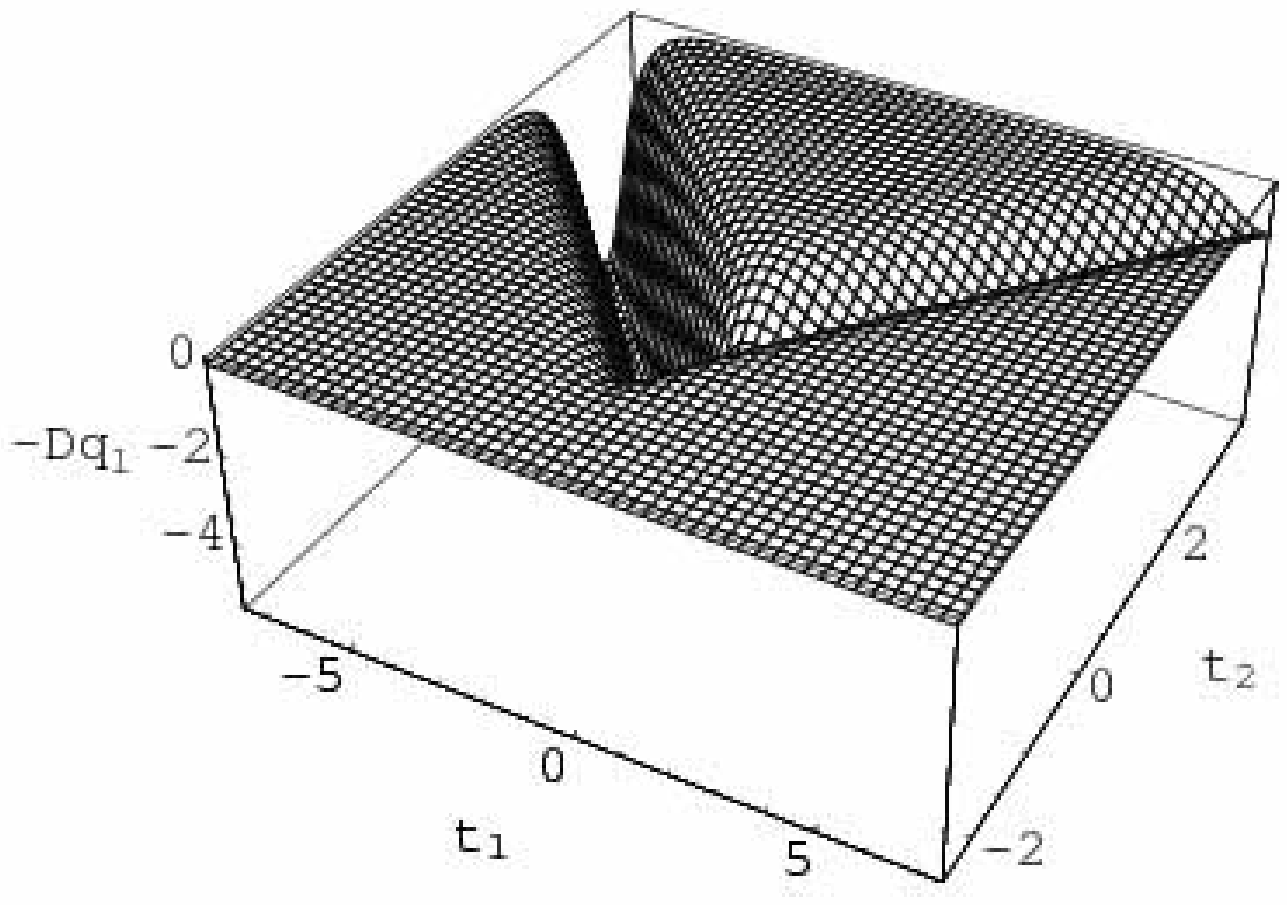}\vspace{-2mm}
  \caption{}
  \label{fig24}
\end{minipage}
\end{figure}

\begin{figure}[t]
\begin{minipage}[b]{7.5cm}
\centering \includegraphics[width=7cm]{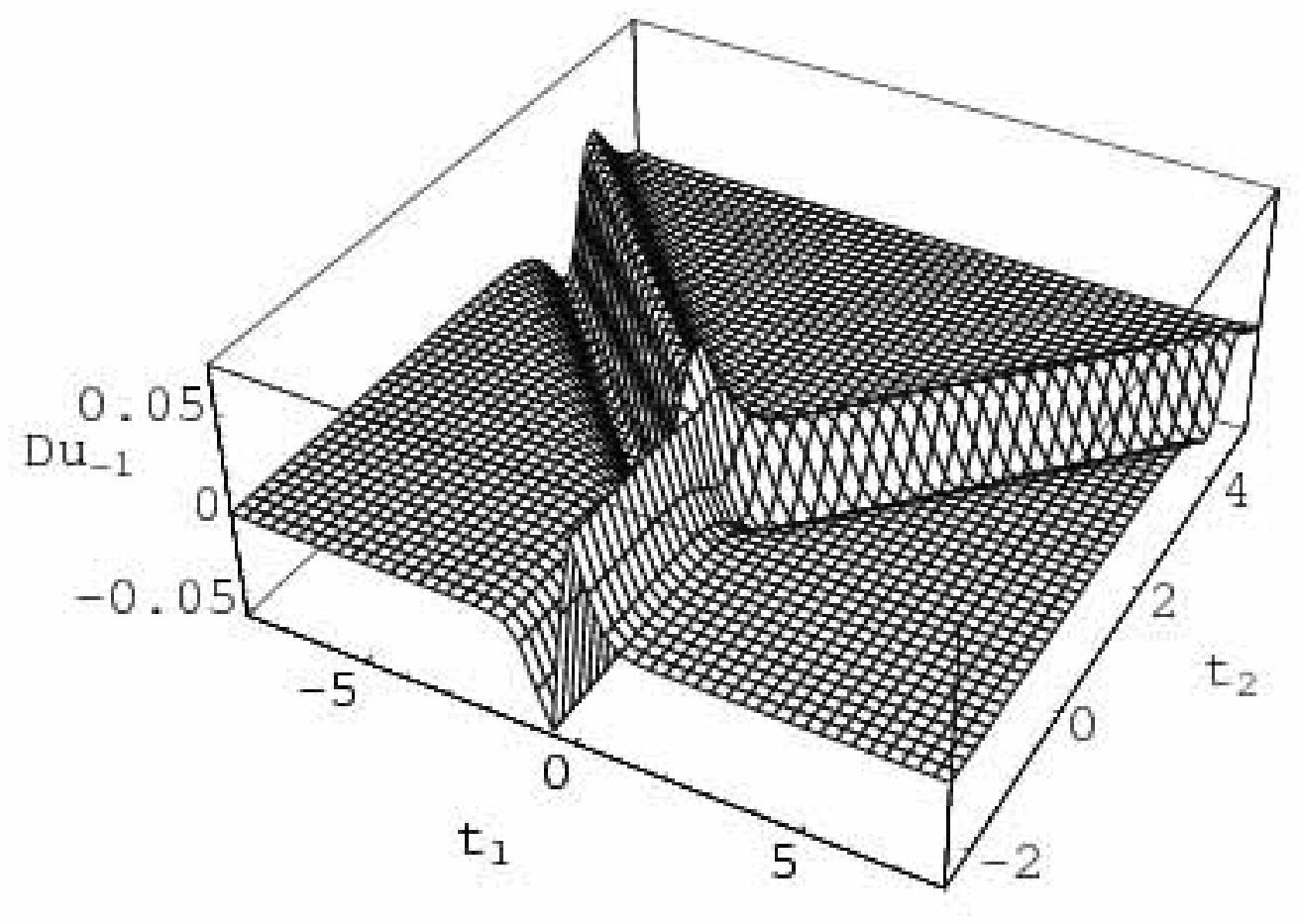}\vspace{-2mm}
\caption{}  \label{fig25}
\end{minipage}\hfill
\begin{minipage}[b]{7.5cm}
  \centering
  \includegraphics[width=7cm]{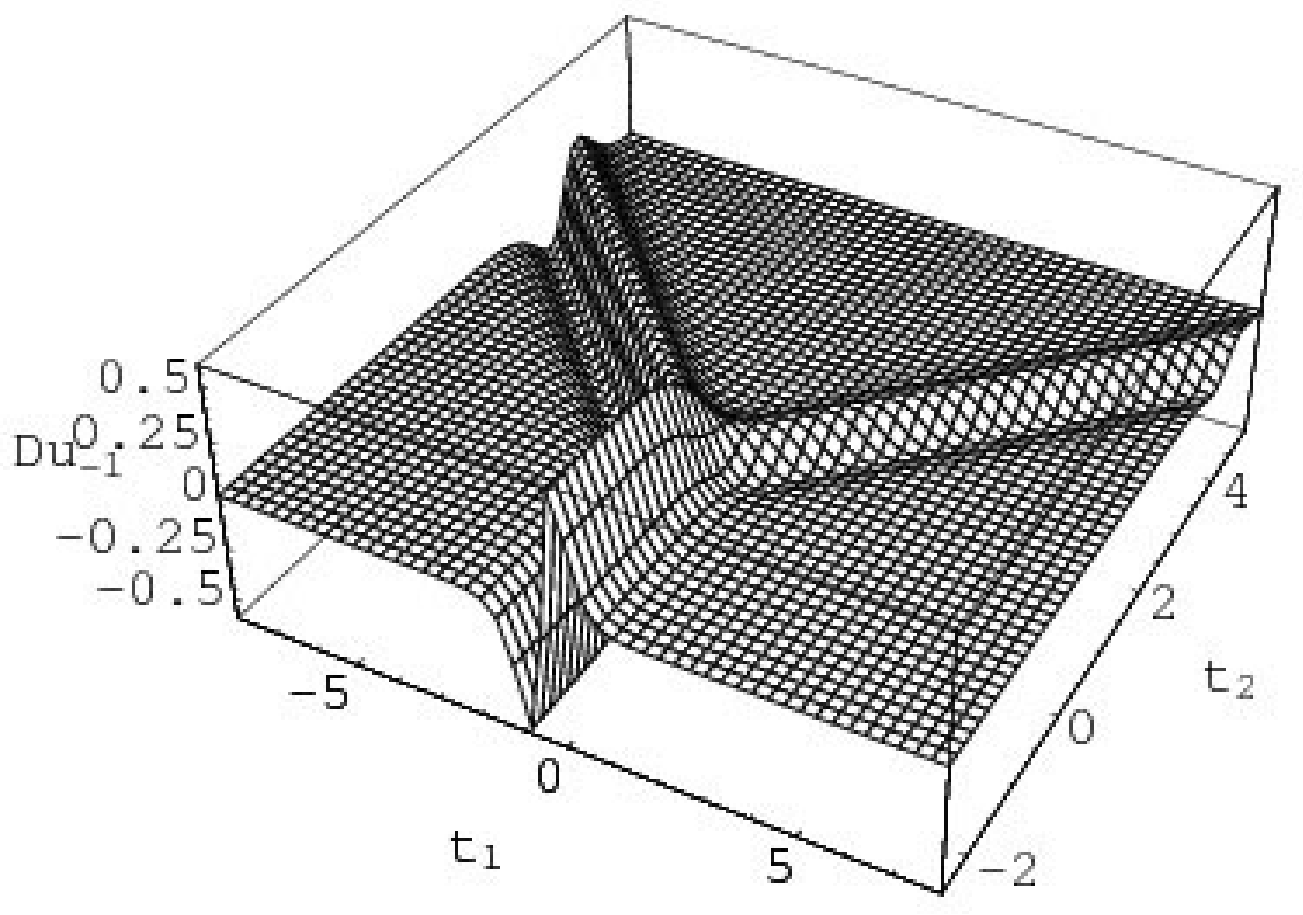}\vspace{-2mm}
  \caption{}
  \label{fig26}
\end{minipage}
\end{figure}

\begin{figure}[t]
\begin{minipage}[b]{7.5cm}
\centering \includegraphics[width=7cm]{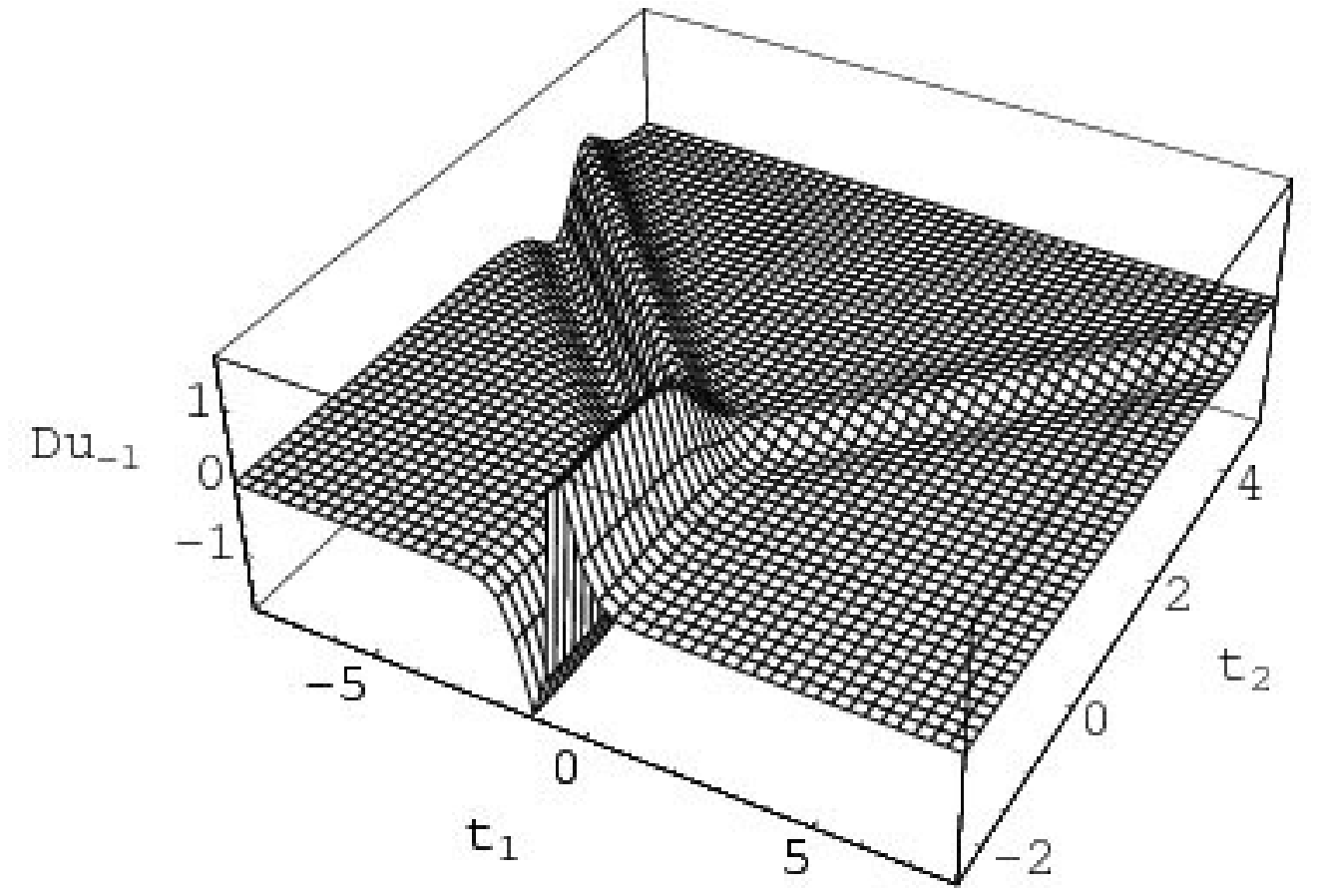}\vspace{-2mm}
\caption{}  \label{fig27}
\end{minipage}\hfill
\begin{minipage}[b]{7.5cm}
  \centering
  \includegraphics[width=7cm]{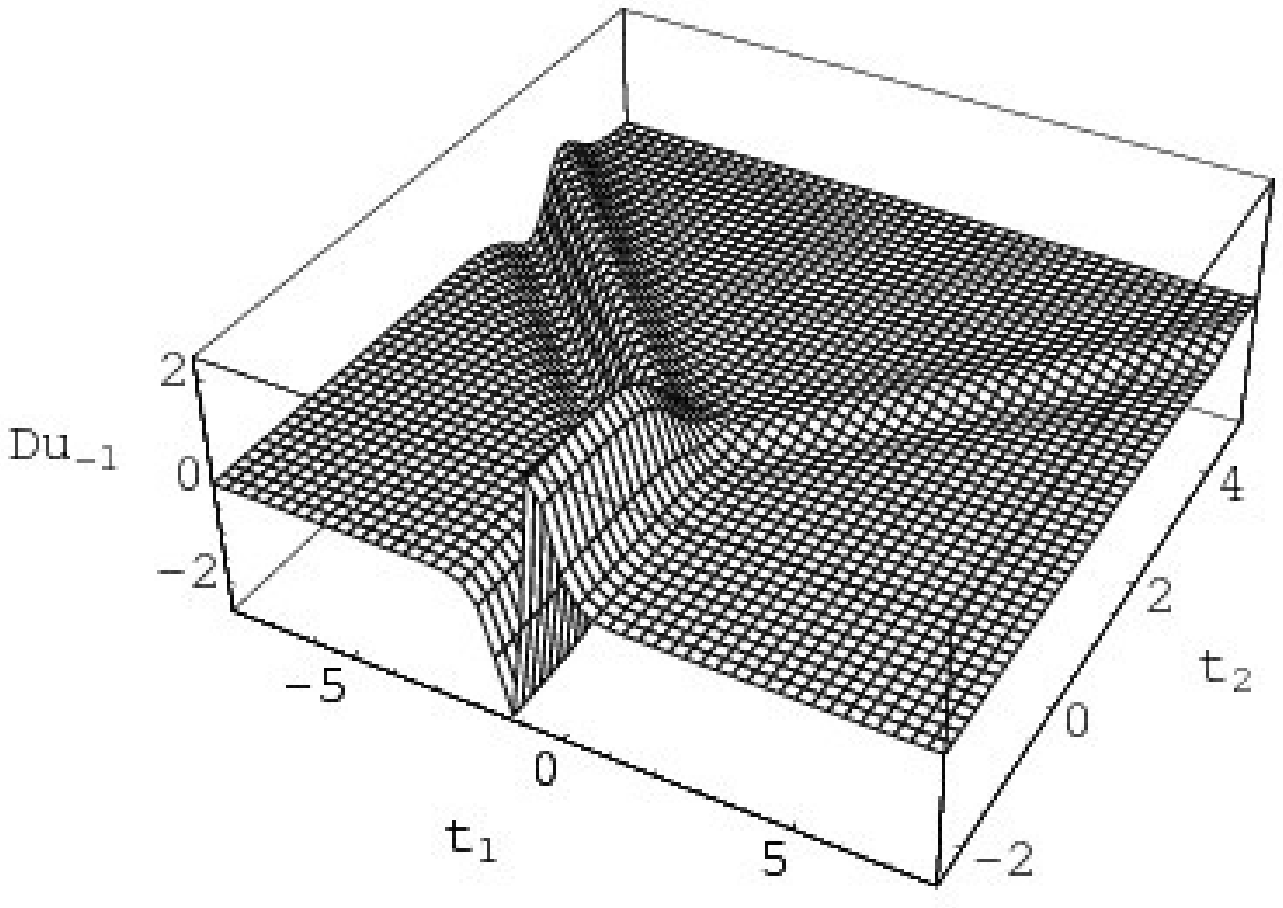}\vspace{-3mm}
  \caption{}
  \label{fig28}
\end{minipage}
\vspace{-3mm}
\end{figure}

\begin{figure}[t]
  \centering
  \includegraphics[width=7cm]{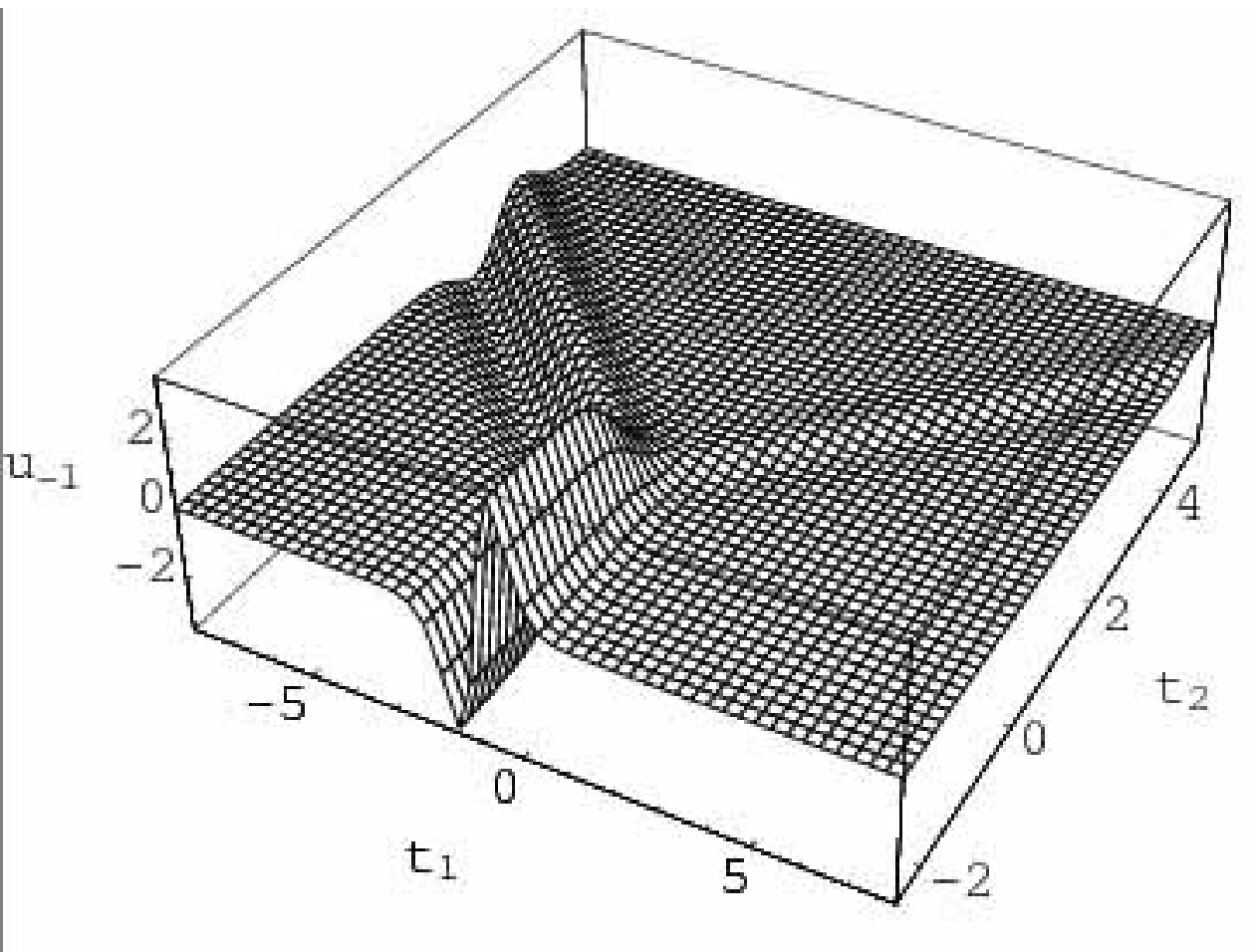}\vspace{-3mm}
  \caption{}
  \label{fig29}
  \vspace{-3mm}
\end{figure}

\section{Conclusions and discussions}
In this paper, we have shown in Theorem~\ref{qgtexistence}  that
there exist two types of elementary gauge transformation operators
for the $q$-KP hierarchy. The changing  rules of $q$-KP under the
gauge transformation are given in Theorems \ref{thmgtI} and
\ref{thmgtII}. We mention that these two types of elementary gauge
transformation operators are introduced f\/irst by Tu
et~al.~\cite{tsl} for $q$-NKdV hierarchy. Considering successive
application of gauge transformation, we
 established the determinant representation of the gauge transformation operator\ of the
$q$-KP hierarchy in Lemma \ref{lemgtdetrep1} and the corresponding
results on the transformed new $q$-KP are given in Theorem
\ref{thmqgtnk}. For the $q$-KP hierarchy generated by $T_{n+k}$
from the ``free'' Lax operator $L=\partial_q$ (i.e.\ the Lax
operator is $L^{(n+k)}=T_{n+k}\circ
\partial_q \circ T_{n+k}^{-1}$),
Corollary~\ref{corzeroinitialtauqkp} shows that the generalized
$q$-Wronskian $IW^q_{k,n}$ of functions $\{\phi_i,\psi_j\}$
$(i=1,2,\ldots,n;\, j=1,2,\ldots,k)$ is a general $\tau$ function
of it, and $q$-Wronskian $W_n^q$ of functions
$\phi_i(i=1,2,\ldots, n)$ is also a special one. Here \{$\phi_i$\}
and \{$\psi_j$\} satisfy special linear $q$-partial dif\/ferential
equations~\eqref{zeroqzslax1+zeroqzslax2}.

The symmetry and symmetry constraint of $q$-KP ($q$-cKP) hierarchy
are discussed in Section~5. On the basis of the representation of
$T_N$ in Lemma \ref{lemTNandInverse}, the $q$-KP hierarchy whose
Lax operator $L^l=T_N\circ \partial^l_q\circ  T_N^{-1}$ is
generated from the ``free'' Lax operator $L=\partial_q$. The
explicit form of its negative part $L^l_-$ is given in
 \eqref{negativepartofqcKP}, which is called $l$-constraint of
the $q$-KP hierarchy. Further we  found necessary and
suf\/f\/icient conditions that are given in Theorem
\ref{thmreducingqkptockp}, reducing a $q$-Wronskian solution in
\eqref{zerointialtaunb} of the $q$-KP hierarchy to solutions of
the multi-component $q$-cKP hierarchy. One example is given in
Section~7 to illustrate the method, i.e., the $q$-KP generated by
$T_N|_{N=2}$ is reduced to one-component $q$-cKP hierarchy.  By
taking f\/inite variables $(t_1,t_2, t_3)$ in $\overline{t}$, the
component $q_1$ and $r_1$ are written out. Our results can be
reduced to the classical results in~\cite{os}.

  As we pointed out in Section~2, $u_{-1}$
is the $q$-analogue of the solution of classical KP equation if we
only consider three variables $(t_1,t_2, t_3)$ in $\overline{t}$.
Therefore, the solution $u_{-1}$ is called $q$-soliton of the
$q$-KP equation, although we do not write out  the $q$-KP equation
on $u_{-1}$. One can f\/ind that the equations of dynamical
variables \{$u_{0}$,$u_{-i}$\} in $q$-KP hierarchy are coupled
with each other and can not get one $q$-partial dif\/ferential
equation associated only with one dynamical variable, like
classical KP equation has one dynamical variable $u_{-1}$. The
reason is that the $q$-Leibnitz rule contains $q$-dif\/ferential
operation  and $\t$ operation, however, the Leibnitz rule of the
standard calculus only contains one dif\/ferential operation. We
get a single $q$-soliton $u_{-1}$ by means of the simplest
$\tau$~function $\tau_q=W_1^q(\phi_1)=\phi_1$ in Section~4.
Meanwhile, the multi-$q$-soliton $u_{-1}$~is obtained from
one-component $q$-cKP hierarchy in Section~7.  Figures of
$q$-ef\/fect $\triangle u_{-1}$ show that $q$-soliton $u_{-1}$
indeed goes to classical soliton of KP equation when $x\rightarrow
0$ and $q\rightarrow 1$ and $q$-deformation does not destroy the
rough prof\/ile of the $q$-soliton. In other worlds, the f\/igure
of $q$-soliton is similar to the classical soliton of KP equation.
We also show the trends of the $q$-ef\/fect $\triangle u_{-1}$
depends on~$x$ and~$q$; $x$ plays a role of the amplif\/ier of
$q$-ef\/fects. In  conclusion, the f\/igures of $q$-ef\/fects
$\triangle u_{-1}$ let us know the process of $q$-deformation in
integrable systems for the f\/irst time. Of course, it is a long
way to explore the physical meaning of $q$ from the soliton
theory.

In comparison with the research of classical soliton
theory~\cite{ac}, in particular, the KP hierarchy~\cite{dkjm,dl1},
the cKP~\cite{kss,aratyn2} hierarchy and the AKNS~\cite{ac}
hierarchy, there exist at least several topics needed to be
discussed in order to research the integrability property of
nonlinear $q$-partial dif\/ferential equations. For instance, the
Hamiltonian structure the $q$-cKP hierarchy and its $q$-W-algebra;
the gauge transformation of the $q$-cKP hierarchy; the $q$-Hirota
equation associated with the bilinear identity of the $q$-KP
hierarchy; the symmetry analysis of $q$-dif\/ferential equation
and $q$-partial dif\/ferential equations; the interaction of
$q$-solitons; the $q$-AKNS hierarchy and its properties. Since the
KP hierarchy has $B$-type and $C$-type sub-hierarchies,  what are
$q$-analogues of them? In particular, we  showed in the previous
sections that convergence of $e_q(x)$ af\/fects the $q$-soliton,
so the analytic property of $e_q(x)$ is a basis for research the
interaction of $q$-solitons. We will try to investigate these
questions in the future.

\subsection*{Acknowledgements}
   This work is supported partly by the 973
project of China for ``Nonlinear Science'',  the National Natural
Science Foundation of China (10301030) and SRFDP of China. The
author (Jingsong He) would like to thank the Centre for
Scientif\/ic Computing and University of Warwick for supporting
him to visit there. Special thanks go to Dr.~Rudolf A.~R\"omer at
Warwick for the numerous helpful discussions, Dr. P.~Iliev for
explaining few parts of his work~\cite{ilievb} and Professor
J.~Mas for answering questions on his paper~\cite{ms}.  Jingsong
He is also grateful to Professors F.~Calogero, A.~Degasperis,
D.~Levi and P.M.~Santini of University of Rome ``La Sapienza'' for
their hospitality during his visit to Rome. We thank anonymous
referees very much  for valuable suggestions and corrections.

\appendix

\section[More explicit expressions of $\partial_q^n\circ f$]{More explicit expressions of $\boldsymbol{\p_q^n\circ f}$}

For $n\geq 1$, we have
\begin{gather*}
\p_q^4\circ f=(\p^4_q f)+(4)_q\theta(\p^3
f)\p_q+\dfrac{(4)_q(3)_q}{(2)_q}\theta^2(\p^2_q
f)\p_q^2+(4)_q\theta^3(\p_q f)\p_q^3+ \theta^4(f)\p_q^4,\\
\p_q^5\circ f=(\p^5_q f)+(5)_q\theta(\p^4_q f)\p_q+
\dfrac{(5)_q(4)_q}{(2)_q}\theta^2(\p^3_q f)\p_q^2 +
\dfrac{(5)_q(4)_q}{(2)_q}\theta^3(\p^2_q f)\p_q^3 \notag
\\
\phantom{\p_q^5\circ f=}{} +(5)_q\theta^4(\p_q f)\p^4_q+
\theta^5(f)\p_q^5.
\end{gather*}
On the other hand, several examples of an explicit expression for
$\p_q^{-n}\circ f$ $(n\geq 1)$ are
\begin{gather*}
\p^{-3}_q\circ
f=\theta^{-3}(f)\p_q^{-3}-\dfrac{(3)_q}{q^3}\theta^{-4}(\p_q
 f)\p_q^{-4}+\dfrac{(3)_q(4)_q}{(2)_q q^{3+4}}\theta^{-5}(\p^2_q
 f)\p_q^{-5}\notag \\
\phantom{\p^{-3}_q\circ  f=}{}
-\dfrac{(4)_q(5)_q}{q^{3+4+5}(2)_q}\theta^{-6}(\p_q^3 f)\p_q^{-6}
+ \dfrac{(5)_q(6)_q}{q^{3+4+5+6}(2)_q}\theta^{-7}(\p_q^4
f)\p_q^{-7}+\cdots \notag \\
\phantom{\p^{-3}_q\circ  f=}{}+
\dfrac{(-1)^k(k+1)_q(k+2)_q}{q^{3+4+5+\cdots+(k+1)+
(k+2)}(2)_q}\theta^{-3-k}(\p_q^k f)\p_q^{-3-k}+\cdots,\\
\p_q^{-4}\circ f=
\theta^{-4}(f)\p_q^{-4}-\dfrac{(4)_q}{q^4}\theta^{-5}(\p_q
f)\p_q^{-5}+\dfrac{(4)_q(5)_q}{q^{4+5}(2)_q}\theta^{-6}(\p_q^2
f)\p^{-6}_q \notag \\
\phantom{\p_q^{-4}\circ f=}{}
-\dfrac{(4)_q(5)_q(6)_q}{q^{4+5+6}(2)_q(3)_q}\theta^{-7}(\p_q^3
f)\p^{-7}_q+\dfrac{(5)_q(6)_q(7)_q}{q^{4+5+6+7}(2)_q(3)_q}\theta^{-8}(\p_q^4
f)\p^{-8}_q+\cdots\notag \\
\phantom{\p_q^{-4}\circ
f=}{}+\dfrac{(-1)^k(k+1)_q(k+2)_q(k+3)_q}{q^{4+5+6+\cdots+(k+2)+(k+3)}(2)_q(3)_q}
\theta^{-4-k}(\p_q^k f)\p^{-4-k}_q+\cdots,\\
\p_q^{-5}\circ
f=\theta^{-5}(f)\p_q^{-5}-\dfrac{(5)_q}{q^5}\theta^{-6}(\p_q
f)\p_q^{-6}+\dfrac{(5)_q(6)_q}{q^{5+6}(2)_q}\theta^{-7}(\p_q^2
f)\p^{-7}_q \notag \\
\phantom{\p_q^{-5}\circ f=}{}
-\dfrac{(5)_q(6)_q(7)_q}{q^{5+6+7}(2)_q(3)_q}\theta^{-8}(\p_q^3
f)\p^{-8}_q+\dfrac{(5)_q(6)_q(7)_q(8)_q}{q^{5+6+7+8}(2)_q(3)_q(4)_q}\theta^{-9}(\p_q^4
f)\p^{-9}_q+\cdots\notag \\
\phantom{\p_q^{-5}\circ
f=}{}+\dfrac{(-1)^k(k+1)_q(k+2)_q(k+3)_q(k+4)_q}{q^{5+6+7+\cdots+(k+3)+(k+4)}(2)_q(3)_q(4)_q}
\theta^{-5-k}(\p_q^k f)\p^{-5-k}_q+\cdots.
\end{gather*}


\section[Positive part of $L^n$ $(n=1,2,3)$]{Positive part of $\boldsymbol{L^n}$ $\boldsymbol{(n=1,2,3)}$}

The f\/irst few of $B_n$ are in the form of
\begin{gather*}
B_1=\p_q+ u_0, \qquad 
B_2=\p^2_q + v_1\p_q+ v_0,\qquad 
B_3=\p^3_q + \tilde{s}_2\p^2_q+ \tilde{s}_1\p_q+ \tilde{s}_0,
\end{gather*}
where
\begin{gather*}
v_1=\theta(u_0)+u_0, \qquad 
v_0= (\p_q u_0)+\theta(u_{-1})+u_0^2+u_{-1},\\ 
v_{-1}=(\p_q u_{-1})+\theta(u_{-2})+u_0u_{-1}+u_{-1}\theta^{-1}(u_0)+u_{-2},
\end{gather*}
and
\begin{gather*}
 \tilde{s}_2=\theta(v_1)+ u_0, \qquad 
 \tilde{s}_1=(\p_q v_1) + \theta(v_0)+ u_0v_1+u_{-1}, \\ 
 \tilde{s}_0=(\p_q v_0) + \theta(v_{-1}) + u_0v_0+u_{-1}\theta^{-1}(v_{1})+u_{-2}. 
\end{gather*}
Note that $v_{-1}$ comes from $L^2=B_2+ v_{-1}\p^{-1}_q+
v_{-2}\p^{-2}_q+\cdots$.

\section[Proof of the $q$-Wronskian identity]{Proof of the $\boldsymbol{q}$-Wronskian identity}

1) The f\/irst $N$ steps. Consider the gauge transformation
generated by the order of $\{\phi_i,\, i=1, 2, \ldots,N\}$
\[
\fun{T}{1}{D}(\phi_1)\longrightarrow
\fun{T}{2}{D}\big(\fun{\phi}{1}{2}\big)\longrightarrow
\cdots\longrightarrow
\fun{T}{i}{D}\big(\fun{\phi}{i-1}{i}\big)\longrightarrow \cdots
\longrightarrow \fun{T}{N}{D}\big(\fun{\phi}{N-1}{N}\big).
\]
 Assume there are
 $l$ functions $\{ \fun{\phi}{N}{N+j}, \, j=1, 2, \ldots, l\}$
 expressed by
\begin{gather*}
\fun{\phi}{N}{N+j}=(T_N\cdot\phi_{N+j})=\dfrac{W_{N+1}^q(\phi_{1},\phi_{2},
\ldots,\phi_{N},\phi_{N+j} )} {W_{N}^q(\phi_{1},\phi_{2}, \ldots,
\phi_{N})},
\end{gather*}
which are generated by $T_{N}$ from $\{\phi_{j} \}$. Here
${\phi_i}$ $(i=1,2,\ldots,N+l)$ are arbitrary functions such that
the gauge transformations\ can be constructed.

2) The last  $l-1$ steps. Let $y_j=\fun{\phi}{N}{N+j}$
$(j=1,2,\ldots,l)$.
 Using  $y_j $ $(j=1,2,\ldots,l-1)$ as the generating functions in order of $T_D$, we can construct
$(l-1)$ steps of gauge transformation operators as
\[
\fun{T}{1}{D}(y_1)\longrightarrow
\fun{T}{2}{D}\big(\fun{y}{1}{2}\big)\longrightarrow
\fun{T}{3}{D}\big(\fun{y}{2}{3}\big)\longrightarrow \cdots
\longrightarrow \fun{T}{l-1}{D}\big(\fun{y}{l-2}{l-1}\big).
\]
According to the determinant of $T_N|_{N=j}$ $(j=1,2,\ldots,l-1)$,
we have
\begin{gather*}
\fun{y}{j}{i} = (T_j\cdot y_i)\left\{ \begin{array}{ll}
    0 & \text{if  $j \geq i$},\vspace{1mm}\\
    \dfrac{W^q_{j+1}(y_{1},y_2,\ldots,
     y_{j}, y_i ) }
    {W^q_j (y_1,y_2,\ldots,
    y_j) }     & \text{if  $j < i $,}
    \end{array}\right.
\end{gather*}
then
\begin{gather}
 y_{1}\cdot \fun{y}{1}{2}\cdot
 \fun{y}{2}{3}\cdots \fun{y}{l-2}{l-1}\fun{y}{l-1}{l}
 =y_1
 \dfrac{W^q_2(y_1,y_2)}
       {W^q_1(y_1)}  \dfrac{W^q_3(y_1,y_2,y_3)}
       {W^q_2(y_1,y_2)}\cdots \nonumber \\
\qquad{} \dfrac{W^q_{l-1}(y_1,y_2,\ldots, y_{l-2},y_{l-1})}
       {W^q_{l-2}(y_1,y_2,\ldots, y_{l-2})}
\dfrac{W^q_l(y_1,y_2,\ldots, y_{l-2},y_{l-1},y_l)}
       {W^q_{l-1}(y_1,y_2,\ldots, y_{l-2},y_{l-1})} \nonumber \\
\qquad{} =W^q_l(y_1,y_2,\ldots,y_l)
=W^q_l\big(\fun{\phi}{N}{N+1},\fun{\phi}{N}{N+2} ,
         \ldots,\fun{\phi}{N}{N+l} \big).
\label{b,1}
\end{gather}

3) Combine two chains of gauge transformations above. In fact, we
can combine two chains into one,
\begin{gather*}
\fun{T}{1}{D}(\phi_1)\longrightarrow
\fun{T}{2}{D}\big(\fun{\phi}{1}{2}\big)\longrightarrow
\cdots\longrightarrow
\fun{T}{i}{D}\big(\fun{\phi}{i-1}{i}\big)\longrightarrow \cdots
\longrightarrow \fun{T}{N}{D}\big(\fun{\phi}{N-1}{N}\big),\nonumber \\
\fun{T}{N+1}{D}\big(\fun{\phi}{N}{N+1}\big)\longrightarrow
\fun{T}{N+2}{D}\big(\fun{\phi}{N+1}{N+2}\big)\longrightarrow
\fun{T}{N+3}{D}\big(\fun{\phi}{N+2}{N+3}\big)\longrightarrow
\cdots \longrightarrow
\fun{T}{N+l-1}{D}\big(\fun{\phi}{N+l-2}{N+l-1}\big).
\end{gather*}
The determinant representation of $T_{N}|_{N+j}$ implies ($1<i$,
$j<l$):
\begin{gather*}
\fun{\phi}{N+j}{N+i}=(T_{N+j}\cdot \phi_{N+i})
=\left\{\begin{array}{ll}
  0&  \text{if $j\geq i$},\vspace{1mm}\\
  \dfrac{W^q_{N+j+1}(\phi_{1},\phi_{2} \cdots,\phi_{N},
        \phi_{N+1},\cdots,\phi_{N+j},
        \phi_{N+i}  )}
        {W^q_{N+j}( \phi_{1},\phi_{2} \cdots,\phi_{N},
        \phi_{N+1},\cdots,\phi_{N+j})}  & \text{if  $j<  i$}.
             \end{array}\right.
\end{gather*}
So
\begin{gather}
 \fun{\phi}{N}{N+1}\cdot \fun{\phi}{N+1}{N+2}\cdot
 \fun{\phi}{N+2}{N+3}\cdots \fun{\phi}{N+l-2}{N+l-1}\fun{\phi}{N+l-1}{N+l}
 =\dfrac{W_{N+1}^q(\phi_1,\phi_2,\ldots,\phi_N,\phi_{N+1})}{W_{N}^q(\phi_1,\phi_2,\ldots,\phi_N)}\times
 \nonumber\\
\qquad{}
\times\dfrac{W_{N+2}^q(\phi_1,\phi_2,\ldots,\phi_{N+1},\phi_{N+2})}{W_{N+1}^q(\phi_1,\phi_2,\ldots,\phi_{N+1})}
\dfrac{W_{N+3}^q(\phi_1,\phi_2,\ldots,\phi_{N+2},\phi_{N+3})}{W_{N+2}^q(\phi_1,\phi_2,\ldots,\phi_{N+2})}
\cdots \nonumber \\
\qquad{} \times
\dfrac{W_{N+l-1}^q(\phi_1,\phi_2,\ldots,\phi_{N+l-2},\phi_{N+l-1})}{W_{N+l-2}^q(\phi_1,\phi_2,\ldots,\phi_{N+l-2})}
\dfrac{W_{N+l}^q(\phi_1,\phi_2,\ldots,\phi_{N+l-1},\phi_{N+l})}{W_{N+l-1}^q(\phi_1,\phi_2,\ldots,\phi_{N+l-1})}
\cdots \nonumber  \\
\qquad{}  =
\dfrac{W_{N+l}^q(\phi_1,\phi_2,\ldots,\phi_{N+l-1},\phi_{N+l})}{W_{N}^q(\phi_1,\phi_2,\ldots,\phi_N)}.
\label{b,2}
\end{gather}
The left hand side of \eqref{b,1} equals  the left hand side of
\eqref{b,2}, which is followed by
\begin{gather*}
\dfrac{W_{N+l}^q(\phi_1,\phi_2,\ldots,\phi_N,\phi_{N+1},\ldots,
\phi_{N+l-1},\phi_{N+l})}{W_{N}^q(\phi_1,\phi_2,\ldots,\phi_N)}
=W^q_l\big(\fun{\phi}{N}{N+1},\fun{\phi}{N}{N+2} ,
         \ldots,\fun{\phi}{N}{N+l}\big).
\end{gather*}
It should be noted that the proof above is independent of the form
of $\phi_k$, so we can replace $\phi_{N+j}$ with
$(\partial_q^l\phi_{N+j})$.  This completes the proof of the
$q$-Wronskian identity.

\LastPageEnding


\begin{thebibliography}{99}
\footnotesize


\bibitem{ks} Klimyk A., Schm\"udgen K., $q$-calculus, in  Quantum Groups and Their
Represntaions, Berlin, Springer, 1997, Chapter 2, 37--52.

\bibitem{kc} Kac V., Cheung P.,  Quantum calculus, New York, Springer-Verlag, 2002.

\bibitem{exton} Exton H., $q$-hypergeometric functions and
applications, Chichester, Ellis Horwood Ltd., 1983.

\bibitem{andrews} Andrews G.E., $q$-series: their development and
application in analysis, number theory, combinatorics, physics,
and computer algebra, Providence, American Mathematical Society,
1986.

\bibitem{jimbo} Jimbo M., Yang--Baxter equation in integrable systems,
{\it Advanced Series in Mathematical Physics}, Vol.~10, Singapore,
World Scientif\/ic, 1990.

\bibitem{connes} Connes A., Noncommutative geometry,
San Diego~-- London, Academic Press, 1994.

\bibitem{majid1} Majid S.,  Free braided dif\/ferential calculus, braided
binomial theorem, and the braided exponential map, {\it  J.~Math.
Phys.}, 1993, V.34, 4843--4856, hep-th/9302076.

\bibitem{majid2} Majid S., Foundations of quantum group theory,
Cambridge, Cambridge University Press, 1995, \S~10.4.


\bibitem{zhang} Zhang D.H., Quantum deformation of KdV hierarchies and their inf\/initely many conservation laws,
{\it J.~Phys.~A: Math. Gen.}, 1993, V.26, 2389--2407.

\bibitem{wzz} Wu Z.Y., Zhang D.H., Zheng Q.R.,  Quantum deformation of KdV hierarchies and their exact solutions:
$q$-deformed solitons, {\it  J. Phys. A: Math. Gen.}, 1994, V.27,
5307--5312.

\bibitem{fr} Frenkel E., Reshetikhin N.,  Quantum af\/f\/ine algebras and deformations of the Virasoro and $W$-algebras,
{\it Comm. Math. Phys.}, 1996, V.178, 237--264, q-alg/9505025.

\bibitem{frenkel} Frenkel E., Deformations of the KdV hierarchy and related soliton equations,
{\it Int. Math. Res. Not.}, 1996, V.2, 55--76, q-alg/9511003.

\bibitem{hl} Haine L., Iliev P., The bispectral property of a $q$-deformation of the Schur polynomials and the $q$-KdV
hierarchy, {\it J. Phys. A: Math. Gen.}, 1997, V.30, 7217--7227.

\bibitem{ahv} Adler M., Horozov E., van Moerbeke P.,  The
solution to the $q$-KdV equation, {\it  Phys. Lett. A}, 1998,
V.242, 139--151, solv-int/9712015.

\bibitem{tsl} Tu M.H., Shaw J.C., Lee C.R.,  On
Darboux--B\"acklund transformations for the $q$-deformed
Korteweg--de Vries hierarchy, {\it   Lett. Math. Phys.}, 1999,
V.49, 33--45, solv-int/9811004.

\bibitem{tl} Tu M.H.,  Shaw J.C.,  Lee C.R.,  On
the $q$-deformed modif\/ied Korteweg--de Vries hierarchy, {\it
Phys. Lett. A}, 2000, V.266, 155--159.


\bibitem{klr}  Khesin B., Lyubashenko V., Roger C., Extensions and contractions of the Lie algebra of
$q$-pseudodif\/ferential symbols on the circle, {\it   J. Funct.
Anal.}, 1997, V.143, 55--97, hep-th/9403189.

\bibitem{ms} Mas J., Seco M., The algebra of $q$-pseudodif\/ferential symbols and the $q$-$W_{\rm KP}^{(n)}$ algebra,
{\it J. Math. Phys.}, 1996, V.37, 6510--6529, q-alg/9512025.

\bibitem{ilieva} Iliev P., Solutions to Frenkel's deformation of the KP hierarchy,
{\it J. Phys. A: Math. Gen.}, 1998, V.31, L241--L244.

\bibitem{ilievb} Iliev P., Tau function solutions to a $q$-deformation of the KP hierarchy,
{\it Lett. Math. Phys.}, 1998, V.44, 187--200.

\bibitem{ilievc} Iliev P., $q$-KP hierarchy, bispectrality and
Calogero--Moser systems, {\it  J. Geom. Phys.}, 2000, V.35,
157--182.

\bibitem{tu} Tu M.H., $q$-deformed KP hierarchy: its additional
symmetries and inf\/initesimal B\"acklund transformations,  {\it
Lett. Math. Phys.}, 1999, V.49,  95--103, solv-int/9811010.


\bibitem{wwwy} Wang S.K., Wu K., Wu X.N., Wu D.L., The $q$-deformation of AKNS-D hierarchy,
{\it J. Phys. A: Math. Gen.}, 2001, V.34, 9641--9651.


\bibitem{hlc2} He J.S., Li Y.H., Cheng Y.,  $q$-deformed
Gelfand--Dickey hierarchy and the determinant representation of
its gauge transformation, {\it  Chinese Ann. Math. Ser. A}, 2004,
V.25, 373--382 (in Chinese).

\bibitem{kss} Konopelchenko B.G., Sidorenko J., Strampp W.,
 $(1+1)$-dimensional integrable systems as
 symmetry constraints of $(2+1)$-dimensional systems, {\it Phys. Lett.~A},
 1991, V.157, 17--21.

\bibitem{cl2} Cheng Y., Li Y.S., The constraint of the Kadomtsev--Petviashvili equation
and its special solutions, {\it  Phys. Lett.~A}, 1991, V.157,
22--26.

\bibitem{os1} Oevel W., Strampp W., Constrained KP hierarchy and bi-Hamiltonian structures,
{\it  Comm. Math. Phys.}, 1993, V.157, 51--81.

\bibitem{cy1} Cheng Y., Constraints of the Kadomtsev--Petviashvili hierarchy, {\it J. Math. Phys.}, 1992,
V.33, 3774--3782.

\bibitem{cy2} Cheng Y., Modifying the KP, the $n$th constrained KP hierarchies and their
    Hamiltonian structures, {\it  Comm. Math. Phys.}, 1995, V.171,
    661--682.

\bibitem{aratyn1} Aratyn H., Ferreira L.A., Gomes J.F., Zimerman A.H.,
 Constrained KP models as integrable matrix hie\-rar\-chies, {\it J. Math. Phys.}, 1997, V.38, 1559--1576, hep-th/9509096.

\bibitem{aratyn2} Aratyn H., On Grassmannian description of the constrained KP
hierarchy, {\it J. Geom. Phys.}, 1999, V.30, 295--312,
solv-int/9805006.


\bibitem{csy} Chau L.L., Shaw J.C., Yen H.C.,  Solving the KP
hierarchy by gauge transformations, {\it Comm. Math. Phys.}, 1992,
V.149, 263--278.

\bibitem{og} Oevel W., Rogers C., Gauge transformations and
reciprocal links in $2+1$ dimensions, {\it Rev. Math. Phys.},
1993, V.5, 299--330.

\bibitem{hlc} He J.S., Li Y.S., Cheng Y., The determinant representation of the gauge transformation
operators,  {\it Chinese Ann. Math. Ser. B}, 2002, V.23, 475--486.

\bibitem{hlc1} He J.S., Li Y.S., Cheng Y., Two choices of the gauge transformation for the
AKNS hierarchy through the constrained KP hierarchy, {\it J. Math.
Phys.}, 2003, V.44, 3928--3960.

\bibitem{dkjm} Date E., Kashiwara M., Jimbo M., Miwa T.,
Transformation group for soliton equations, in Bosonization,
Editor  M.~Stone, Singapore, World Scientif\/ic, 1994, 427--507.

\bibitem{dl1} Dickey L.A., Soliton equations and Hamiltonian systems,
 Singapore, World Scientif\/ic, 1991.

\bibitem{os} Oevel W., Strampp W., Wronskian solutions of the constrained Kadomtsev--Petviashvili
       hierarchy, {\it J. Math. Phys.}, 1996, V.37, 6213--6219.

\bibitem{ostt} Ohta Y., Satsuma J., Takahashi D., Tokihiro T., An elementary introduction to
Sato theory, {\it Progr. Theoret. Phys. Suppl.}, 1988, N~94,
210--241.

\bibitem{ac} Ablowitz M.J., Clarkson P.A., Solitons, nonlinear
evolution equations and inverse scattering, Cambridge, Cambridge
University Press, 1991.


\end{thebibliography}
\end{document}